\documentclass{aa}
\usepackage{graphicx}
\usepackage{natbib}
\usepackage{amsmath}
\usepackage{txfonts}
\usepackage{color}
\usepackage{lscape}
\usepackage{xspace}
\usepackage{longtable}
\def\ein{{\it Einstein}}
\def\asca{{\it ASCA}}
\def\ros{{\it ROSAT}}
\def\chandra{{\it Chandra}}
\def\sax{{{\it Beppo}SAX}}
\def\xmm{{\it XMM-Newton}}
\def\n253{NGC~253}
\def\m82{M82}
\def\la{\mathrel{\hbox{\rlap{\hbox{\lower4pt\hbox{$\sim$}}}\hbox{$<$}}}}
\def\ga{\mathrel{\hbox{\rlap{\hbox{\lower4pt\hbox{$\sim$}}}\hbox{$>$}}}}
\def\subsun{\mbox{$_{\odot}$}}
\def\deg{\hbox{$^\circ$}}
\def\ha{H$\alpha$}
\newcommand{\ergs}[1]{$\times10^{#1}$~\hbox{erg~s$^{-1}$}}
\newcommand{\oergs}[1]{$10^{#1}$~erg~s$^{-1}$}
\newcommand{\hcm}[1]{$\times10^{#1}$~cm$^{-2}$}

\newcommand{\oexpo}[1]{$10^{#1}$}

\def\cm-2{cm$^{-2}$}

\newcommand{\nh}{\hbox{$N_{\rm H}$}}

\newcommand{\ct}{ct~s$^{-1}$}

\newcommand{\ie}{i.e.\xspace}
\newcommand{\eg}{e.g.\xspace}
\newcommand{\colblue}[1]{#1}
\newcommand{\comment}[1]{}
\def\specwidth{6cm}
\newcommand{\fig}[1]{#1}
\newcommand{\bold}[1]{#1}

\begin{document}
   \title{XMM-Newton observations of the diffuse X-ray emission in the starburst galaxy \n253\thanks{Based on observations obtained with XMM-Newton, an ESA science mission with instruments and contributions directly funded by ESA Member States and NASA}}
%   \titlerunning{}

   \author{M.~Bauer\inst{1}, W.~Pietsch\inst{1}, G.~Trinchieri\inst{2}, D.~Breitschwerdt\inst{3}, M.~Ehle\inst{4}, M.J.~Freyberg\inst{1} \and  A.\bold{\colblue{M.}}~Read\inst{5} 
          }
   \authorrunning{M. Bauer et al.}

   \offprints{M. Bauer, \email{mbauer@mpe.mpg.de}}

   \institute{Max-Planck-Institut f\"ur extraterrestrische Physik, Giessenbachstra\ss e, 85741 Garching, Germany
   \and
   INAF Osservatorio Astronomico di Brera, via Brera 28, 20121 Milano, Italy
   \and
   Institut f\"ur Astronomie der Universit\"at Wien, T\"urkenschanzstr. 17, A-1180 Wien, Austria
   \and
   XMM-Newton Science Operations Centre, ESA, P.O.\ Box 78, 28691 Villanueva de la Ca\~nada, Madrid, Spain
   \and
   Department of Physics \& Astronomy, University of Leicester, Leicester LE1 7RH, UK
    }

   \date{Received 29 October 2007; accepted 15 July 2008.}

% \abstract{}{}{}{}{}
% 5 {} token are mandatory

  \abstract
   {}
   {We present a study of the diffuse X-ray emission in the halo and the disc of the starburst galaxy \n253.
   }
   {After removing point\bold{\colblue{-like}} sources, we analysed \xmm\ images, hardness ratio maps and spectra from several regions in the halo and the disc.
   We introduce a method to produce vignetting corrected images from the EPIC~pn data, and we developed a procedure that allows a correct background treatment for low surface brightness spectra, using a local background, together with closed filter observations.
   }
  {Most of the emission from the halo is at energies below 1~keV. 
   In the disc, also emission at higher energies is present.
   The extent of the diffuse emission along the major axis of the disc is 13.6~kpc.
   The halo resembles a horn structure and reaches out to $\sim$9~kpc perpendicular to the disc.
   Disc regions that cover star forming regions, like spiral arms, show harder spectra than regions with lower star forming activity.
   Models for spectral fits of the disc regions need at least three components: two thermal plasmas with solar abundances plus a power law and galactic foreground absorption.
   Temperatures are between 0.1 and 0.3~keV and between 0.3 and \bold{0.9}~keV for the soft and the hard component, respectively.
   The power law component may indicate an unresolved contribution from X-ray binaries in the disc.
   The halo emission is not uniform, neither spatially nor spectrally.
   The southeastern halo is softer than the northwestern halo.
   To model the spectra in the halo, we needed two thermal plasmas with solar abundances plus galactic foreground absorption.
   Temperatures are around 0.1 and 0.3~keV.
   A comparison between X-ray and UV emission shows that both originate from the same regions.
   The UV emission is more extended in the southeastern halo, where it seems to form a shell around the X-ray emission.
   }
   {}


   \maketitle

%%%%%%%%%%%%%%%%%%%%%%%%%%%%%%%%%%%%%%%%%%%%%%%%%%%%%%%%%%%%%%%%%%%%%%%%%%%%%%%%%%%%%%%%%%%%%%%%%%%%%%%%%%%%%%%%%%%%%%
%%%%%%%%%%%%%%%%%%%%%%%%%%%%%%%%%%%%%%%%%%%%%%%%%%%%%%%%%%%%%%%%%%%%%%%%%%%%%%%%%%%%%%%%%%%%%%%%%%%%%%%%%%%%%%%%%%%%%%

\section{Introduction}
The diffuse X-ray emission of starburst galaxies can be quite prominent.
Especially in galaxies that we see edge-on, we can find very complex emission from galactic halos.
One famous example is the starburst galaxy \n253\ in the Sculptor Group.
It is close enough \citep[2.58~Mpc, 1\arcmin=750~pc,][]{PC1991} to resolve structures in the disc and halo, and to separate the detected point\bold{\colblue{-like}} sources from the diffuse emission.
Also, it is seen almost edge-on \citep[78.5\deg,][]{P1980}, so an unobscured analysis of the halo emission is possible.
\n253 has been observed in X-rays many times.
There are observations with \ein\ \citep[e.g.][]{FT1984}, \ros\ \citep[e.g.][]{P1992,RP1997,DW1998,VP1999, PV2000}, \asca\ \citep[e.g.][]{PS1997}, \sax\ \citep[e.g.][]{CP1999}, \xmm\ \citep[e.g.][]{PR2001, BP2007}, and \chandra\ \citep[e.g.][]{WH2002, SH2002, SH2004A, SH2004B}.
While with some instruments one was not able to separate emission from point sources and diffuse emission, other instruments, especially \ros, \xmm, and \chandra, do have a narrow enough point spread function to do so.
We here report on the first extensive analysis of the diffuse emission in \n253\ with \xmm.

%%%%%%%%%%%%%%%%%%%%%%%%%%%%%%%%%%%%%%%%%%%%%%%%%%%%%%%%%%%%%%%%%%%%%%%%%%%%%%%%%%%%%%%%%%%%%%%%%%%%%%%%%%%%%%%%%%%%%%
%%%%%%%%%%%%%%%%%%%%%%%%%%%%%%%%%%%%%%%%%%%%%%%%%%%%%%%%%%%%%%%%%%%%%%%%%%%%%%%%%%%%%%%%%%%%%%%%%%%%%%%%%%%%%%%%%%%%%%

\section{Observations and data reduction}

\n253 was observed with \xmm\ \citep{JL2001} during three orbits in June 2000 and June 2003, using all of the European Photon Imaging Camera (EPIC) instruments \citep{SBD2001,TA2001}, the two co-aligned RGS spectrometers, RGS1 and RGS2 \citep{HB2001}, and the Optical Monitor \citep{MB2001}, for a total of about 216~ks.
The revolution number, observation identifier, observing date, pointings and orientation of the satellite (P.A.), and the total exposure times for the EPIC~pn camera (T$_{\rm{exp}}$) are shown in Table~\ref{tab:Exposures}.
\begin{table*}
  \caption{\xmm\ EPIC~pn \n253\ observation log.}
  \label{tab:Exposures}
  \begin{center}
    \begin{tabular}{ccccccccccccc}
      \hline\noalign{\smallskip}
      \hline\noalign{\smallskip}
      \multicolumn{1}{c}{Nr.} & Revolution & \multicolumn{1}{c}{Obs. ID.} &\multicolumn{1}{c}{Obs. Dates} & \multicolumn{2}{c}{Pointing Direction} & \multicolumn{1}{c}{P. A.}& Filter & \multicolumn{1}{c}{T$_{\rm{exp}}$}& \multicolumn{1}{c}{T$_{\rm{exp,\ clean}}$}\\
      \noalign{\smallskip}
       & Number & & & \multicolumn{2}{c}{RA/DEC (J2000)} & \multicolumn{1}{c}{(deg)}& & \multicolumn{1}{c}{(ks)}& \multicolumn{1}{c}{(ks)}\\
      \noalign{\smallskip}
      \multicolumn{1}{c}{(1)} & \multicolumn{1}{c}{(2)} & \multicolumn{1}{c}{(3)} & \multicolumn{1}{c}{(4)} & \multicolumn{1}{c}{(5)} & \multicolumn{1}{c}{(6)} & (7) & \multicolumn{1}{c}{(8)} & \multicolumn{1}{c}{(9)} & (10)\\
      \noalign{\smallskip}
      \hline
      \noalign{\smallskip}
      1 & 89  & 0125960101 & 2000-06-03 & 00:47:36.74 & -25:17:49.2 & 56.9  & Medium & 60.8  & 24.3 \\
      2 & 89  & 0125960201 & 2000-06-04 & 00:47:36.57 & -25:17:48.7 & 57.0  & Thin   & 17.5  & 3.1  \\
      3 & 186 & 0110900101 & 2000-12-14 & 00:47:30.20 & -25:15:53.2 & 233.8 & Thin   & 24.4  & 4.4  \\
      4 & 646 & 0152020101 & 2003-06-19 & 00:47:36.89 & -25:17:57.3 & 53.8  & Thin   & 113.0 & 47.9 \\
      \hline
    \end{tabular}
  \end{center}
\end{table*}

Throughout the following analysis, we used the MOS data {\it only} to detect and remove point\bold{\colblue{-like}} sources.
We were especially interested in low-surface brightness diffuse emission at energies below 1~keV, where the MOS detectors have a lower sensitivity than the EPIC~pn.
By not utilising the MOS data for the analysis of the diffuse emission, we avoided a higher background noise level.

We analysed the data using the Science Analysis System ({\tt SAS}), version 7.0.0.
In a first step, we cleaned the EPIC~pn and MOS observations by excluding times, where the count rate over the whole detector above 10~keV exceeded the count rate during quiescent times.
This cleaning was done to avoid times with high particle background, which result in a higher background level. 
The observations showed no obvious additional times with solar wind charge exchange \citep[\eg][]{SC2004}, which would result in times with high background at energies below 1~keV, so no further exclusion of exposure time was necessary.
The exposure times after screening for high background (T$_{\rm{exp,\ clean}}$) are shown in Table~\ref{tab:Exposures}.
Summing up over the final good time intervals, we ended up with 80~ks of exposure time in total.
This means only about 37\% of the original exposure time could be used for the analysis presented in this paper.
This number is quite small, compared to typical exposure time fractions of usable times after screening of 60--70\%\footnote{see the \xmm\ EPIC Background Working Group webpage http://www.star.le.ac.uk/$\sim$amr30/BG/BGTable.html}.
Next, we screened for bad pixels that were not detected by the pipeline. 
In order to be able to merge images later on, we calculated sky coordinates ({\it X,Y}) for the events in all observations with respect to the centre reference position $\alpha_{2000}$=00$^{\rm h}$47$^{\rm m}$33\fs3, $\delta_{2000}$=-25\deg17\arcmin18\arcsec.
For the following analysis we split the data set into five energy bands: 0.2--0.5~keV, 0.5--1.0~keV, 1.0--2.0~keV, 2.0--4.5~keV and 4.5--12~keV as bands 1 to 5.

\begin{figure*}
\centering
\begin{minipage}[t]{17.4cm}
\centering
\fig{\frame{\includegraphics[width=8.6cm]{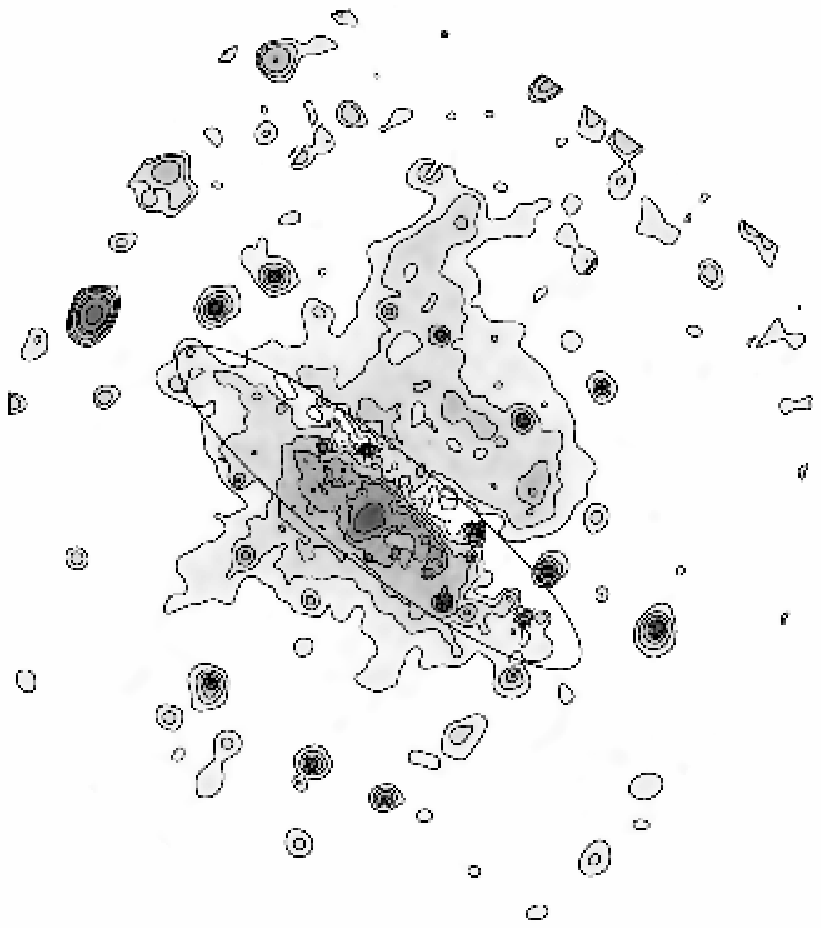}}}
\fig{\frame{\includegraphics[width=8.6cm]{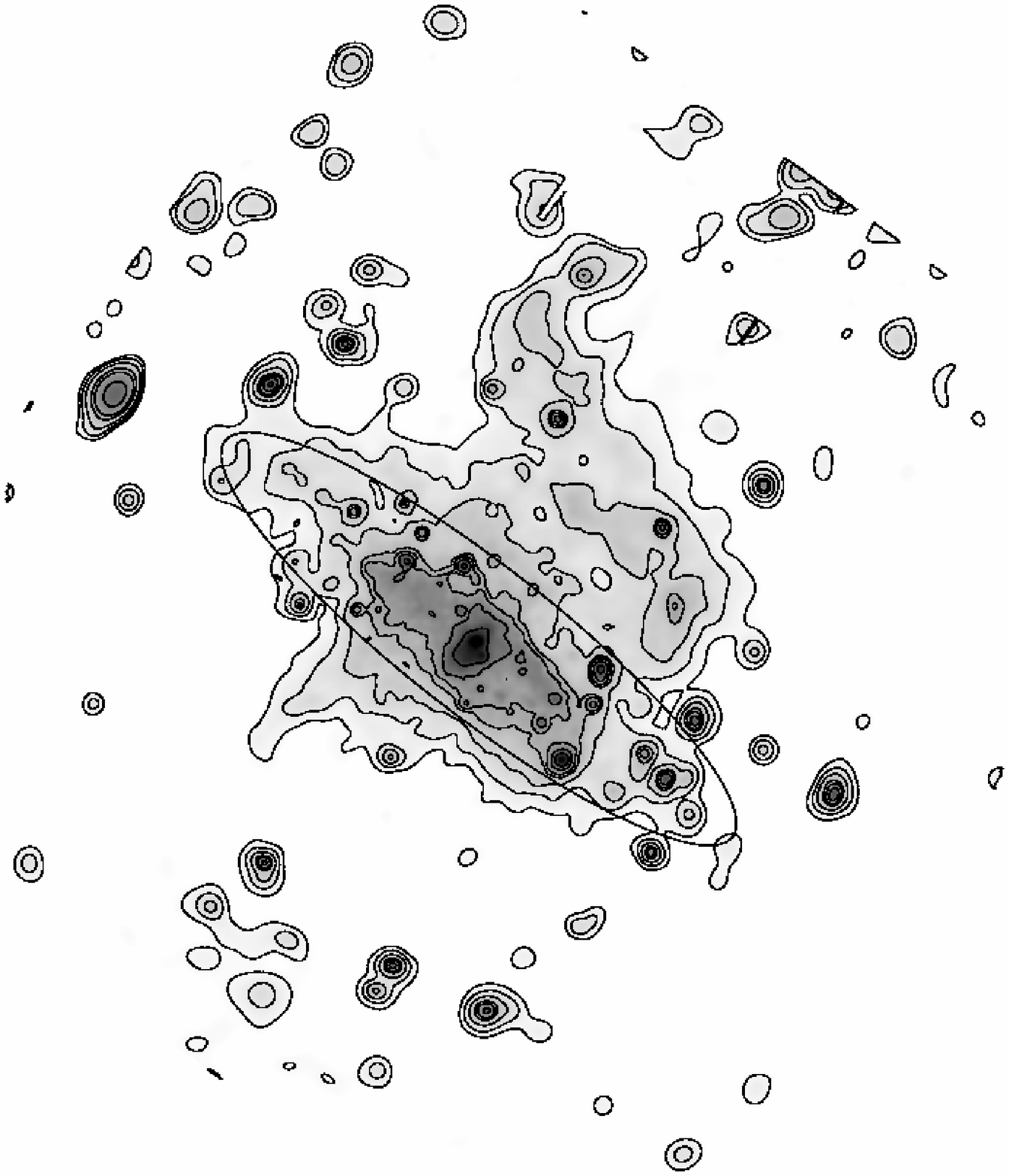}}}
\fig{\frame{\includegraphics[width=8.6cm]{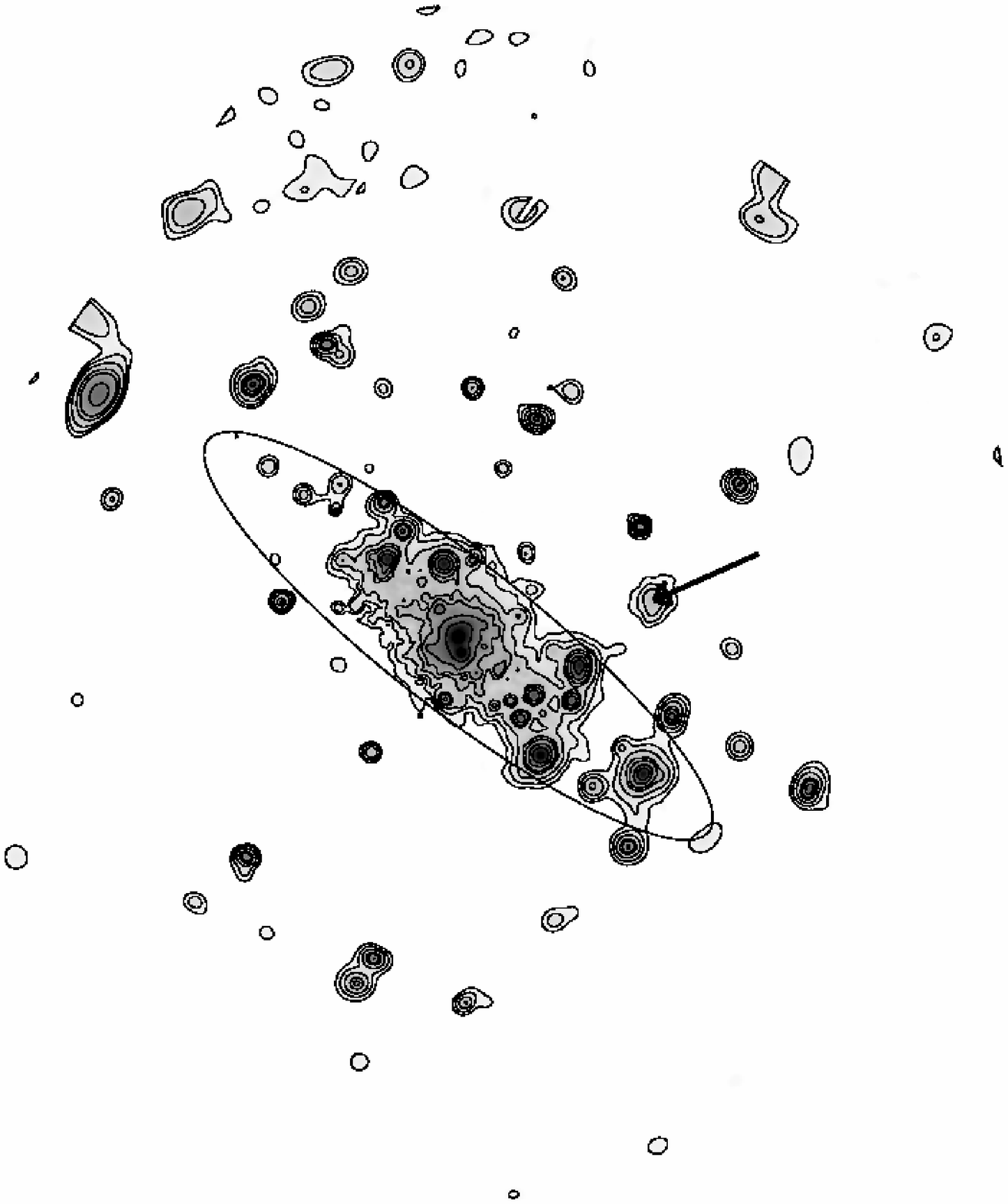}}}
\fig{\frame{\includegraphics[width=8.6cm]{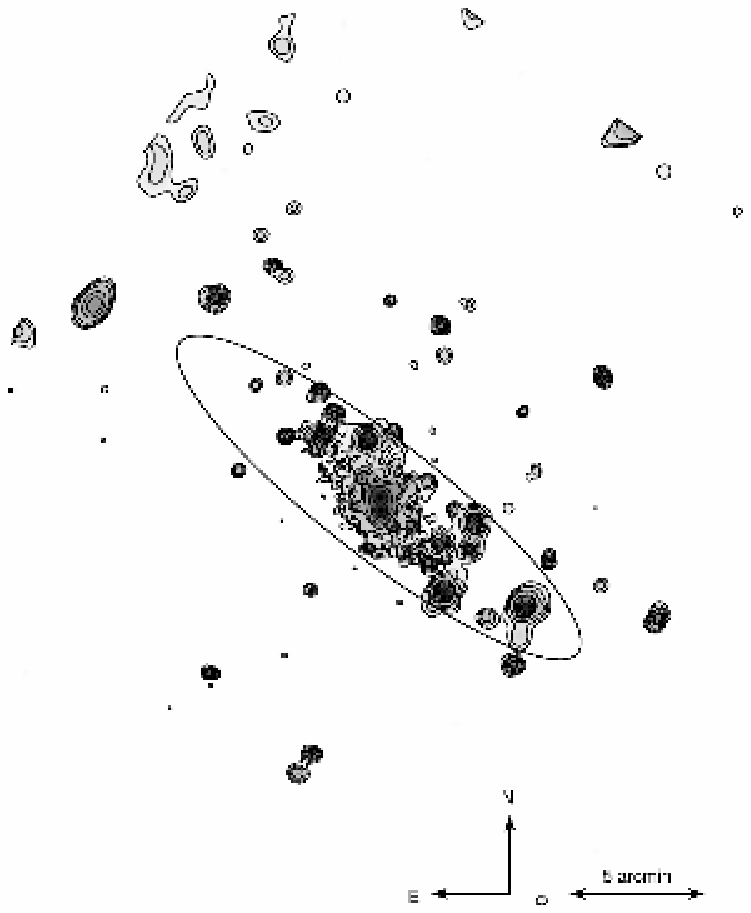}}}
\end{minipage}
\caption{
Adaptively smoothed EPIC~pn images with contours in the lower 4 energy bands: 
(top-left) 0.2--0.5~keV, (top-right) 0.5--1.0~keV, (bottom-left) 1.0--2.0~keV, and (bottom-right) 2.0--4.5~keV. 
Contours are at (0.35, 0.50, 0.80, 1.6, 2.5, 6.0, 20, 100) $\times$ $10^{-5}$~\ct~pix$^{-1}$. 
Additionally we show the inclination corrected optical $D_{25}$ ellipse in black. 
\bold{The black arrow in the lower left image indicates the position of a possible background galaxy cluster}.
}
\label{fig:PN_images}
\end{figure*}

\subsection{Point source removal}
In this paper we did not study the population of the point sources, but we focused on the diffuse emission in the halo and the disc of the galaxy.
To do so, we had to remove contributions from point\bold{\colblue{-like}} sources.
In order to run the source detection algorithm of the {\tt SAS}-software package, we created images for the EPIC~pn, selecting only single events (PATTERN=0) in energy band 1, and single and double events (PATTERN$\leq$4) for the other bands.
For MOS we used single to quadruple events (PATTERN$\le$12) in all bands.
To avoid differences in the background over the EPIC~pn detector, we omitted the energy range between 7.2~keV and 9.2~keV, where the detector background shows strong spatially variable fluorescence lines \citep{FB2004}.
We created images, background images and exposure maps, and masked them to an acceptable detector area.
The binning for all images is 2\arcsec.

\begin{figure*}
  \centering
  \begin{minipage}[t]{17.4cm}
  \fig{\frame{\includegraphics[width=8.8cm]{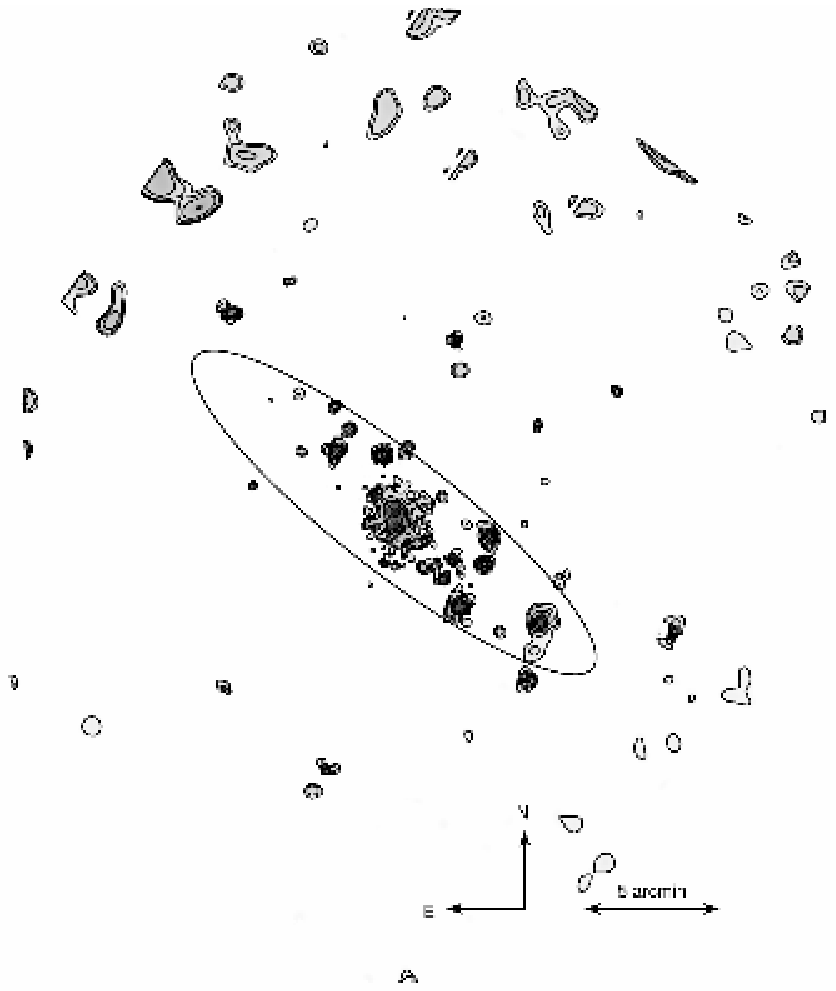}}}
  \fig{\frame{\includegraphics[width=8.8cm]{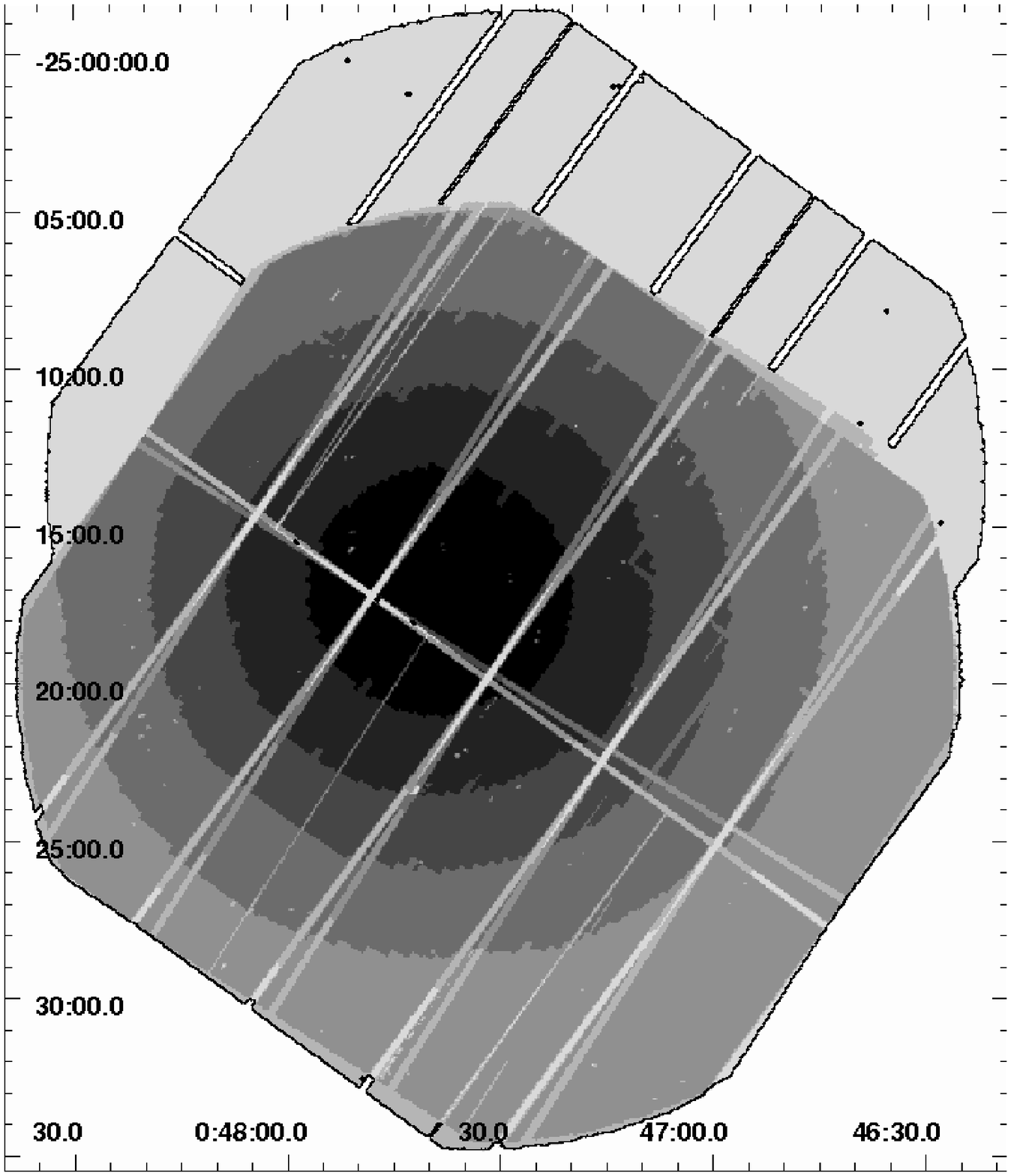}}}
  \end{minipage}
  \addtocounter{figure}{-1}
  \caption{continued. {\it (left):} Adaptively smoothed EPIC~pn image with contours in the highest different energy band   (4.5--12~keV). \comment{Contour levels are the same as in Fig.~\ref{fig:PN_images}.}
    {\it (right):} The vignetting corrected exposure map of the merged four observations. 
    The outer contour indicates 0~ks and the exposure increases linearly towards the centre by one seventh of the maximum (80~ks) per gray-scale level (0--11.4~ks, 11.4--22.9~ks, 22.9--34.3~ks, 34.3--45.7~ks, 45.7--57.1~ks, 57.1--68.6~ks, 68.6--80.0~ks). Except for a few pixels, all the detector gaps are covered by at least 4.4~ks in the central region.
  }
\end{figure*}

\bold{Using the {\tt SAS}-tasks {\tt eboxdetect}, version 4.19, and {\tt emldetect}, version 4.60}, we searched for point\bold{\colblue{-like}} sources in the field of view (FOV), simultaneously in the 5 energy bands and three detectors. 
First, we searched in each observation separately, to correct for inaccuracies in the pointing positions.
The resulting source lists were correlated to catalogues from USNO \citep{ML2003}, SIMBAD\footnote{http://simbad.u-strasbg.fr/simbad/}, and \chandra\ \citep{SH2002}.
Offsets were determined and applied to each observation.
With the position corrected event files, we again created images on which we executed the final point source detection.
We searched in the merged images from observations 1, 2, and 4, and separately in the images from observation 3.
The reason for merging only observations 1, 2 and, 4 is that observation 3 has a pointing offset ($\sim$6\arcmin) into the northwestern halo and therefore we would have different point spread functions on the same sky coordinates.

Additionally, we created a point source catalogue for the \chandra\ observations.
Point\bold{\colblue{-like}} sources in \chandra ObsID 3931 were identified using the Wavelet-Based detection Algorithm \citep[{\tt wavdetect} in the {\tt ciao} software, version 3.4,][]{FK2002}, in the 0.5--5.0~keV energy band using scales of 1\arcsec, 2\arcsec, 4\arcsec, 8\arcsec, and 16\arcsec.
For ObsID 969 and ObsID 790 we adopted the published source list from \cite{SH2002}.
The combined \xmm\ and \chandra\ source list will be published and further discussed in a forthcoming paper.

The combined source list was used to remove the point\bold{\colblue{-like}} sources from the data sets.
The {\tt SAS}-task {\tt region} was used to produce elliptical regions that approximate the PSF with an analytical model at a given detector position and flux value (0.5 times the background flux at this position).
Sources that were not detected in the \xmm\ data sets, but are known from \chandra\ observations, were excluded with a circular region with a diameter of 8\arcsec.
One might argue that these sources contribute only little to the overall emission.
However, we took up a conservative position and also excluded these sources to keep any unwanted interference at a minimum.

\begin{figure}
\centering
\fig{\includegraphics[width=6cm]{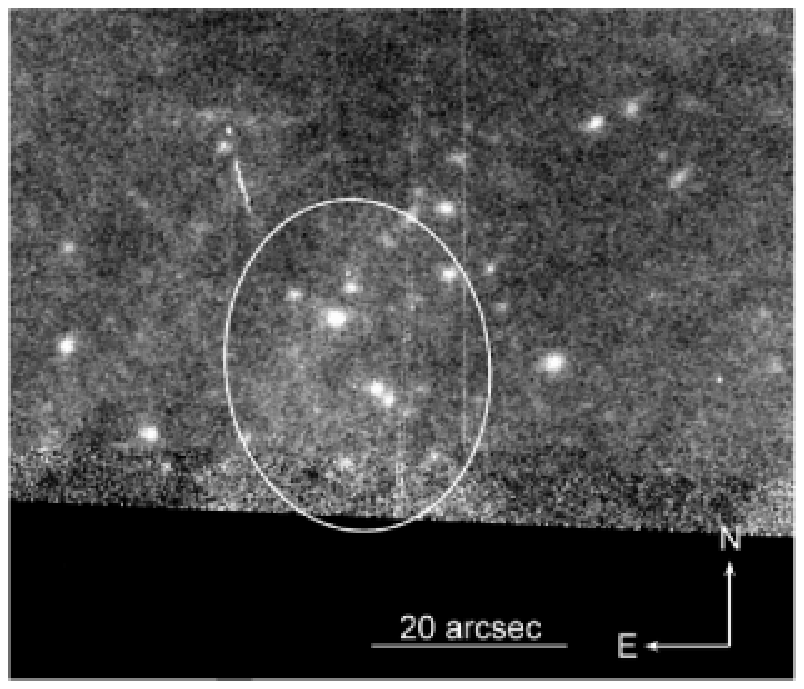}}
\caption
  {
  \bold{ESO/MPG 2.2m Wide Field Imager \colblue{(WFI)} I-band image of a possible galaxy cluster in the background of \n253.
  The white ellipse marks the source extent of the X-ray source as found in the source detection.
  The bottom quarter of the image \colblue{is not covered by the WFI detector}}.
  }
\label{fig:X10_image}
\end{figure}

\bold{\colblue{One of the sources in the XMM-Newton list }in the northwestern halo at $\alpha_{2000}$=00$^{\rm h}$47$^{\rm m}$07\fs7, $\delta_{2000}$=-25\deg16\arcmin21\arcsec\ \bold{\colblue{(indicated in Fig.~\ref{fig:PN_images}) was flagged as extended by {\tt emldetect}}}.
In a combined I-band image produced from several observations with the Wide Field Imager on the 2.2m ESO/MPG telescope between July 1999 and August 2000, we also find an enhancement of sources in this region (see Fig.~\ref{fig:X10_image}).
These findings indicate that this source could be a galaxy cluster in the background.
For the further analysis of \bold{\colblue{the \n253 diffuse emission the source was removed using a}} circular region of 1.5\arcmin\ diameter}.

\subsection{Images}
\begin{figure}
\fig{\includegraphics[width=88mm]{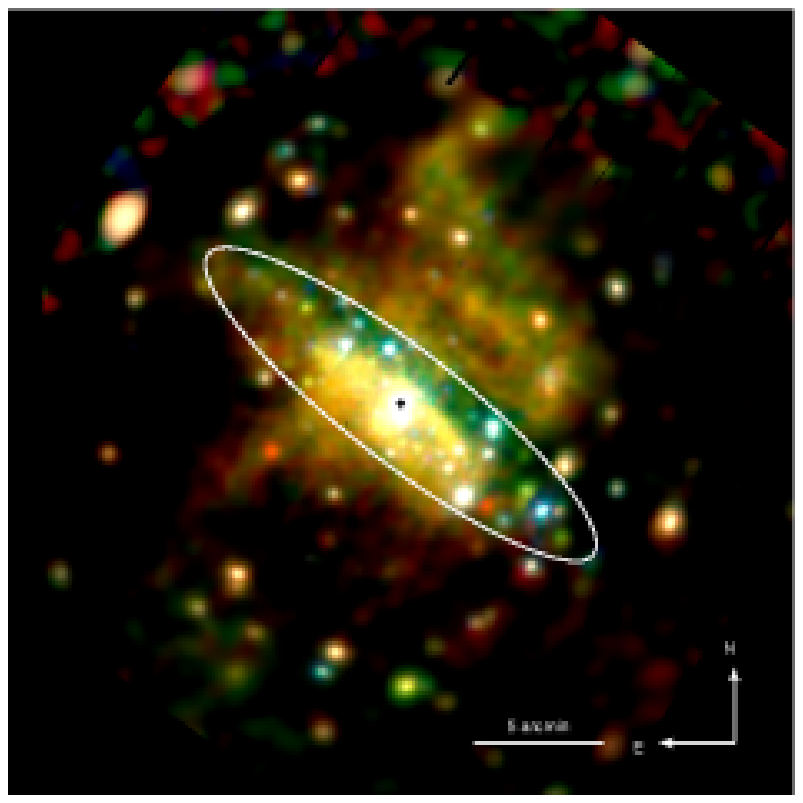}}
\caption{
Adaptively smoothed EPIC~pn image of \n253. 
The colours correspond to the energy bands (0.2--0.5~keV, red), (0.5--1.0~keV, green) , and (1.0--2.0~keV, blue). 
Overplotted in white is the inclination corrected optical $D_{25}$ ellipse of \n253. 
\bold{The centre of the galaxy is marked with a black cross}.
Scale and orientation are indicated.  
}
\label{fig:PN_RGB_image}
\end{figure}

\begin{figure}
  \fig{\includegraphics[width=88mm]{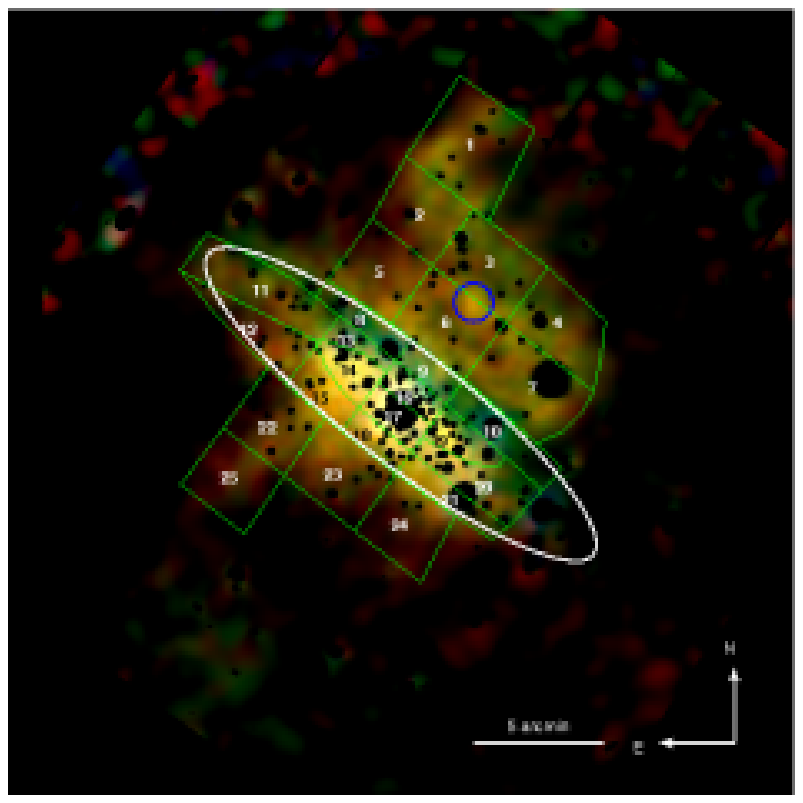}}
  \caption
  {
    Adaptively smoothed EPIC~pn image of the diffuse emission of \n253.
    Point\bold{\colblue{-like}} sources have been removed.
    Overplotted in green are the regions that were used for extracting hardness ratios and spectra.
    \bold{\colblue{The bright knot in the northwestern halo (see Sect.~\ref{sec:HaloDiffuseEmission}) is indicated by a blue circle.}\comment{Indicated in blue is the brighter knot in the northwestern halo.}}
    The inclination corrected optical $D_{25}$ ellipse is shown in white.
  }
  \label{fig:PN_cheese_image}
\end{figure}

We used all 4 observations to produce images.
The observations have different pointing directions and position angles, so we obtained images where almost all the CCD gaps are filled.
The single images, from the energy bands 1 to 5, were corrected for the detector background (electronic noise, high energy particles) by subtracting the surface brightness of the detector corners, that are outside of the field of view. 
\comment{\bold{To avoid background variability over the EPIC~pn images, we omitted the energy range 7.2--9.2~keV in band 5 where strong fluorescent lines cause higher background \citep{FB2004}}}
The images were exposure and vignetting corrected, and adaptively smoothed with a Gaussian kernel, with sizes between 10\arcsec\ and 47\arcsec\ (Fig.~\ref{fig:PN_images}).
For a detailed description of this method see App.~\ref{app:images}.
A false-colour image was produced by combining \bold{the images in} the three lowest energy bands 1, 2, and 3, as channels red, green, and blue, respectively (Fig.~\ref{fig:PN_RGB_image}).
The image, after the point source removal is shown in Fig.~\ref{fig:PN_cheese_image}.

\subsection{Hardness ratio maps and spectra}
As a big advantage, compared to the observations by \ros\ and \chandra, the higher count rates in \xmm\ allowed us to extract spectra with reasonable statistics from smaller regions in the disc and the halo.
For the hardness ratios and spectra we again restricted ourselves to the EPIC~pn data.
We did not use observations 2 and 3 for hardness ratios and spectra, because after good time interval screening only little exposure was left (cf.\ Table~\ref{tab:Exposures}).

Energy spectra of several regions (cf.\ Fig.~\ref{fig:PN_cheese_image}) were extracted from the event files, \bold{using the {\tt SAS}-task {\tt evselect}, version 3.60}, after removal of the point\bold{\colblue{-like}} sources.
To calculate the area of these regions, we used the task {\tt backscale}, version 1.4.2.
We produced background spectra using a region at the southwestern border of the FOV, together with observations where the filter wheel was closed.
A detailed description of this procedure, which also handles the binning of the spectra, can be found in Appendix~\ref{app:bkgspec}.
The final, background subtracted source spectrum for each region has a significance of at least 3~$\sigma$ in each data bin.
\bold{The advantage of using this method is a more accurate background subtraction which also accounts for vignetting and the particle background. 
As shown in Appendix~\ref{app:bkgspec}, this can have a large effect (of up to $\sim$20\%) on the flux values and \colblue{it can eliminate the need of introducing a power law component to account for residuals at high energies }in the spectral fits}.

\begin{figure*}
  \begin{minipage}[t]{17.4cm}
  \fig{\frame{\includegraphics[width=8.9cm]{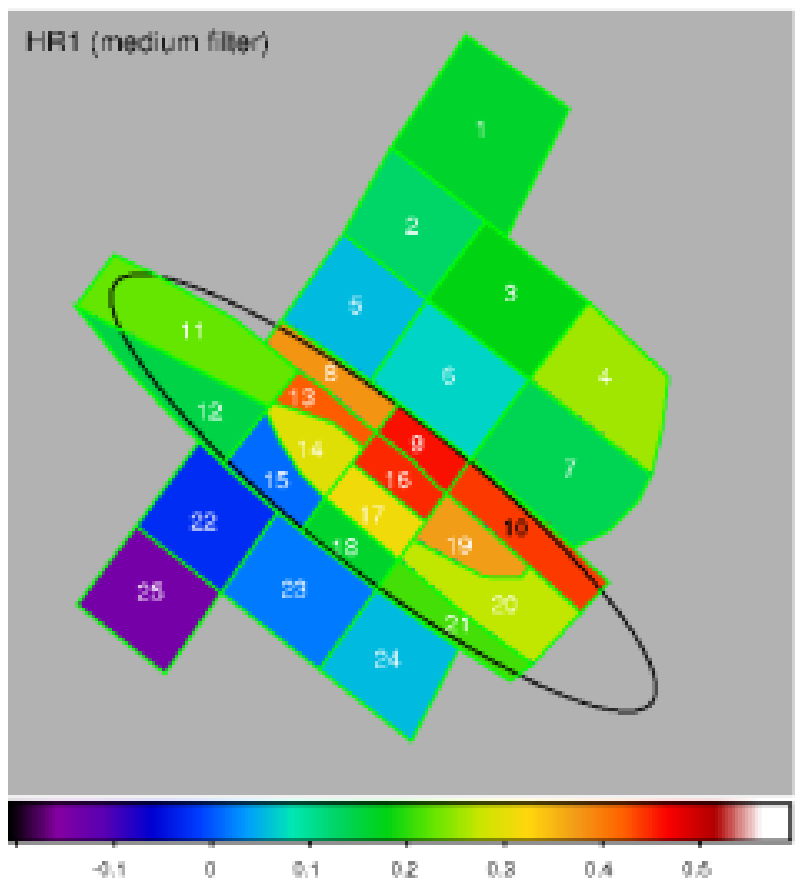}}}
  \fig{\frame{\includegraphics[width=8.9cm]{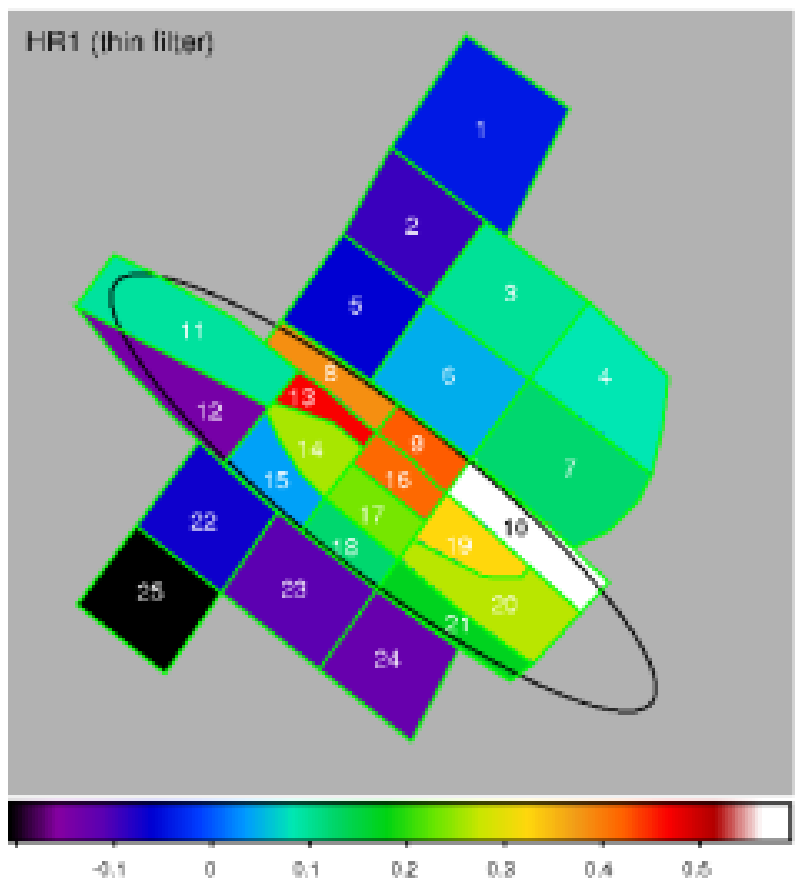}}}
  \fig{\frame{\includegraphics[width=8.9cm]{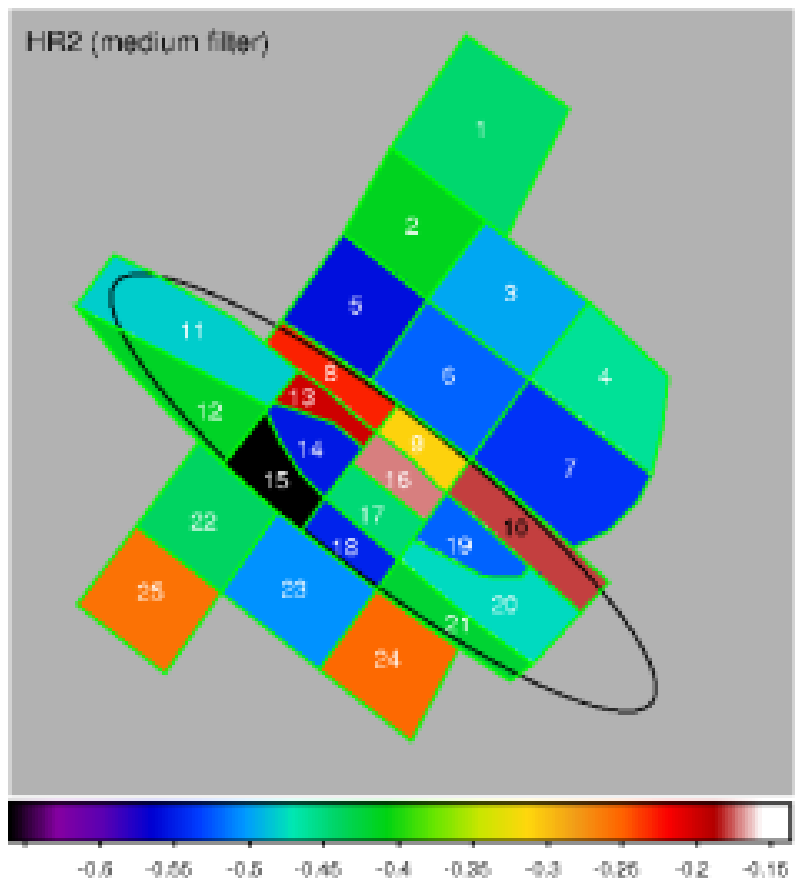}}}
  \fig{\frame{\includegraphics[width=8.9cm]{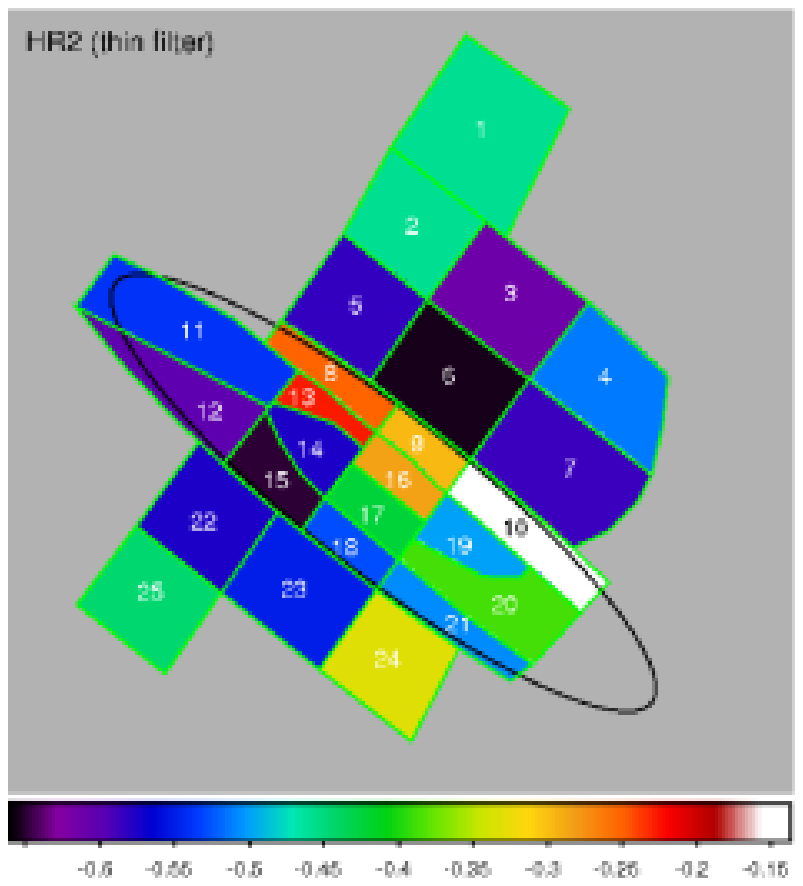}}}
  \end{minipage}
  \caption
  {
  \bold{Hardness ratio maps of observations 1 {\it (left)} and 4 {\it (right)}}: 
  The map was binned to the same regions as in Fig.~\ref{fig:PN_cheese_image}. 
  The higher the index, the harder the spectrum. 
  The background colour corresponds to an artificially set value. 
  We also show the inclination corrected optical $D_{25}$ ellipse. 
  {\it (top):} HR1=(B$_2$-B$_1$)/(B$_2$+B$_1$), where B$_1$ and B$_2$ are the count rates in the energy bands 0.2--0.5~keV and 0.5--1.0~keV, respectively. 
  {\it (bottom):} HR2=(B$_3$-B$_2$)/(B$_3$+B$_2$), where B$_2$ and B$_3$ are the count rates in the energy bands 0.5--1.0~keV and 1.0--2.0~keV, respectively.
  }
  \label{fig:HR_C}
\end{figure*}

Since the emission is mostly confined to energies between 0.2 and 2.0~keV, we only calculated the hardness ratios HR1 and HR2, where HR1=(B$_2$-B$_1$)/(B$_2$+B$_1$), and HR2=(B$_3$-B$_2$)/(B$_3$+B$_2$).
B$_1$, B$_2$, and B$_3$ are the count rates in the energy bands 1 to 3, \ie 0.2--0.5~keV, 0.5--1.0~keV, and 1.0--2.0~keV, respectively.
They were obtained by summing up the background subtracted counts in the spectra in the energy bands.
\bold{\colblue{Observations 1 and 4 were performed for EPIC~pn using the medium and thin filter, respectively.
As these filters have a significantly different throughput}, we did not calculate the combined hardness ratios, but show the results for the two observations independently (Fig.~\ref{fig:HR_C} and Table~\ref{tab:HR}).
The errors on the hardness ratios were propagated from the errors on the counts in the spectra}.

In order to fit the spectra with physical models, we created the proper response and anxiliary response files for extended sources for each spectrum.
In XSPEC~11.3.2, we linked the model parameters between the two observations and included a global renormalisation factor to account for differences between the observations to fit the spectra from observations 1 and 4 simultaneously.
\bold{To fit the spectra with physical models we used the chi-squared statistics in XSPEC. 
\colblue{To assess the significance of additional model components, the Bayesian Information Criterion \citep[BIC, introduced by][ an astronomical application can be found, e.g., in \citealp{L2004}]{S1978} was applied:
a difference of 2 for the BIC is regarded as positive evidence, and of 6 or more as strong evidence, against the model with the larger value \citep{J1961,MF1998}.
We did not use the F-test, since this method is not valid to test if an additional absorption component is justified \citep{PD2002}}, since the null value of the additional parameter is on the boundary of the set of possible parameter values, i.e.\ zero, and negative column density values are not supported by the tbabs \citep{WA2000} model.
\colblue{For the BIC, these restrictions do not apply}.
The criteria for our best-fit models are the following: (i) the $\chi^2_{\nu}$ is below 1.4, \colblue{(ii) the BIC difference between a given model and a model with more parameters is greater than 2}, (iii) the errors on all model parameter are constrained, and (iv) the model is physically \colblue{meaningful}}.

\begin{table*}
\begin{minipage}[t]{\textwidth}
\caption{\bold{Comparison between hardness ratios, for the individual observations (obs 4: thin filter; obs 1: medium filter)}}
\label{tab:HR}
\centering
\renewcommand{\footnoterule}{}  % to avoid a line before footnotes
\begin{tabular}{lrrrr}
\hline
\hline
Region                                  & \multicolumn{2}{c}{HR1}                                & \multicolumn{2}{c}{HR2} \\
                                        & \multicolumn{1}{c}{Obs 1}  & \multicolumn{1}{c}{Obs 4} & \multicolumn{1}{c}{Obs 1}  & \multicolumn{1}{c}{Obs 4} \\
\hline
\multicolumn{3}{l}{Northwestern halo}\\
\hline
1                                       & $ 0.16\pm 0.06$ & $-0.04\pm 0.05$ & $-0.44\pm 0.04$ & $-0.46\pm 0.04$ \\
2                                       & $ 0.13\pm 0.06$ & $-0.09\pm 0.05$ & $-0.41\pm 0.04$ & $-0.46\pm 0.05$ \\
3                                       & $ 0.18\pm 0.06$ & $ 0.10\pm 0.04$ & $-0.50\pm 0.04$ & $-0.61\pm 0.03$ \\
4                                       & $ 0.26\pm 0.06$ & $ 0.08\pm 0.05$ & $-0.46\pm 0.04$ & $-0.51\pm 0.04$ \\
5                                       & $ 0.05\pm 0.05$ & $-0.06\pm 0.04$ & $-0.56\pm 0.04$ & $-0.58\pm 0.03$ \\
6                                       & $ 0.07\pm 0.04$ & $ 0.05\pm 0.03$ & $-0.52\pm 0.03$ & $-0.65\pm 0.04$ \\
7                                       & $ 0.14\pm 0.05$ & $ 0.12\pm 0.04$ & $-0.54\pm 0.03$ & $-0.59\pm 0.03$ \\
2\dots4 \comment{40}                    & $ 0.19\pm 0.03$ & $ 0.04\pm 0.03$ & $-0.46\pm 0.02$ & $-0.54\pm 0.02$ \\
5\dots7 \comment{39}                    & $ 0.09\pm 0.03$ & $ 0.05\pm 0.02$ & $-0.53\pm 0.02$ & $-0.62\pm 0.02$ \\
1\dots7 \comment{36}                    & $ 0.13\pm 0.02$ & $ 0.03\pm 0.02$ & $-0.50\pm 0.01$ & $-0.58\pm 0.01$ \\
\hline
\multicolumn{3}{l}{Disc}\\
\hline
8 \comment{C37}                         & $ 0.38\pm 0.07$ & $ 0.39\pm 0.05$ & $-0.23\pm 0.06$ & $-0.25\pm 0.05$ \\
9 \comment{C10}                         & $ 0.46\pm 0.06$ & $ 0.42\pm 0.05$ & $-0.31\pm 0.05$ & $-0.30\pm 0.04$ \\
10 \comment{C11}                        & $ 0.44\pm 0.08$ & $ 0.60\pm 0.07$ & $-0.18\pm 0.06$ & $-0.14\pm 0.05$ \\
11 \comment{C13}                        & $ 0.23\pm 0.05$ & $ 0.10\pm 0.04$ & $-0.48\pm 0.04$ & $-0.54\pm 0.03$ \\
12 \comment{C14}                        & $ 0.15\pm 0.06$ & $-0.14\pm 0.05$ & $-0.41\pm 0.04$ & $-0.60\pm 0.04$ \\
13 \comment{C15}                        & $ 0.43\pm 0.07$ & $ 0.47\pm 0.05$ & $-0.20\pm 0.06$ & $-0.23\pm 0.04$ \\
14  \comment{C16}                       & $ 0.30\pm 0.03$ & $ 0.26\pm 0.02$ & $-0.55\pm 0.03$ & $-0.57\pm 0.02$ \\
15  \comment{C17}                       & $ 0.01\pm 0.04$ & $ 0.04\pm 0.03$ & $-0.66\pm 0.03$ & $-0.65\pm 0.03$ \\
16  \comment{C18}                       & $ 0.45\pm 0.04$ & $ 0.42\pm 0.03$ & $-0.17\pm 0.04$ & $-0.28\pm 0.03$ \\
17  \comment{C31}                       & $ 0.31\pm 0.03$ & $ 0.23\pm 0.02$ & $-0.45\pm 0.03$ & $-0.42\pm 0.02$ \\
18  \comment{C32}                       & $ 0.17\pm 0.04$ & $ 0.12\pm 0.03$ & $-0.55\pm 0.03$ & $-0.53\pm 0.03$ \\
19 \comment{C28}                        & $ 0.37\pm 0.04$ & $ 0.32\pm 0.03$ & $-0.52\pm 0.03$ & $-0.50\pm 0.03$ \\
20 \comment{C21}                        & $ 0.27\pm 0.04$ & $ 0.27\pm 0.03$ & $-0.48\pm 0.03$ & $-0.39\pm 0.03$ \\
21 \comment{C22}                        & $ 0.21\pm 0.06$ & $ 0.17\pm 0.05$ & $-0.42\pm 0.04$ & $-0.51\pm 0.03$ \\
\hline
\multicolumn{3}{l}{Southeastern halo}\\
\hline
22 \comment{C24}                        & $-0.02\pm 0.06$ & $-0.07\pm 0.05$ & $-0.44\pm 0.04$ & $-0.58\pm 0.04$ \\
23 \comment{C25}                        & $ 0.02\pm 0.05$ & $-0.11\pm 0.04$ & $-0.50\pm 0.04$ & $-0.55\pm 0.04$ \\
24 \comment{C26}                        & $ 0.05\pm 0.08$ & $-0.12\pm 0.07$ & $-0.25\pm 0.05$ & $-0.33\pm 0.06$ \\
23+24 \comment{C34}                     & $ 0.03\pm 0.04$ & $-0.12\pm 0.03$ & $-0.41\pm 0.03$ & $-0.48\pm 0.03$ \\
25 \comment{C27}                        & $-0.14\pm 0.07$ & $-0.21\pm 0.06$ & $-0.26\pm 0.06$ & $-0.44\pm 0.06$ \\
22\dots24 \comment{C38}                 & $ 0.01\pm 0.03$ & $-0.10\pm 0.03$ & $-0.42\pm 0.02$ & $-0.51\pm 0.03$ \\
22\dots25 \comment{C41}                 & $-0.01\pm 0.03$ & $-0.12\pm 0.02$ & $-0.39\pm 0.02$ & $-0.50\pm 0.02$ \\
\hline
\end{tabular}
\end{minipage}
\end{table*}

%%%%%%%%%%%%%%%%%%%%%%%%%%%%%%%%%%%%%%%%%%%%%%%%%%%%%%%%%%%%%%%%%%%%%%%%%%%%%%%%%%%%%%%%%%%%%%%%%%%%%%%%%%%%%%%%%%%%%%
%%%%%%%%%%%%%%%%%%%%%%%%%%%%%%%%%%%%%%%%%%%%%%%%%%%%%%%%%%%%%%%%%%%%%%%%%%%%%%%%%%%%%%%%%%%%%%%%%%%%%%%%%%%%%%%%%%%%%%

\section{Results}
To characterise the diffuse emission in the disc and the halo, we analysed images in different energy bands, and hardness ratios and spectra from several regions.
\bold{The size of these regions was chosen to provide enough counts for a spectral analysis.
As for the shape, in} the disc the regions were chosen in a way that they follow the spiral arm structure.
In the halo, we chose plane-parallel regions above the galactic disc.
The projected heights of the halo regions are 2~kpc, with exception of the region furthest to the northwest (region~1), which has a projected height of 3~kpc.
The regions are overplotted on top the false-colour X-ray image in Fig.~\ref{fig:PN_cheese_image}.
The hardness ratios in the different regions are given in Table~\ref{tab:HR} and shown graphically in Fig.~\ref{fig:HR_C}. 

\subsection{Disc diffuse emission}
The disc shows diffuse emission in energies up to $\sim$10~keV, where the harder emission is located close to the centre of \n253.
The soft emission ($<$1~keV) shows the largest extent along the major axis.
From the nucleus, it reaches $\sim$7.0~kpc to the northwest and $\sim$6.4~kpc to the southeast.

A prominent feature in the disc is the lack of very soft emission northwest of the major axis.
This is already known from \ros\ observations \citep[e.g.][]{PV2000} and can be explained by the geometry of the system:
The galaxy's disc is oriented so that we see the underside of the disc.
The emission from the northwestern halo behind the disc is therefore absorbed by the intervening disc material.

\begin{figure*}
\begin{minipage}[t]{17.4cm}
\fig{\includegraphics[width=\specwidth]{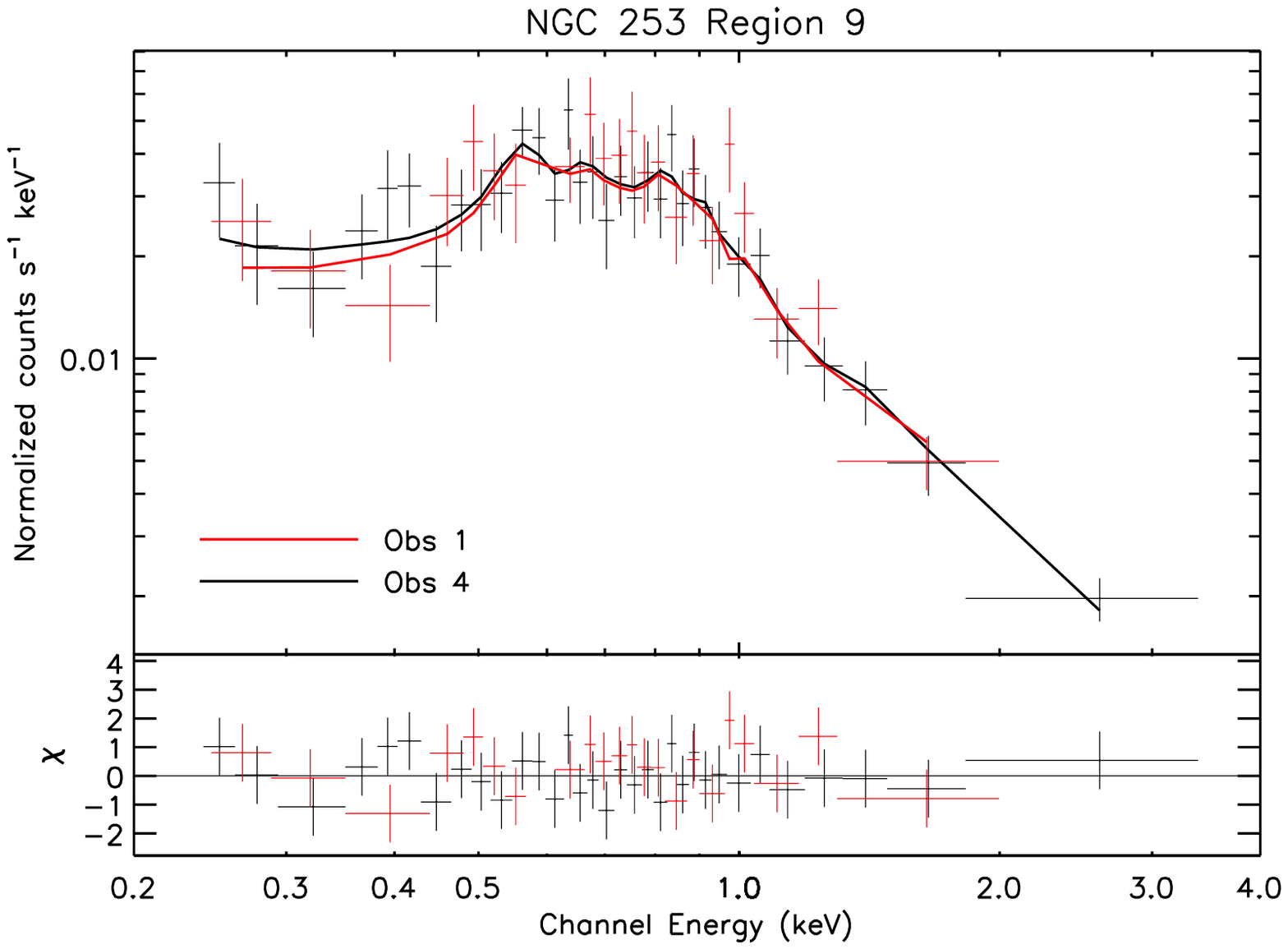}
\includegraphics[width=\specwidth]{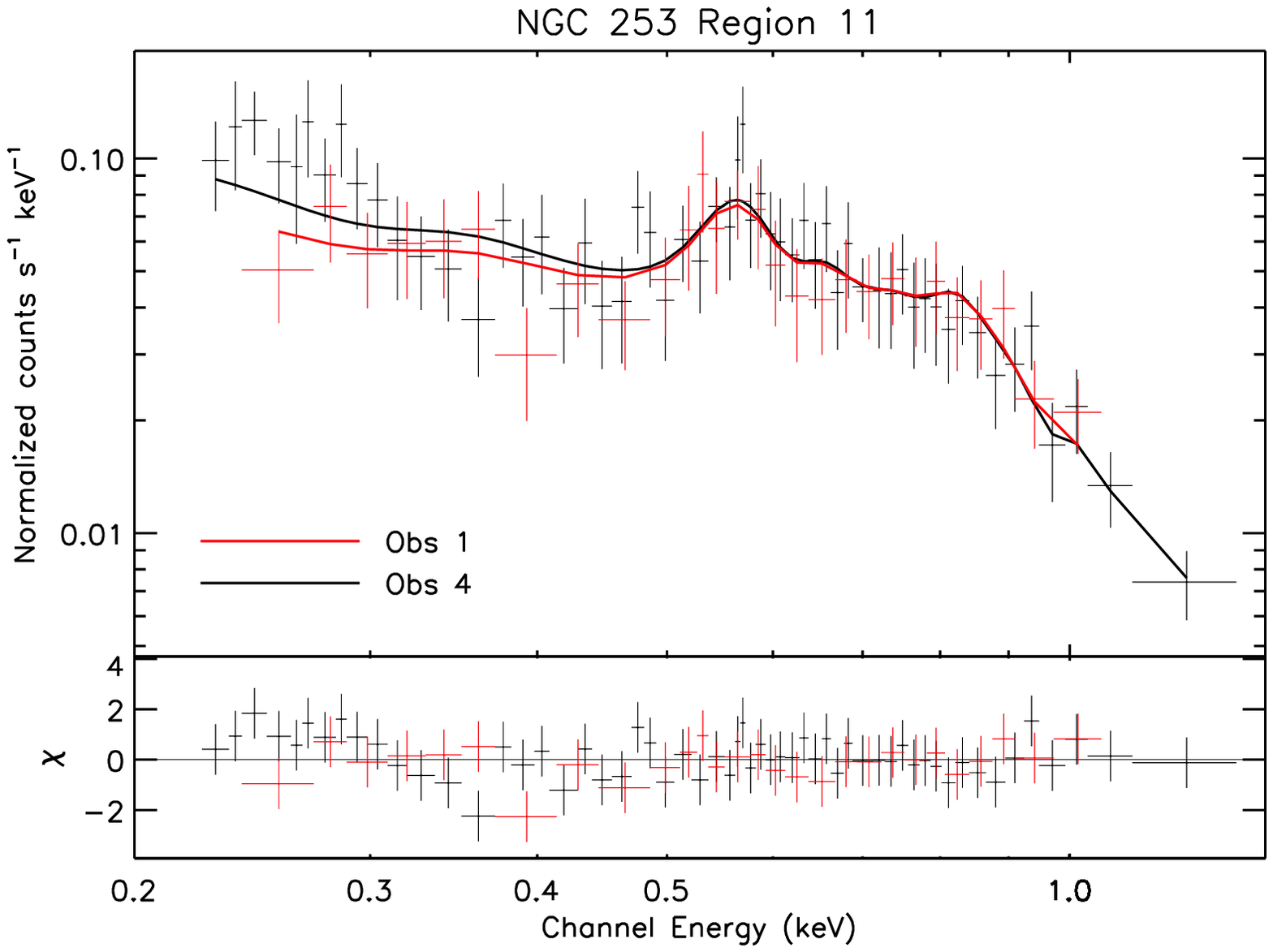}
\includegraphics[width=\specwidth]{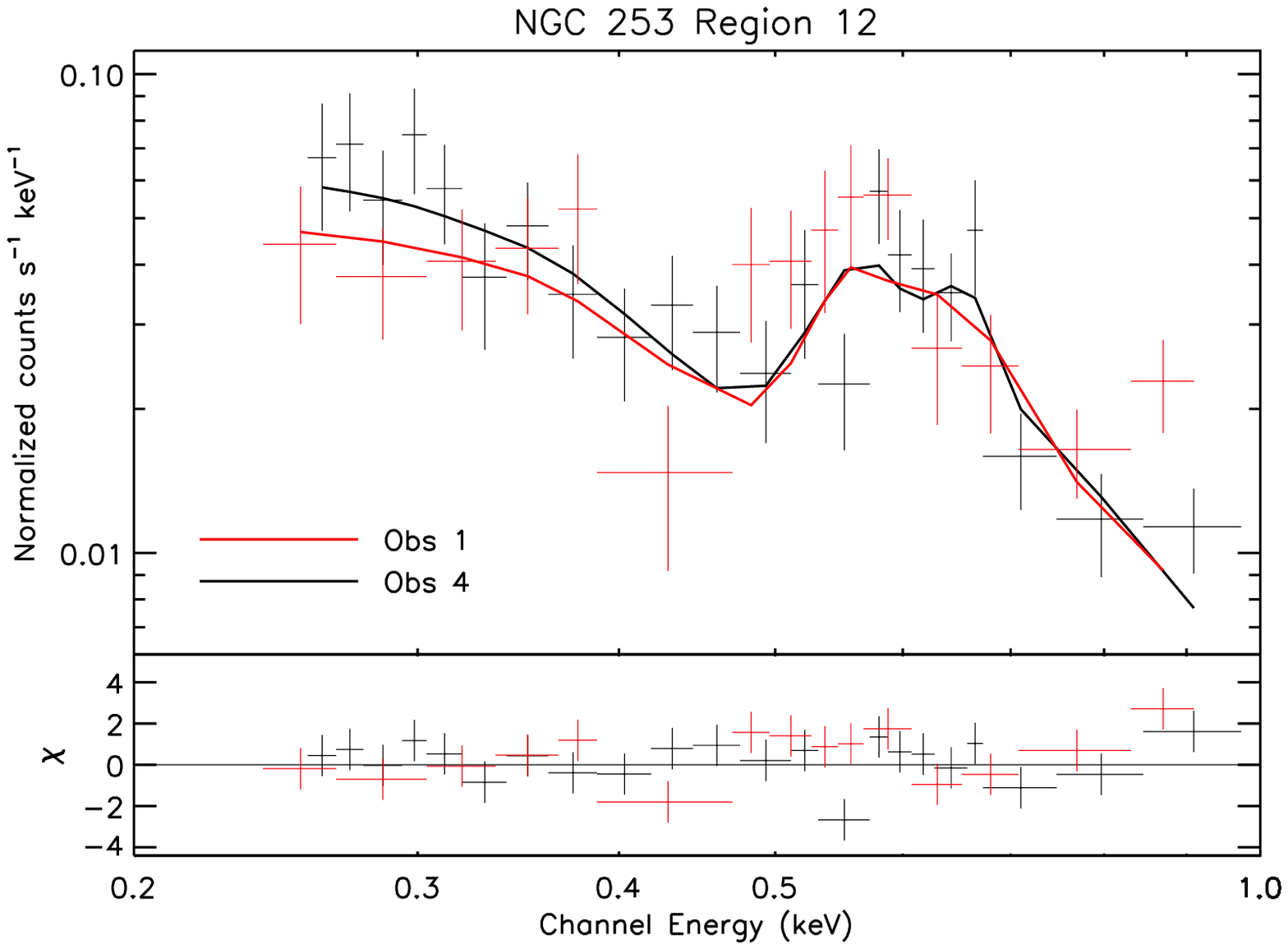}
\includegraphics[width=\specwidth]{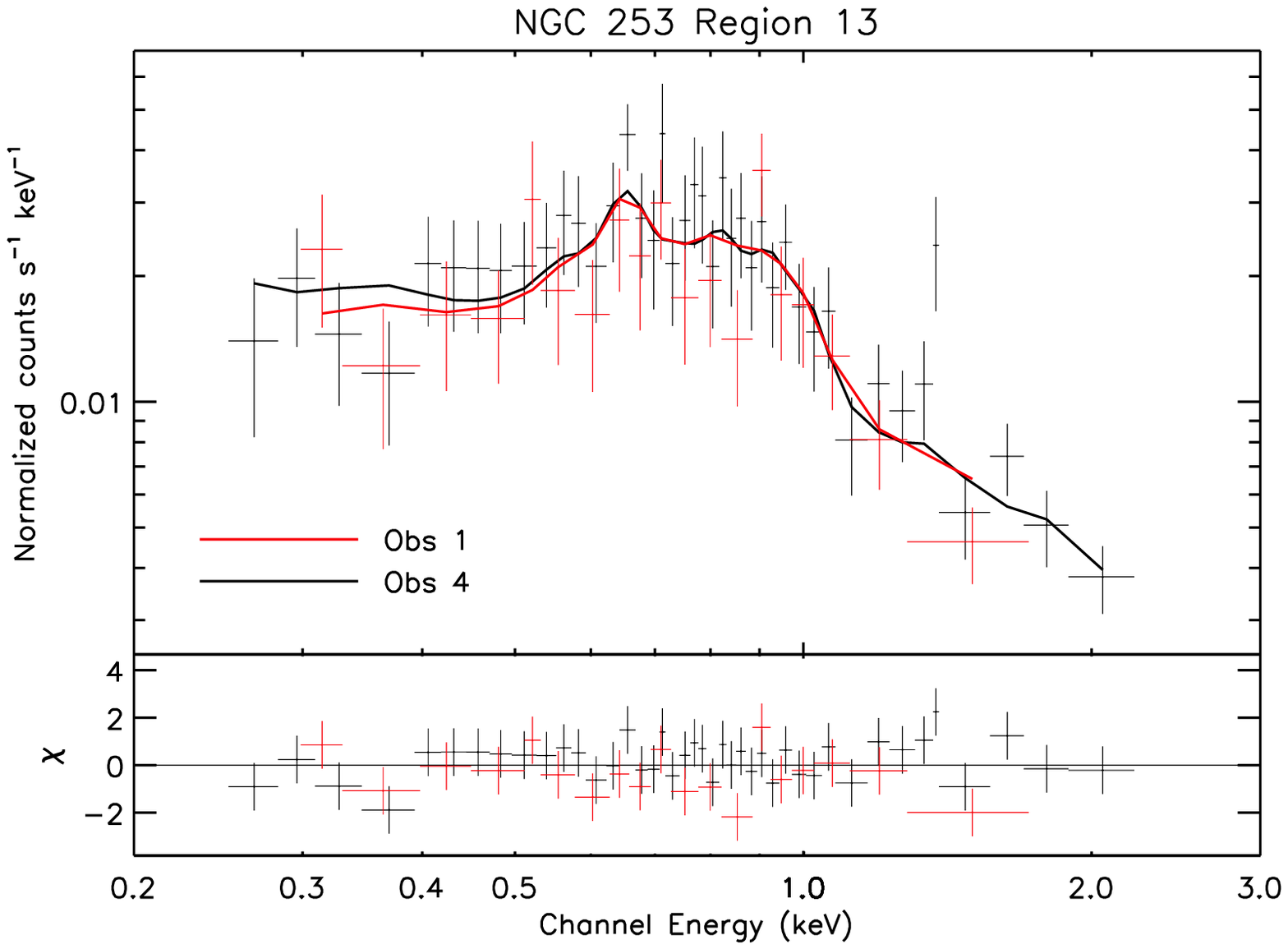}
\includegraphics[width=\specwidth]{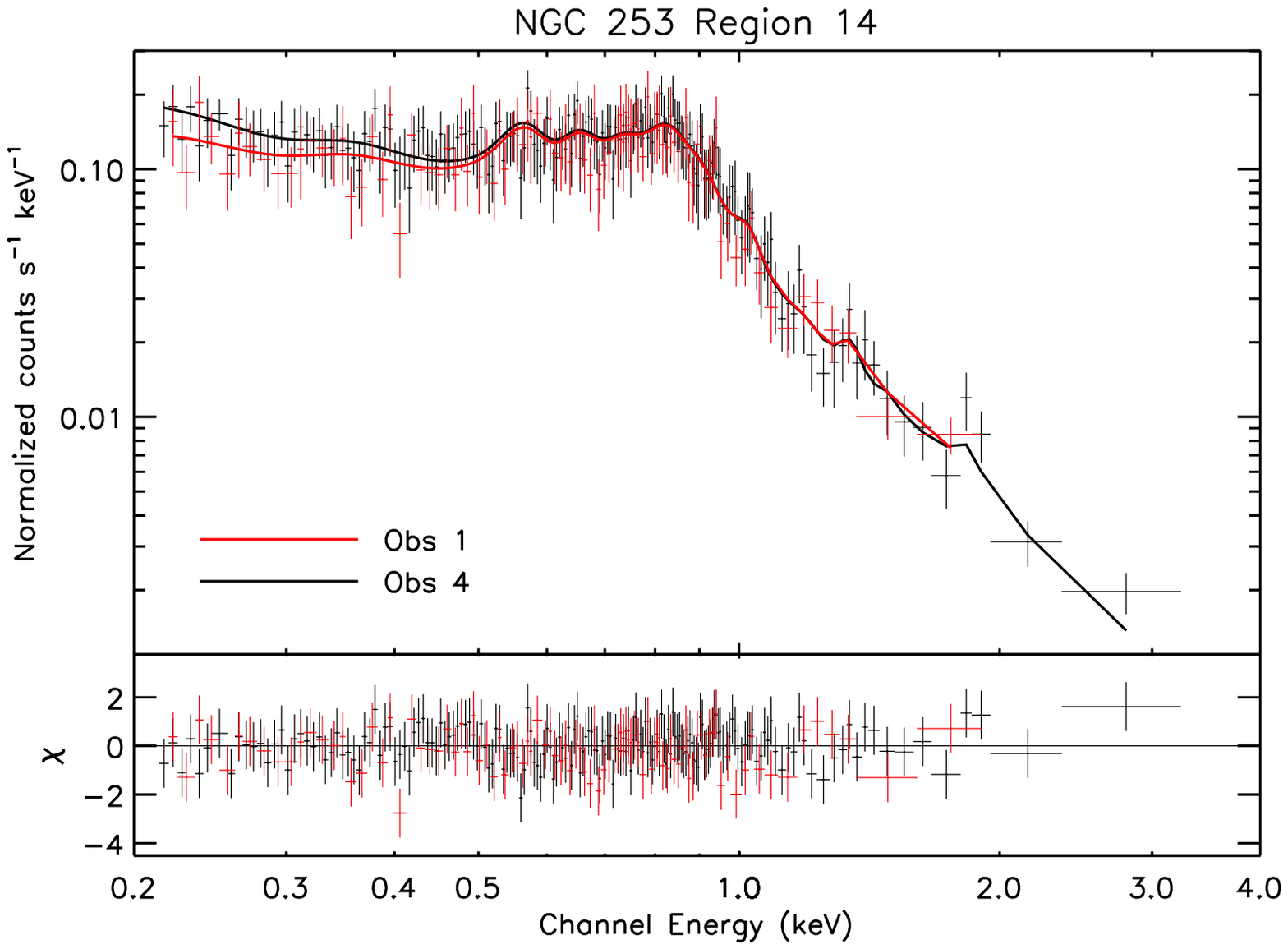}
\includegraphics[width=\specwidth]{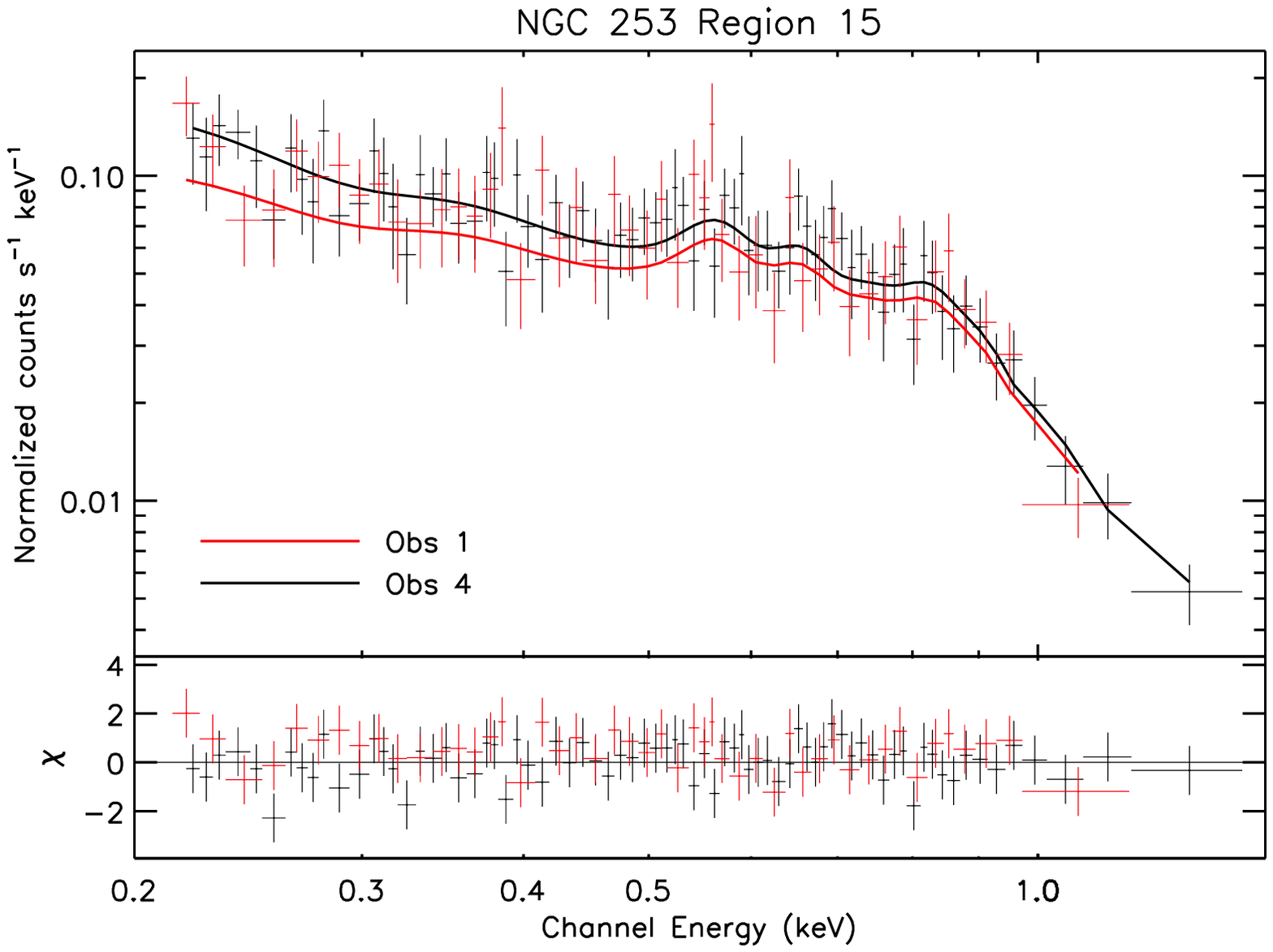}
\includegraphics[width=\specwidth]{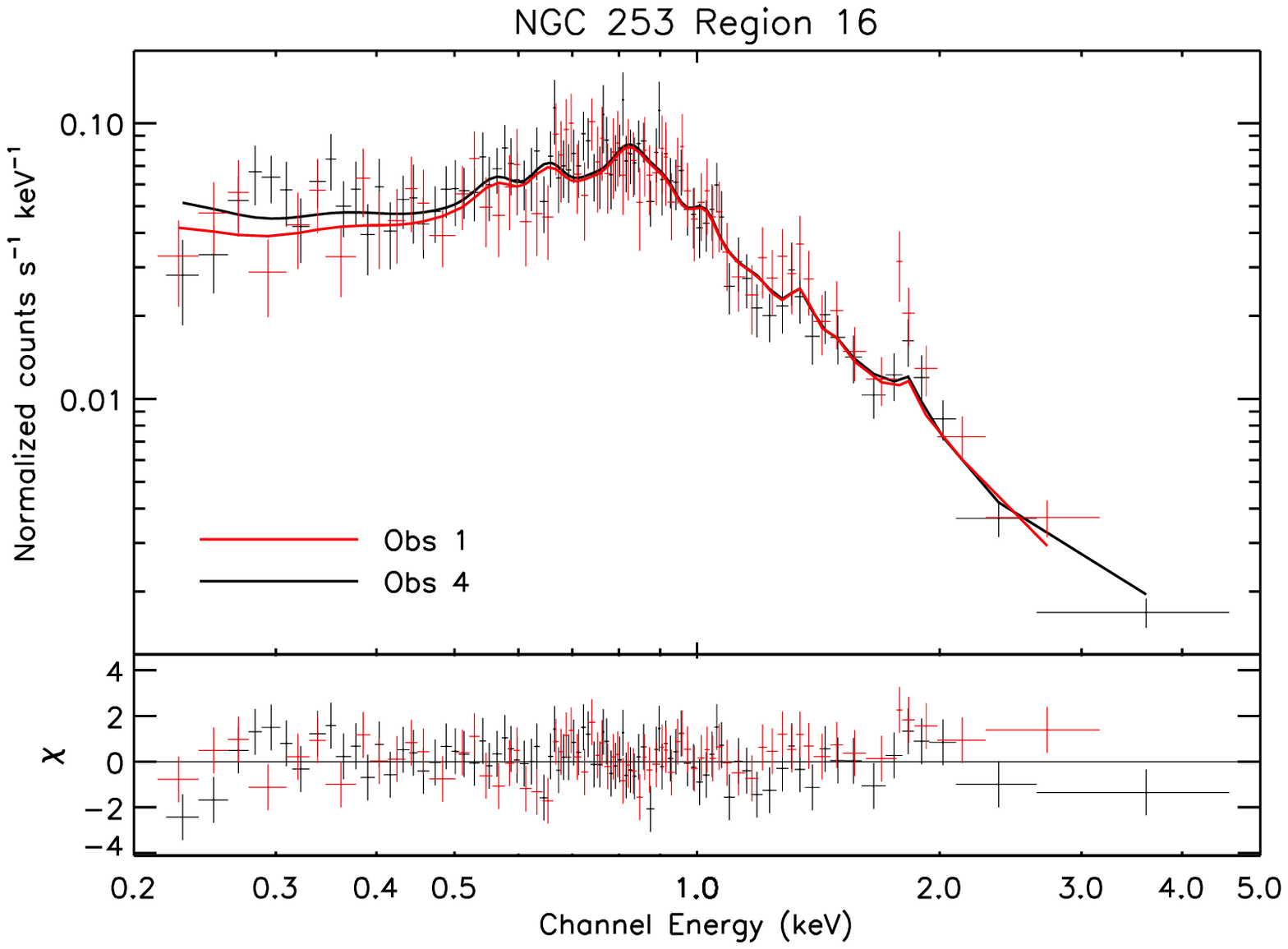}
\includegraphics[width=\specwidth]{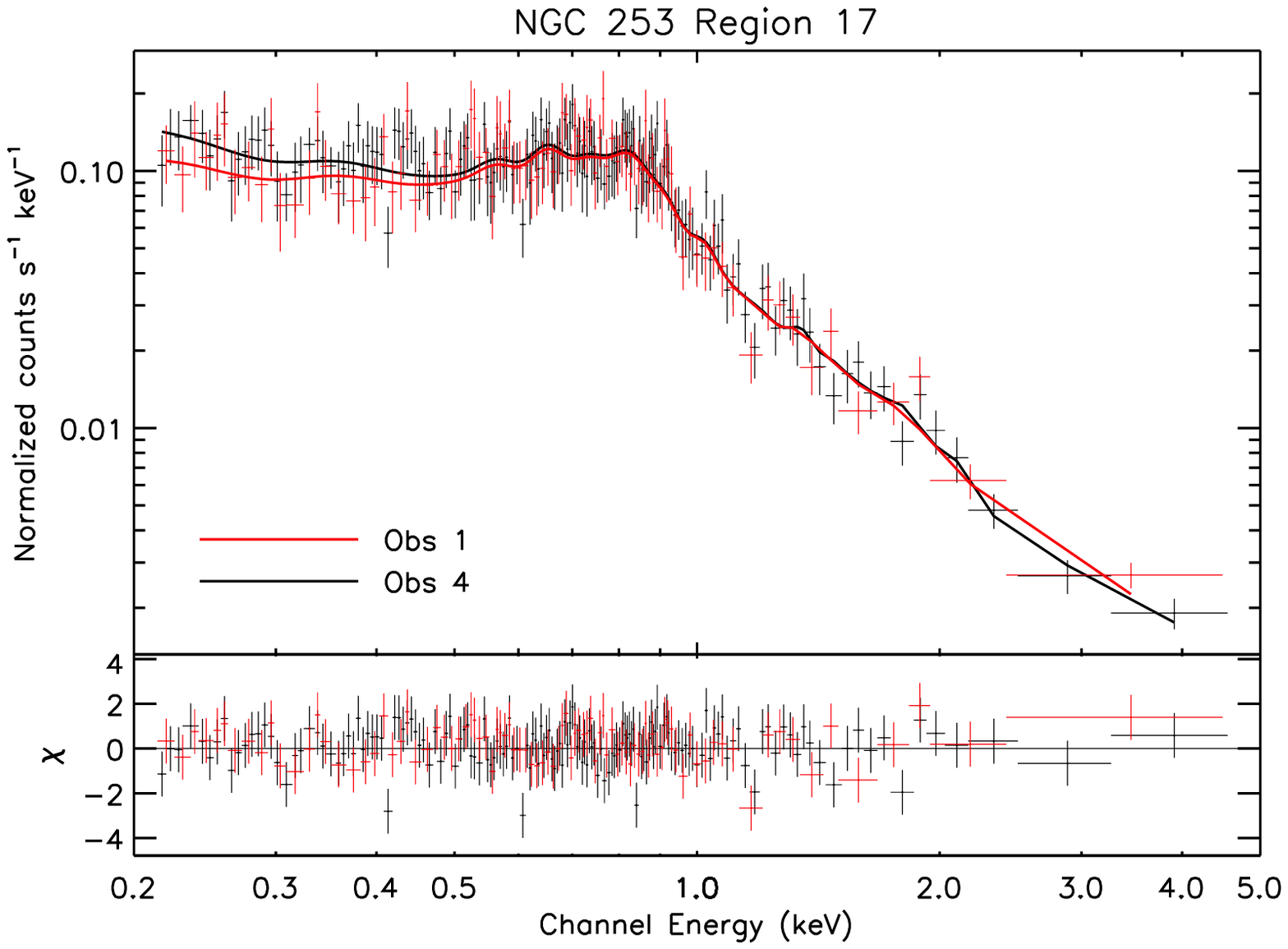}
\includegraphics[width=\specwidth]{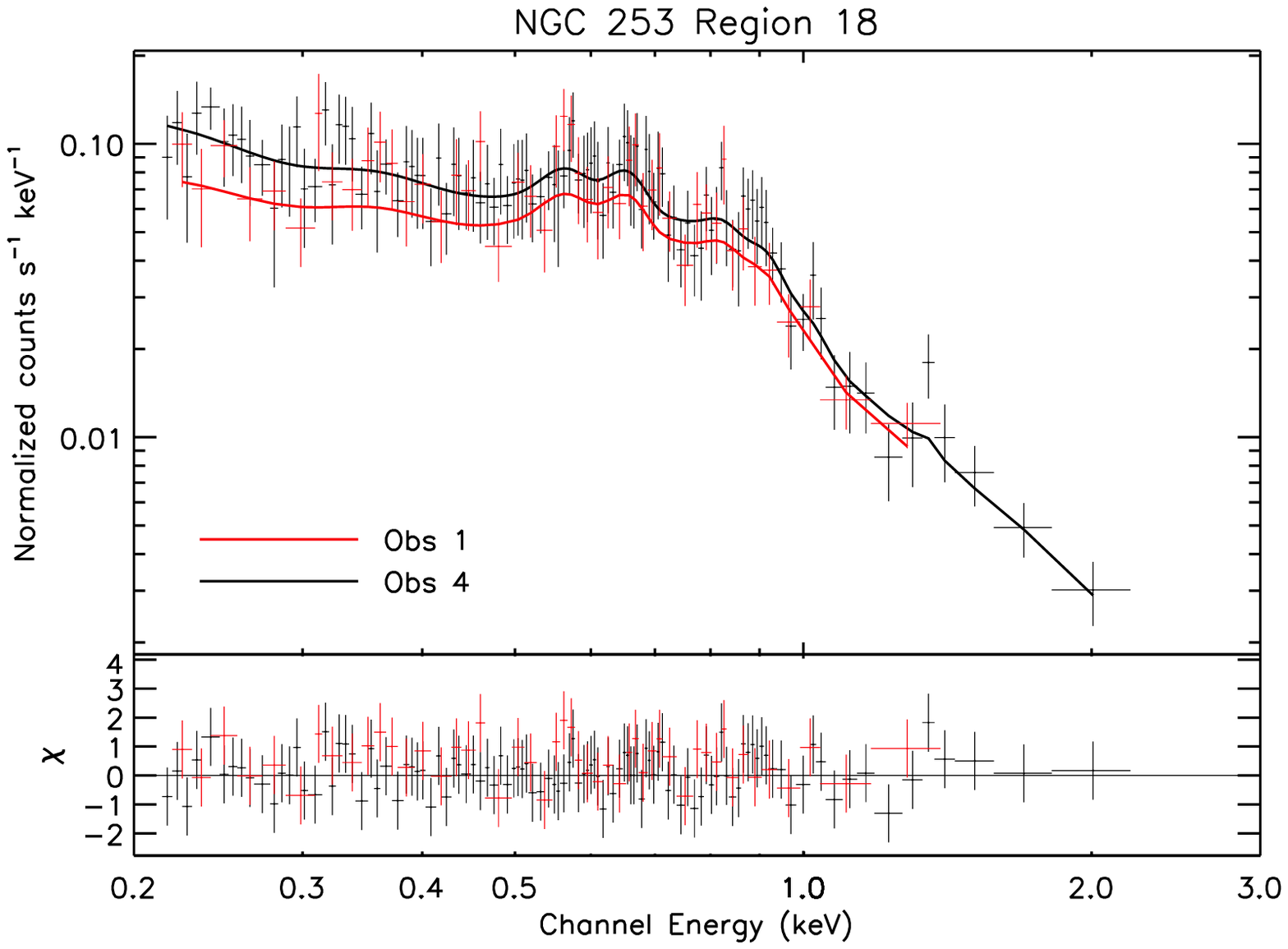}
\includegraphics[width=\specwidth]{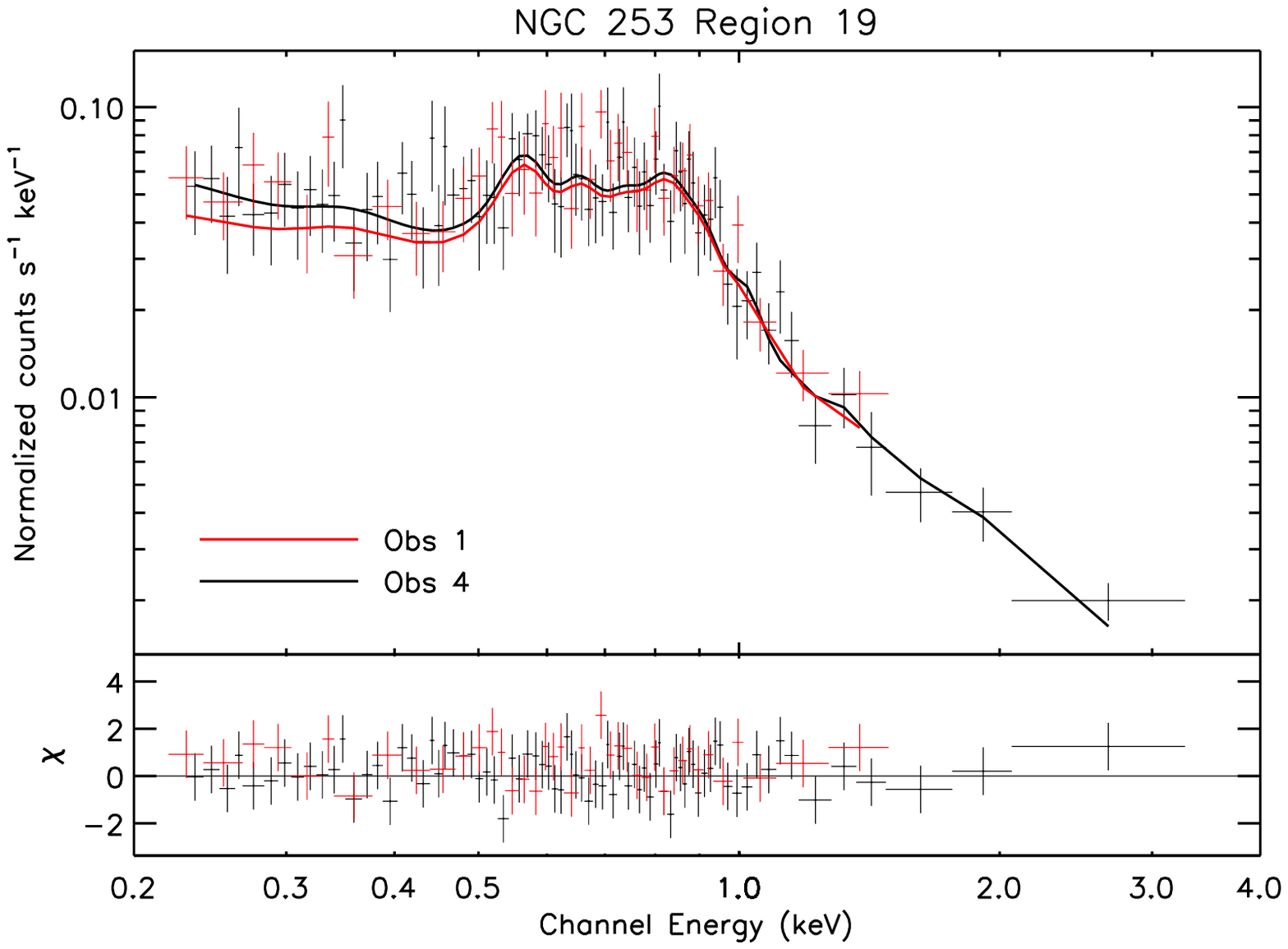}}
\end{minipage}
\caption{\bold{Spectra of different regions in the disc.  The red and the black data points and model fits are from observations 1 and 4, respectively. The lower panel shows the residuals of the fits.}}
\label{fig:specDisk}
\end{figure*}

The spectral properties in different regions of the disc are summarised in Table~\ref{tab:DiskFits}.
\bold{The spectra of all disc regions are shown in Fig.~\ref{fig:specDisk}.
A representative example (region~14) of one of these disc spectra shows lines from \ion{O}{VII} ($\sim$0.57~keV), \ion{O}{VIII} (0.65~keV), \ion{Fe}{XVII} (0.73--0.83~keV), \ion{Ne}{IX} ($\sim$0.91~keV), \ion{Ne}{X} (1.0~keV), \ion{Mg}{XI} ($\sim$1.3~keV), and \ion{Si}{XIII} ($\sim$1.9~keV)}.

To fit the spectra, we tried several different models, which all contain an absorption model tbabs for the Galactic foreground \nh\ of $1.3\times 10^{20}$~\cm-2 \citep{DL1990}.
Also the abundances were fixed to solar values (see also Sec.~\ref{sec:abund}) from \cite{WA2000}.
A simple one-temperature thin thermal plasma model \citep[apec,][]{SBL2001} did not result in a good fit (i.e.\ $\chi^2_{\nu}\le1.4$) in any case.
Similarly, a power law model did not give good fits \bold{(see Table~\ref{tab:DiskFits})}.
At least three components were necessary for most of the regions: two thin thermal plasmas plus a power law component.
The power law was needed to account for the emission above $\sim$1~keV and probably results from point sources below the point source detection limit, or incomplete source removal due to too small extraction radii.

The obtained temperatures are quite uniform throughout the disc and vary from 0.1 to 0.3~keV and from 0.3 to \bold{0.9}~keV for the soft and the hard component, respectively.
The intrinsic luminosity (corrected for Galactic absorption) of the diffuse emission within the inclination corrected optical $D_{25}$ ellipse is 2.4\ergs{39} (0.2--10.0~keV), or 8.5\ergs{38} (2.0--10~keV).
Both values were corrected for the area of cut-out point\bold{\colblue{-like}} sources.

The spectra decrease in hardness from the northwest to the southeast parallel to the minor axis, which can easily be seen in the hardness ratio maps (\bold{HR1}, Fig.~\ref{fig:HR_C}).
This is not an effect caused by different temperatures, but by the increasing strength of the soft spectral component towards the southeast (compared to the hard component), as the optical depth through the halo on the near side of the disc increases.

\begin{figure*}
\begin{minipage}[t]{17.4cm}
\fig{\includegraphics[width=\specwidth]{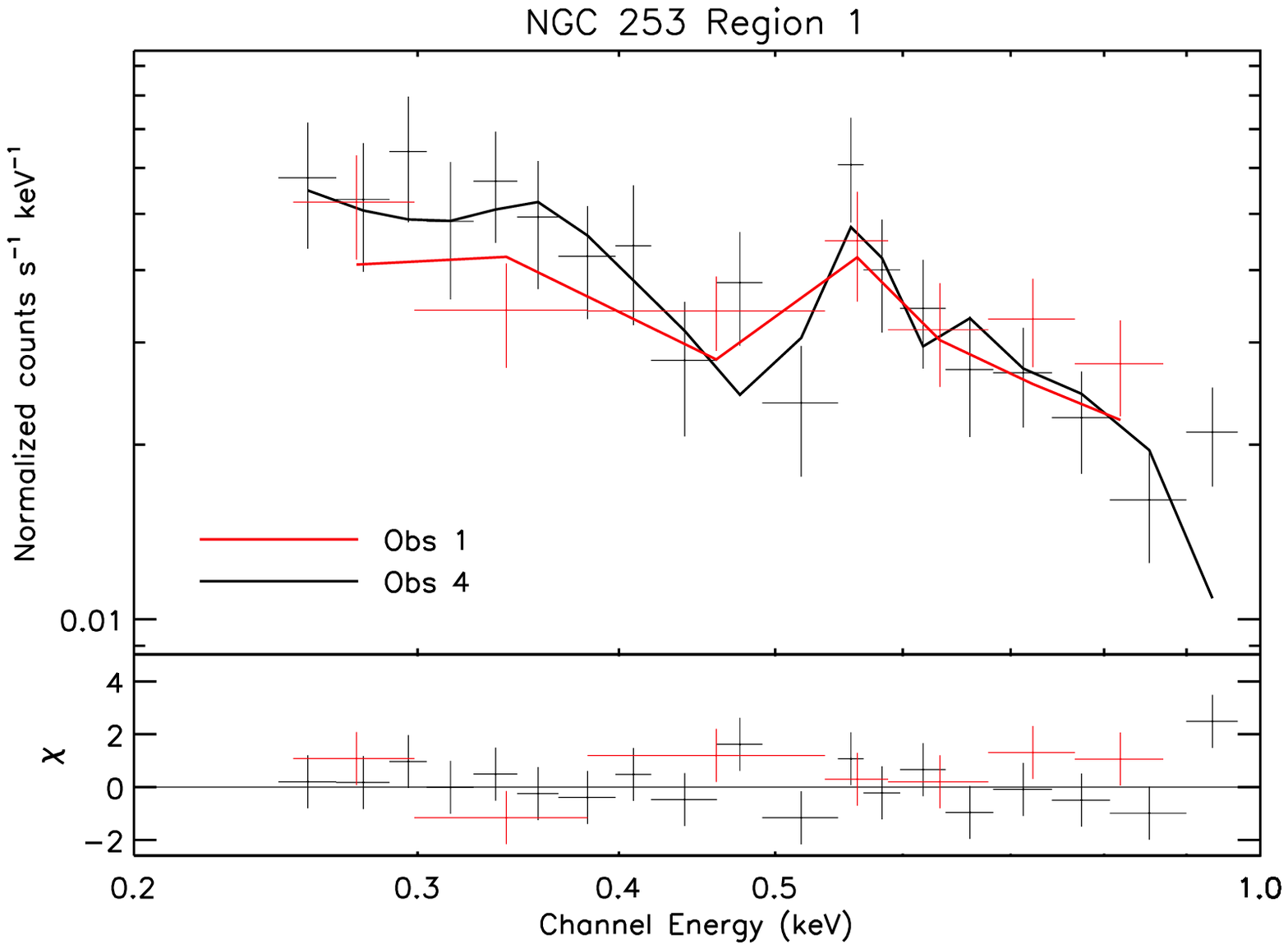}
\includegraphics[width=\specwidth]{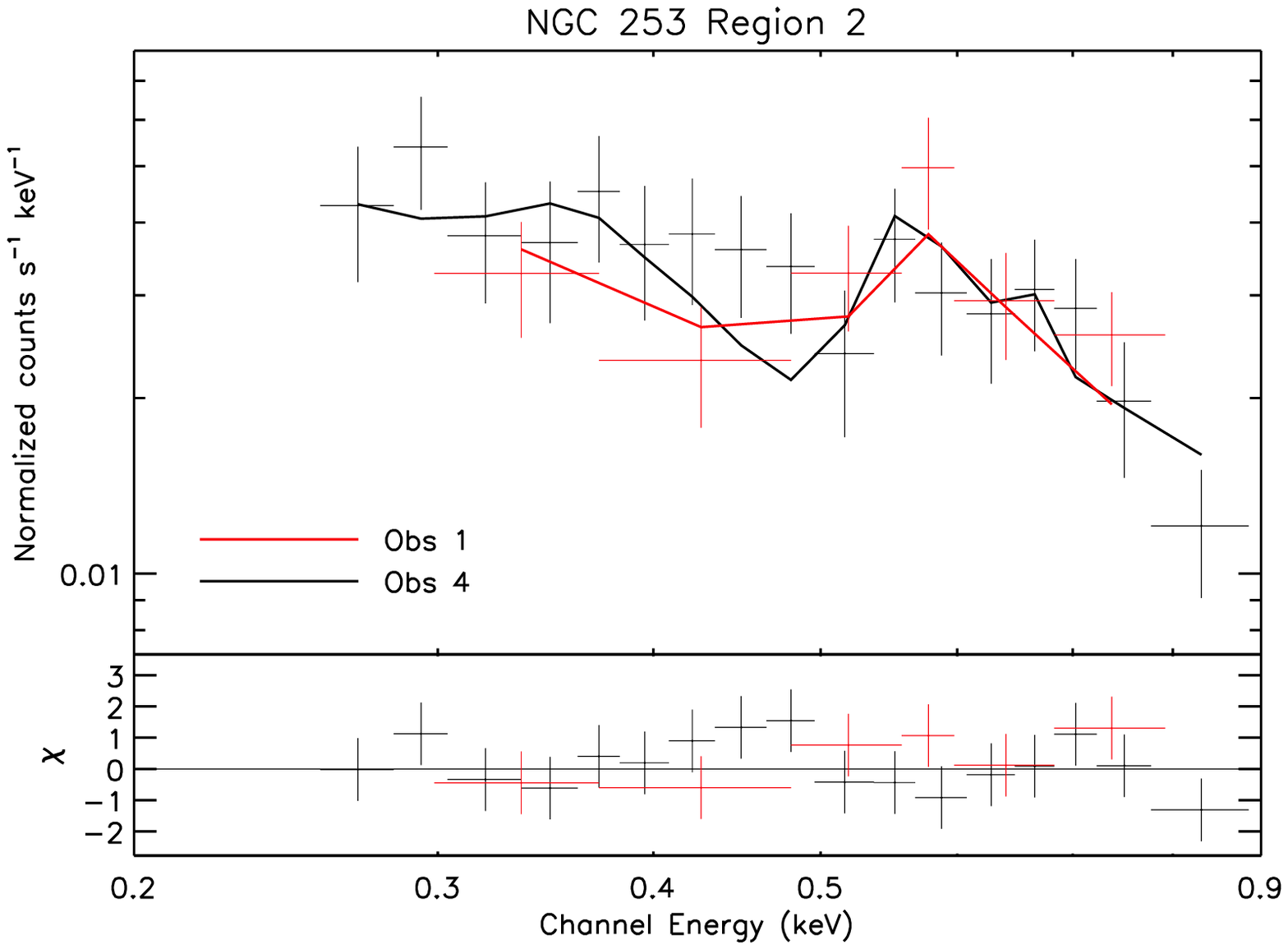}
\includegraphics[width=\specwidth]{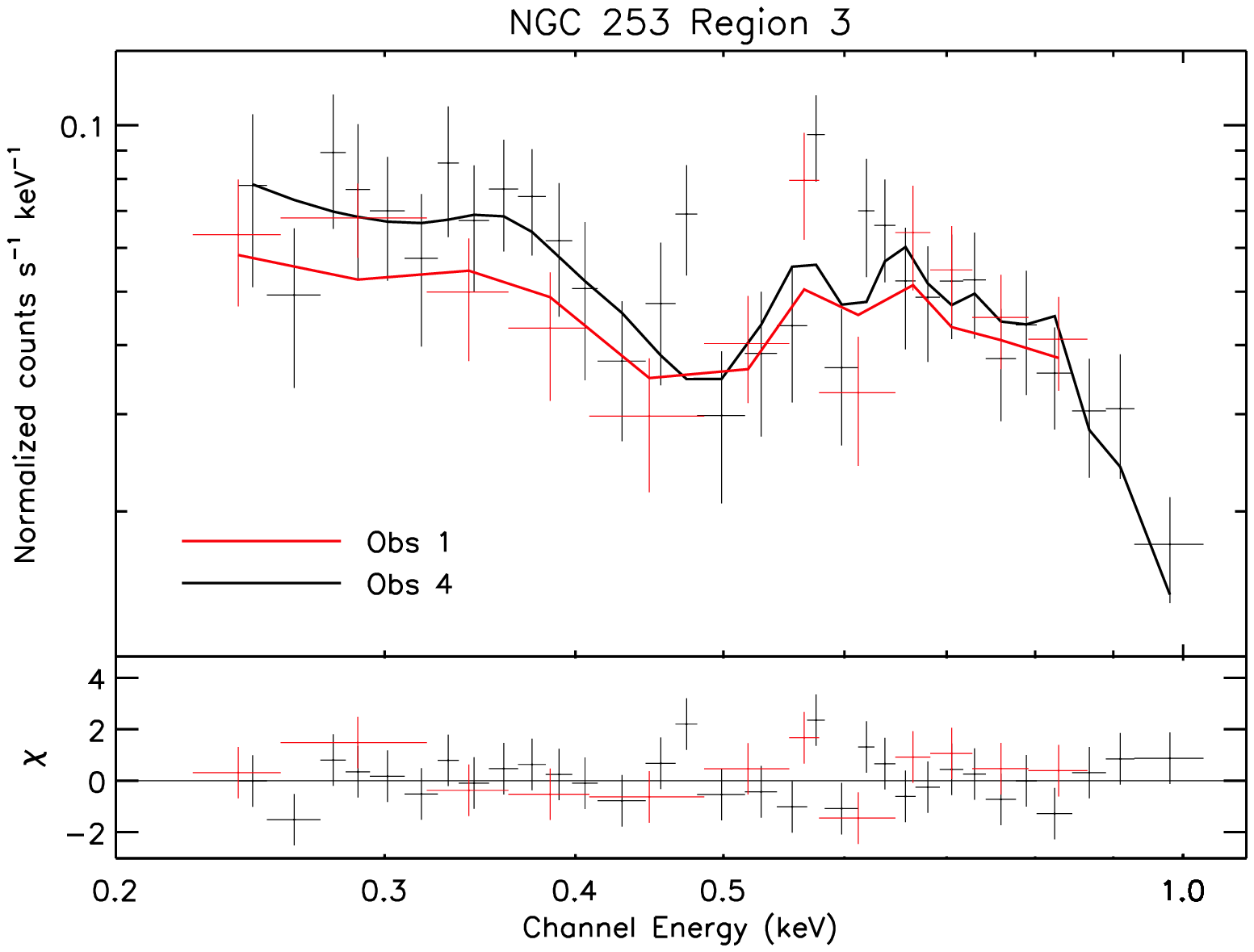}
\includegraphics[width=\specwidth]{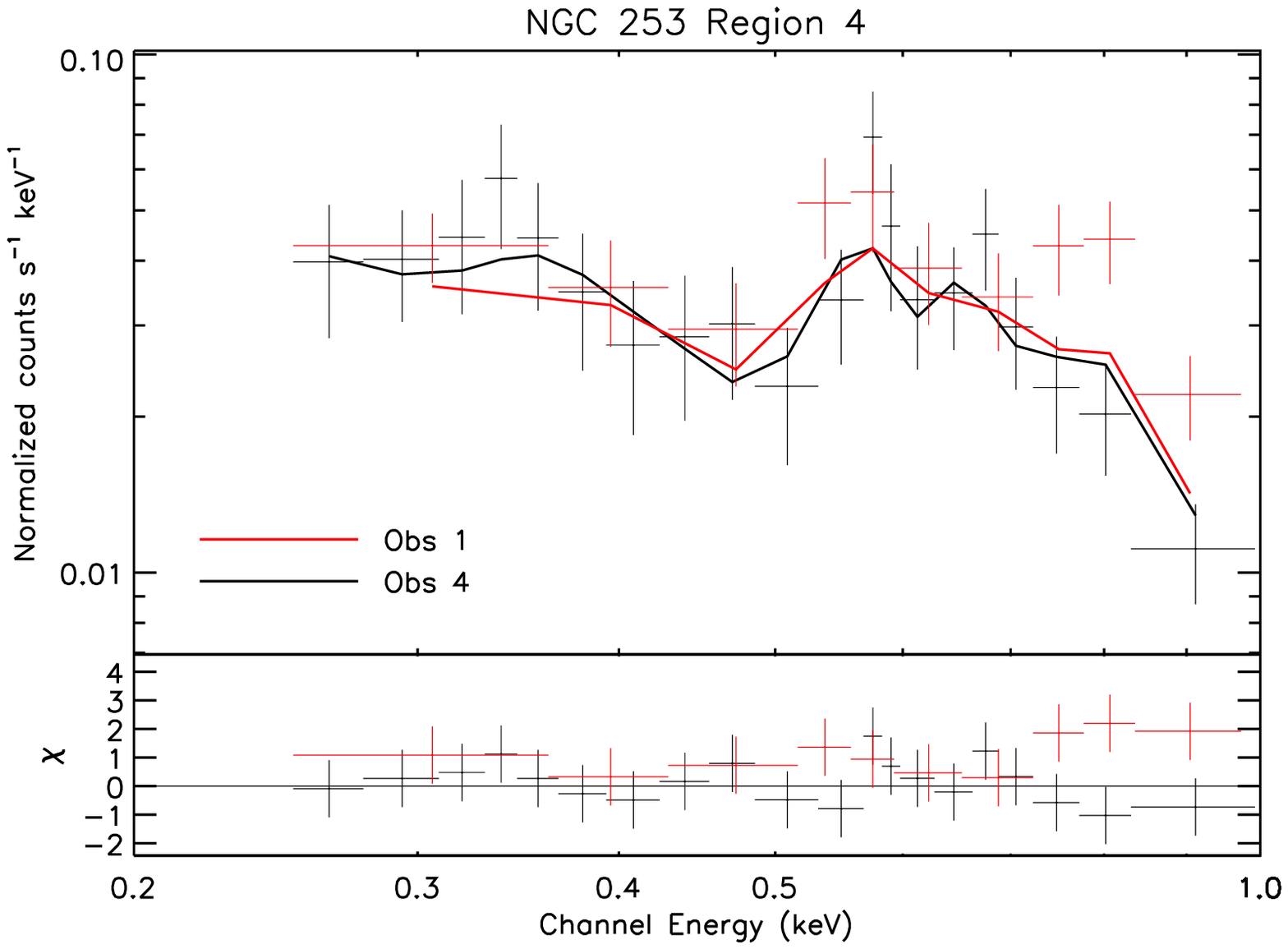}
\includegraphics[width=\specwidth]{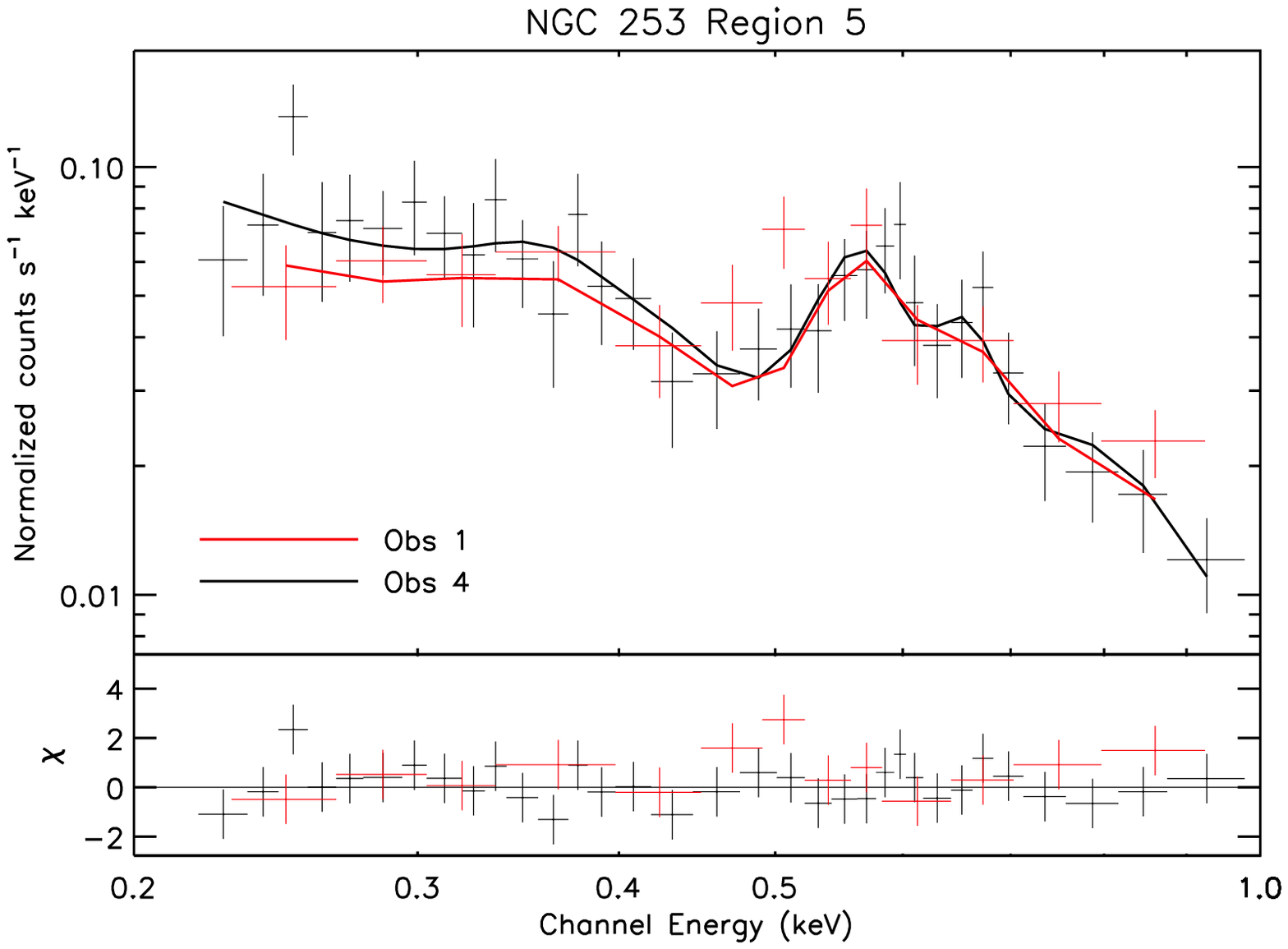}
\includegraphics[width=\specwidth]{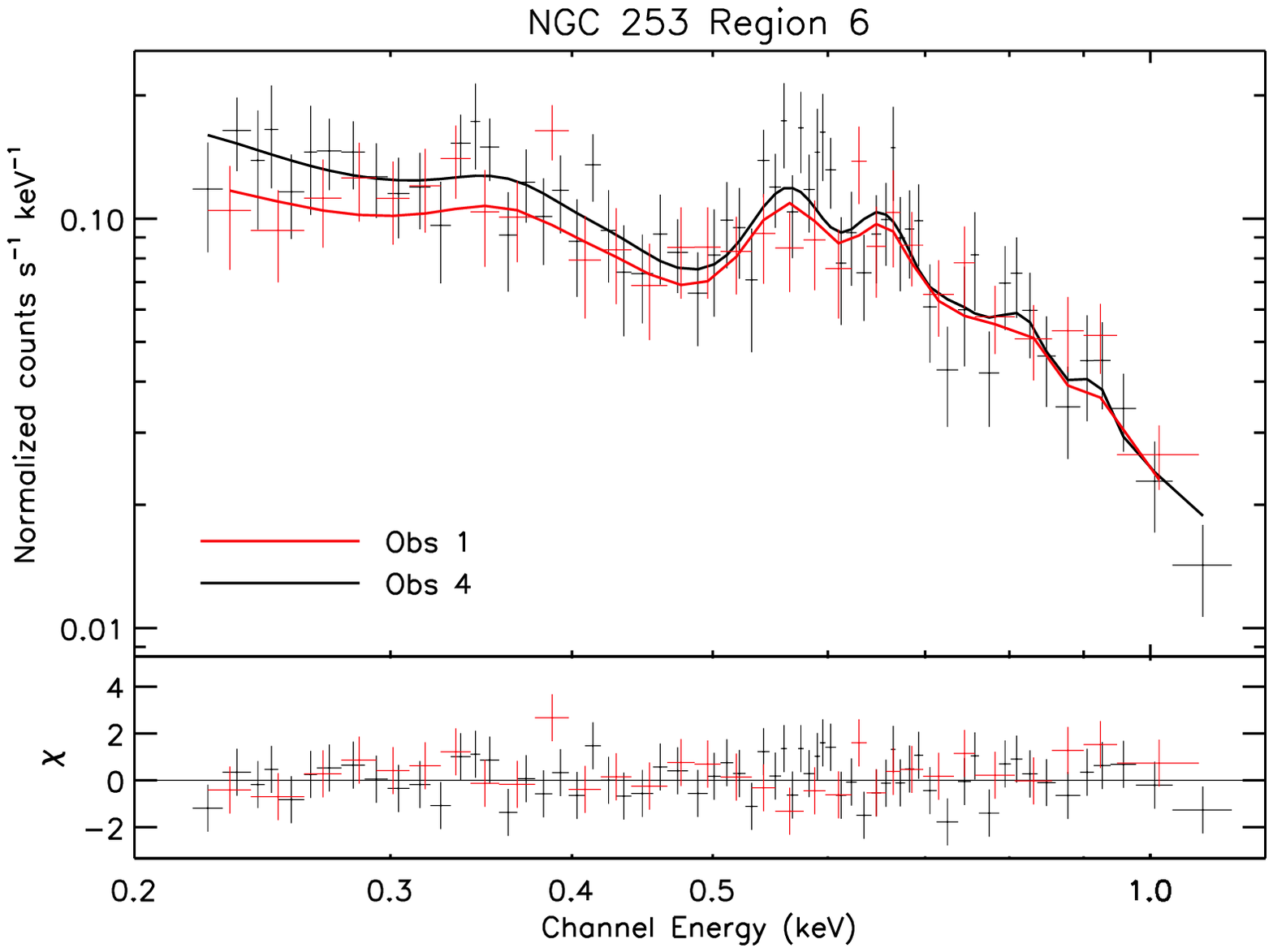}
\includegraphics[width=\specwidth]{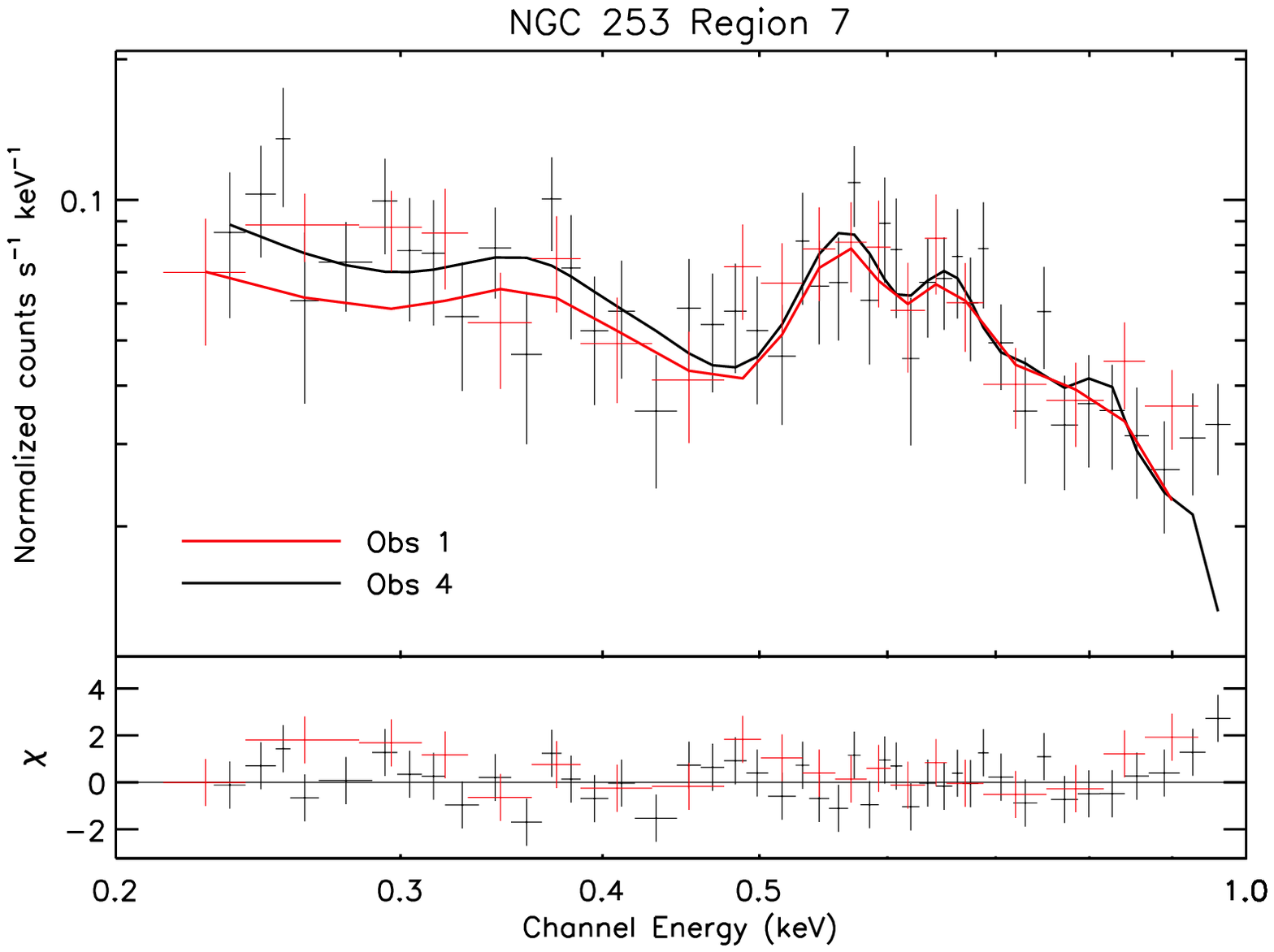}
\includegraphics[width=\specwidth]{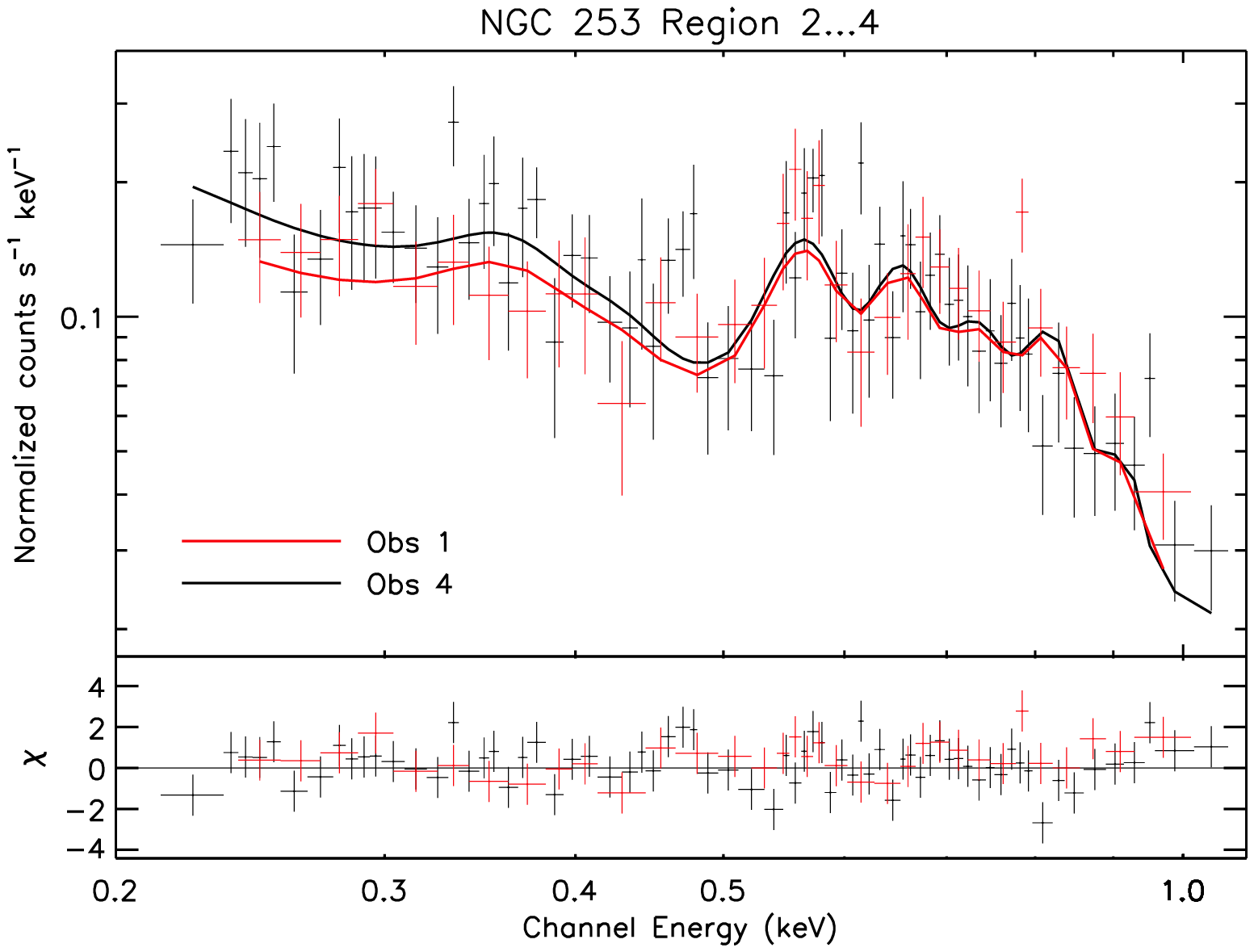}
\includegraphics[width=\specwidth]{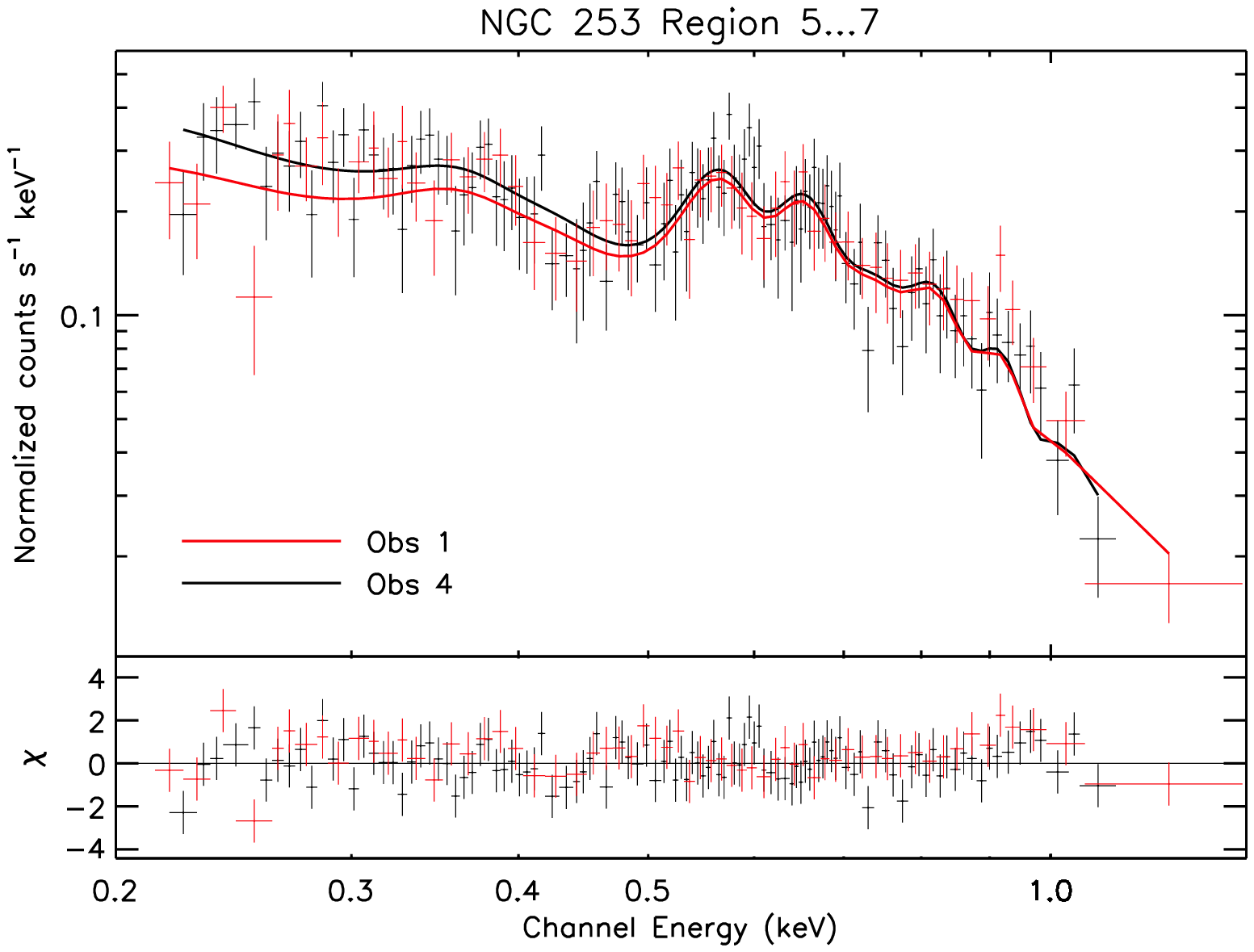}
\includegraphics[width=\specwidth]{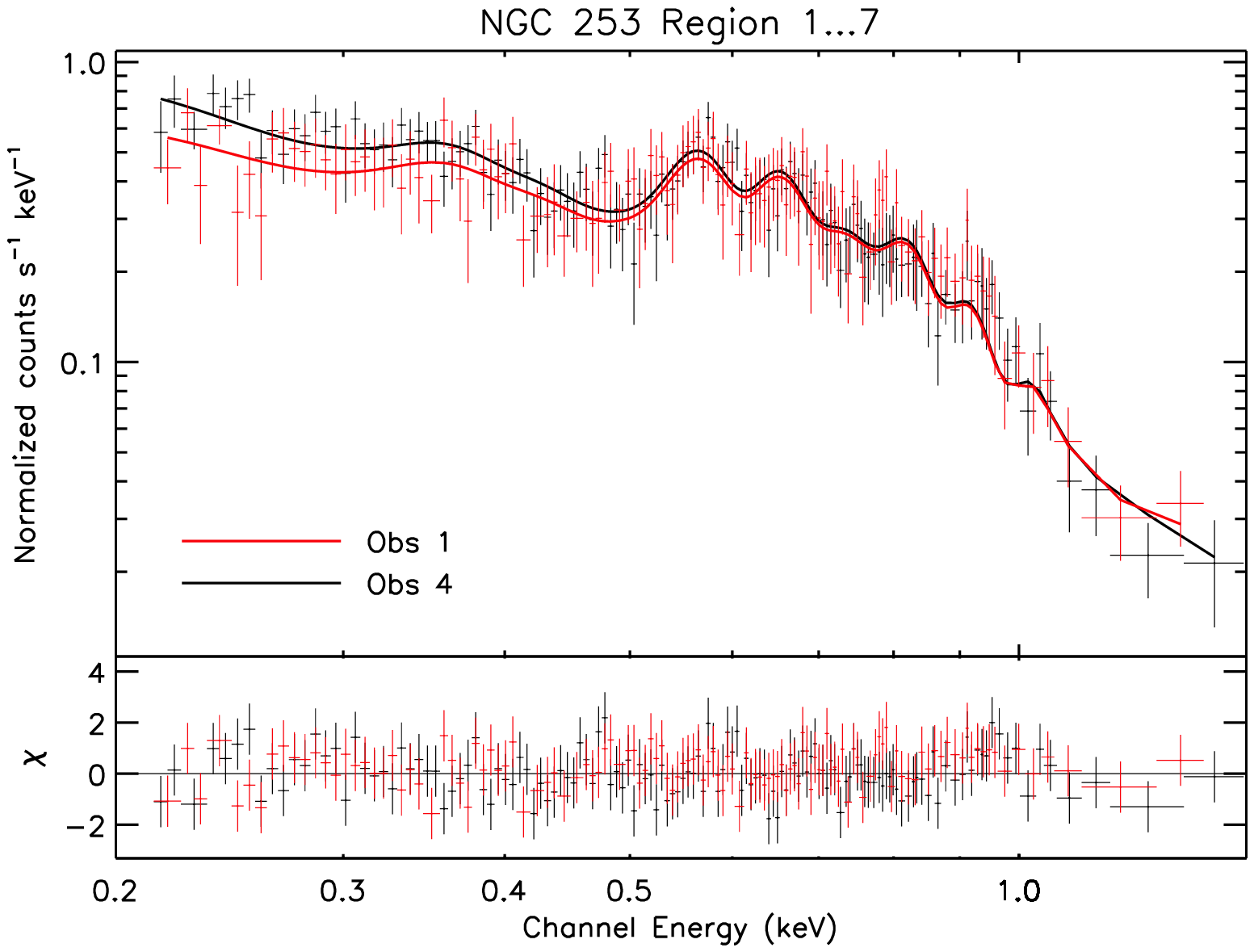}}
\end{minipage}
\caption{\bold{Spectra of regions in the northwestern halo (as Fig.~\ref{fig:specDisk}).}}
\label{fig:specNWhalo}
\end{figure*}

\begin{figure*}
\begin{minipage}[t]{17.4cm}
\fig{\includegraphics[width=\specwidth]{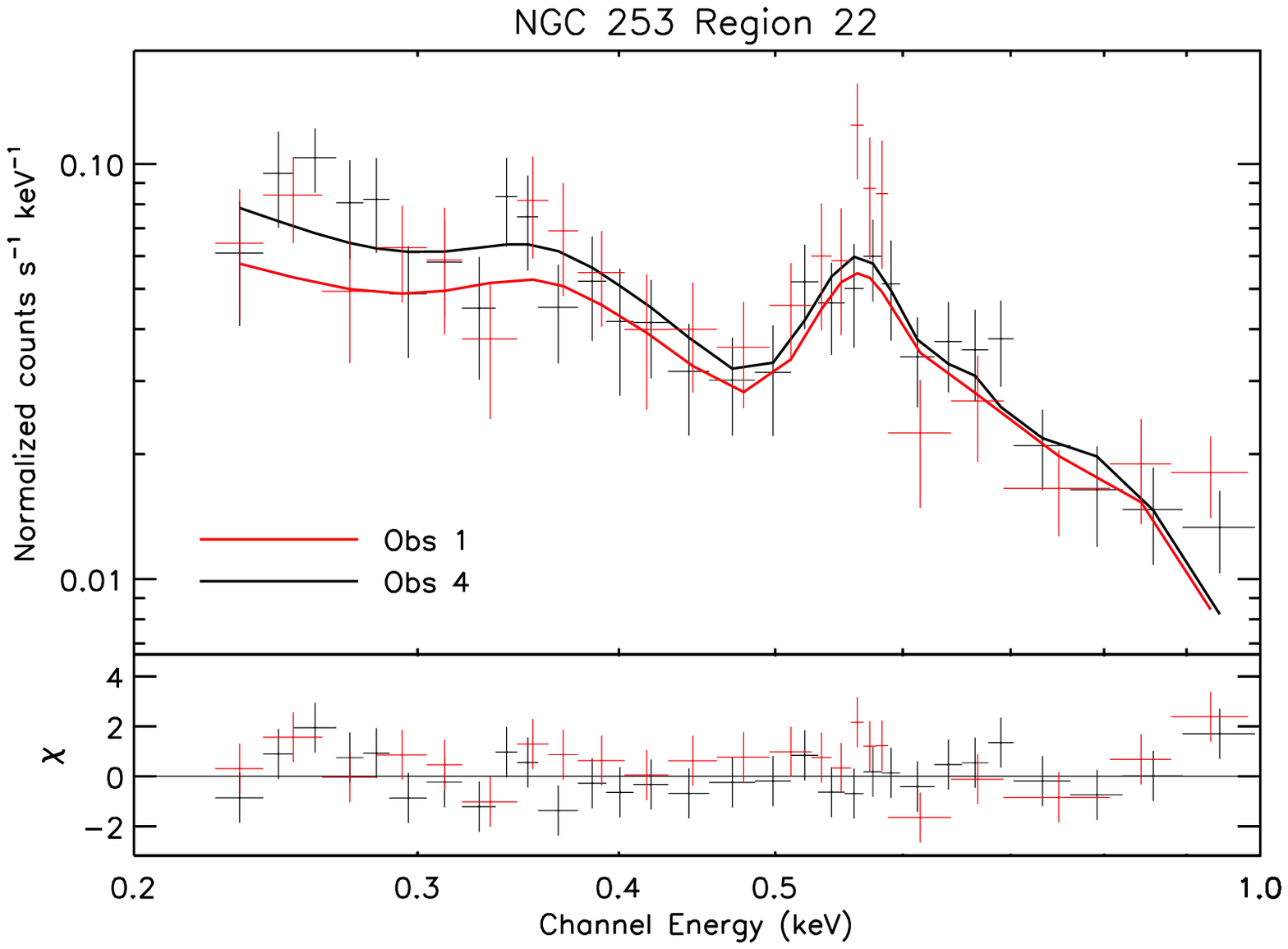}
\includegraphics[width=\specwidth]{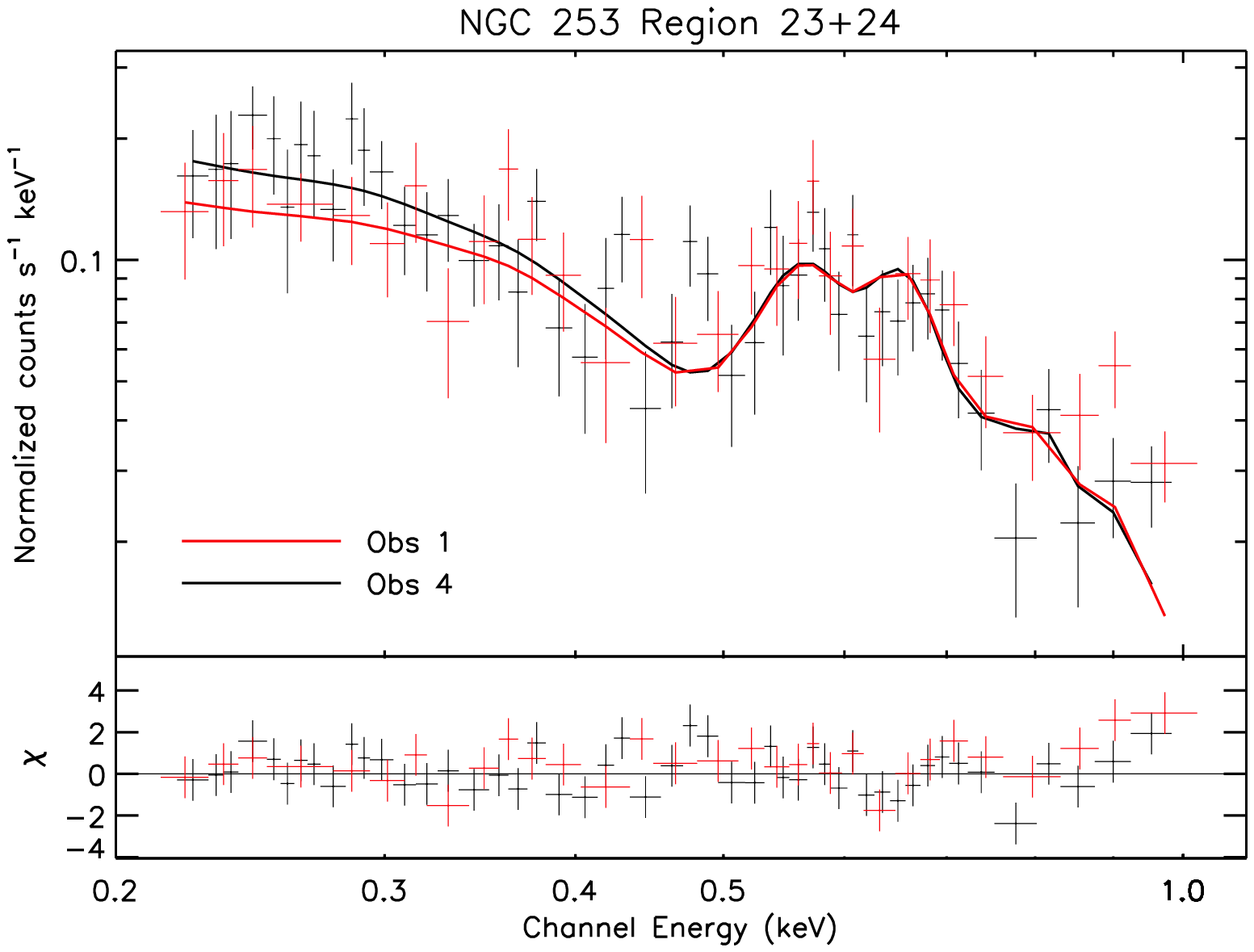}
\includegraphics[width=\specwidth]{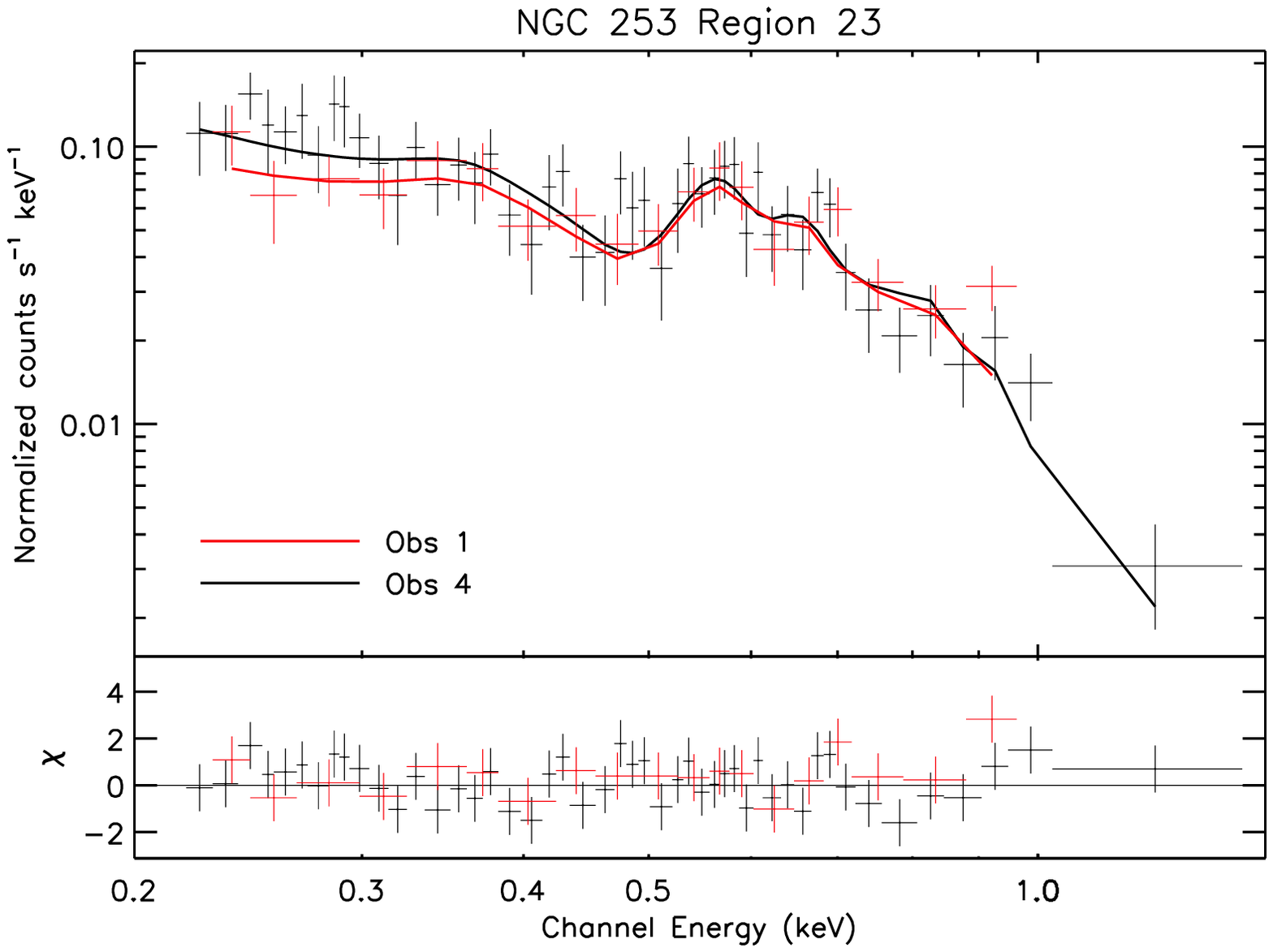}
\includegraphics[width=\specwidth]{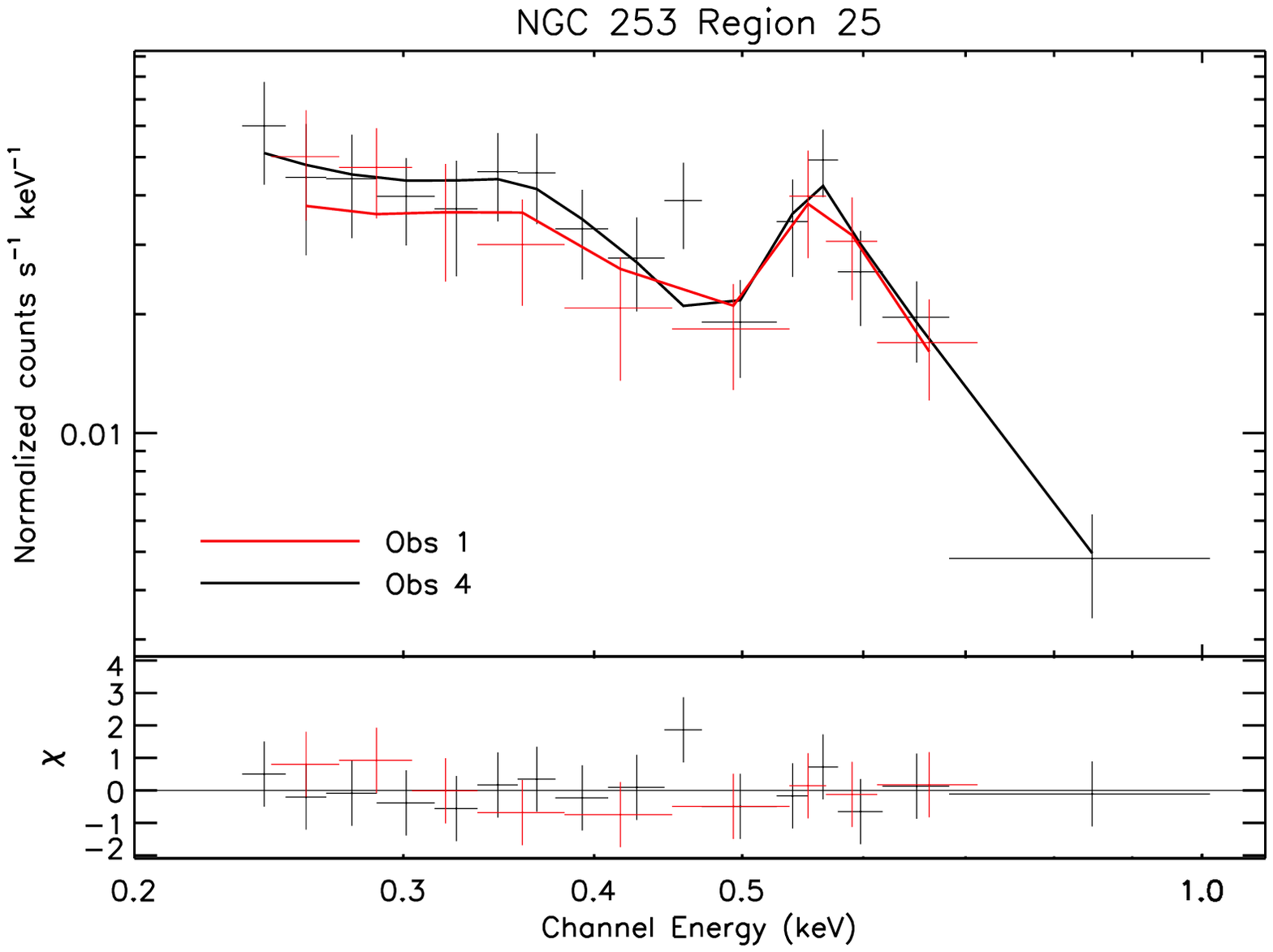}
\includegraphics[width=\specwidth]{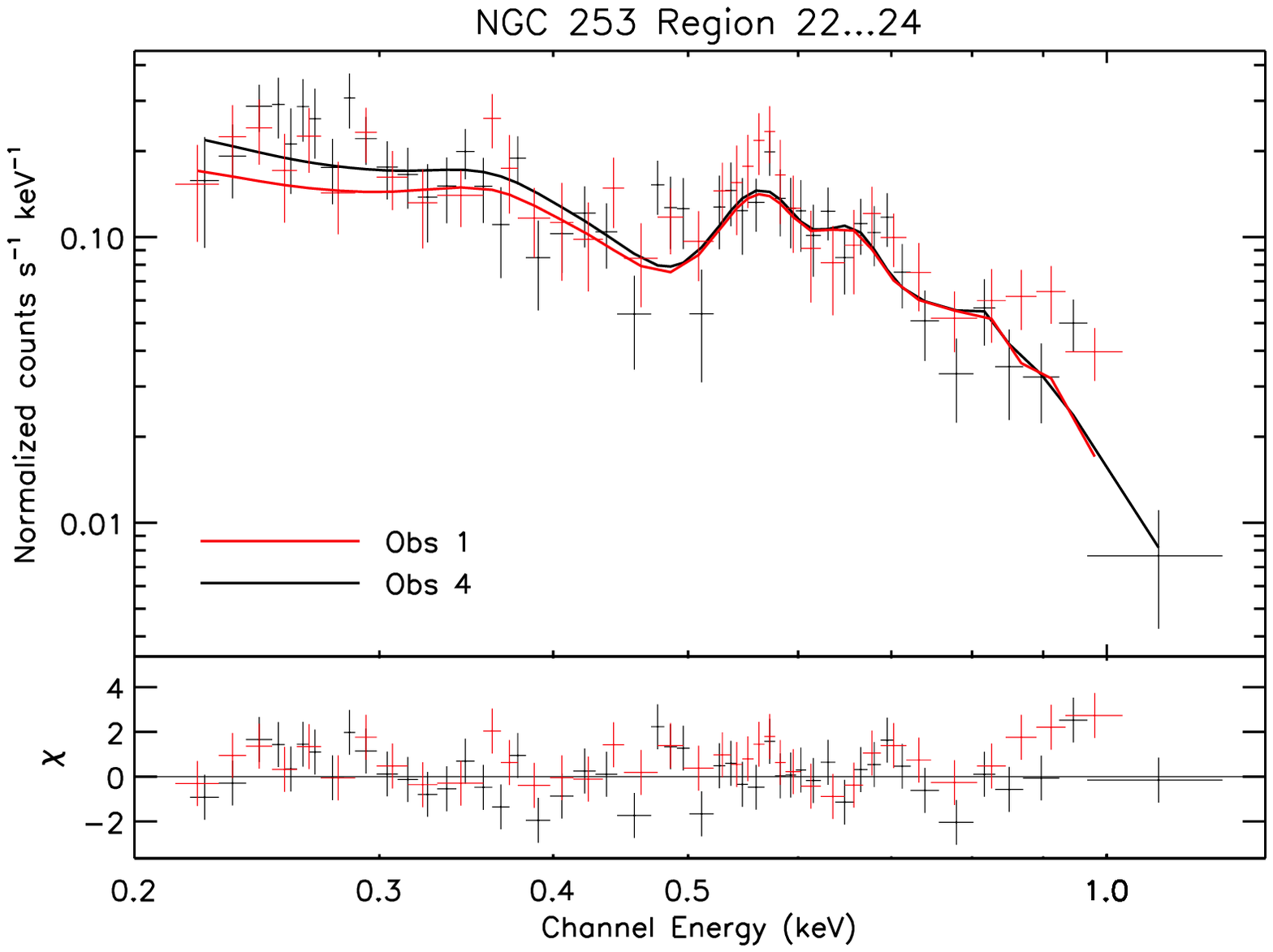}
\includegraphics[width=\specwidth]{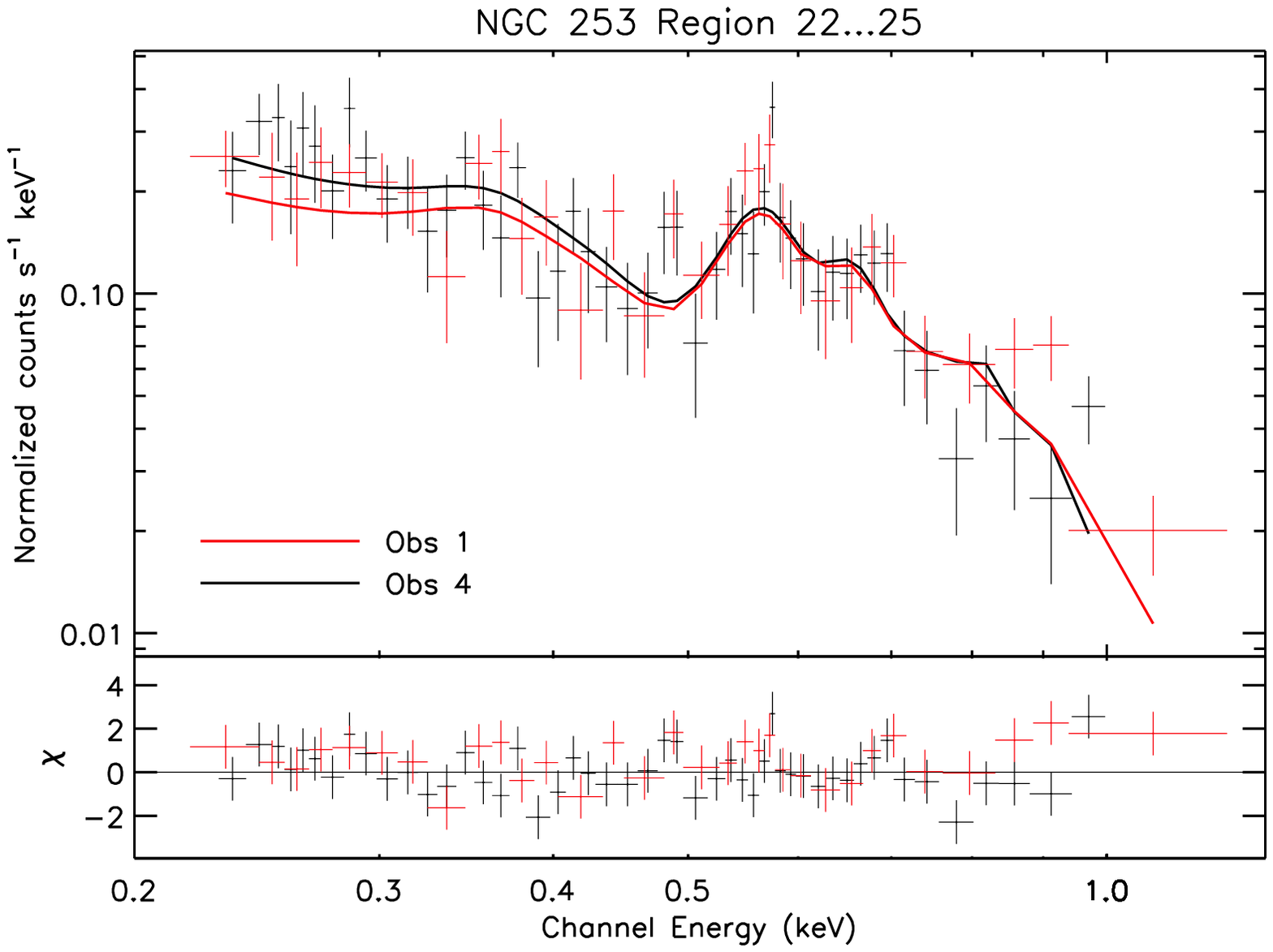}}
\end{minipage}
\caption{\bold{Spectra of different combined regions in the southeastern halo (as Fig.~\ref{fig:specDisk}).}}
\label{fig:specSEhalo}
\end{figure*}

\bold{For the disc regions that lie in the absorption band (region 9) and along the nuclear outflow (regions 16, 17, and 18) we tested for additional absorption components.
In other regions we do not expect a measurable absorber in the foreground with a sufficiently bright source in the background.
In region 9, we obtain a column density of $0.50_{-0.17}^{+0.39}$\hcm{22} for the hard component.
The region covering the northwestern outflow (region 16) closest to the centre shows values of $0.076_{-0.053}^{+0.086}$\hcm{22} and $0.58_{-0.18}^{+0.35}$\hcm{22} \colblue{for the soft and hard component, respectively.
Regions 17 and 18, covering the centre of \n253 and }the southeastern outflow do not seem to require an additional absorption component.}

In a few cases, the two thermal plasma plus a power law component model did not give the best fit.
The southern region on the northeastern end of the disc (region~12) did not require a power law component.
It does not cover a spiral arm of \n253, thus the contribution from point\bold{\colblue{-like}} sources below the detection limit may not be significant.
The region furthest to the south in the disc (region~21) was well fit with one thermal plasma and a power law.
A second thermal plasma was not required.

We found that spectra are harder in regions which cover spiral arms.
The northeastern regions of the disc (regions~11 \& 12) are the best example for this.
The region which covers the spiral arm (region~11) shows temperatures of $0.18^{+0.05}_{-0.04}$ and $0.58^{+0.16}_{-0.18}$~keV, whereas the region adjacent to the spiral arm (region 12) is significantly cooler with temperatures of $0.07\pm0.01$ and $0.25^{+0.04}_{-0.03}$~keV.
The latter spectrum is actually more similar to the typical halo spectrum (see next section).

\comment{\begin{figure}
  \centering
\fig{  \includegraphics[width=8.9cm]{8935f10a.eps}
  \includegraphics[width=8.9cm]{8935f10b.eps}}
  \caption
  {
  Representative spectra of a region in the disc ({\it top}, region~14) and of a region in the halo ({\it bottom}, region~7).
  The red and the black data points and model fits are from observations 1 and 4, respectively (see Table~\ref{tab:Exposures}).
  The lower panel shows the residuals of the fits.
  }
  \label{fig:spectra}
\end{figure}}

\subsection{Halo diffuse emission}
\label{sec:HaloDiffuseEmission}
The halo shows emission only below $\sim$1~keV.
Its projected maximum extent is $\sim$9.0~kpc to the northwest, and $\sim$6.3~kpc to the southeast, perpendicular to the major axis.
The general shape resembles a horn structure.
This was already seen with \ros\ \citep[e.g.][]{PV2000}, and \chandra\ \citep[e.g.][ \bold{\colblue{using the term "X-shaped morphology"}}]{SH2002}.
In the northwestern halo, the EPIC~pn images only show the eastern horn.
In the southeastern halo, both the eastern and the western horn are visible in the energies between 0.2 and 0.5~keV.
At higher energies, the western horn is not visible.

On smaller scales the halo emission seems not to be uniformly distributed.
It shows a \bold{patchy} structure, as was seen before in the \ros\ data.
One notable feature is a brighter knot \bold{(see Fig.~\ref{fig:PN_cheese_image})}, which coincides with the nuclear outflow axis in the northwestern halo at a height of about 3.5~kpc above the disc.
It is bright in energies between 0.2 and 1.0~keV and appears yellow in the false-colour image (Fig.~\ref{fig:PN_RGB_image}).
We checked whether any of the detected structures coincide with chip gaps of the detector, and could therefore be artificial, but no correlation was found.

The spectral properties in different regions in the halo are summarised in Table~\ref{tab:HaloFits}.
To fit the spectra, we applied the same approach as for the disc.
Again, simple models cannot describe the spectra.
A model with two thin thermal plasmas gave a good fit in all regions in the halo.
\bold{In region 6, and all regions containing region 6, \colblue{adding a }power law component results in a better fit, possibly pointing at a cosmic ray contribution.}

The southeastern halo is softer than the northwestern halo, which results in redder colours in the southeastern halo in the EPIC~pn false-colour image (Fig.~\ref{fig:PN_RGB_image}), and also in lower values in HR1.
A fit to the spectrum of the whole northwestern halo gave temperatures of $0.10\pm0.01$ and $0.31\pm0.02$~keV.
The spectrum in the southeastern halo is similar, with temperatures of $0.09\pm0.02$ and $0.29^{+0.03}_{-0.04}$~keV.
The difference in hardness is because the two plasma components contribute different amounts.
Compared to the normalisation of the hotter plasma, the normalisation of the cooler plasma is about \bold{2.1} times stronger in the southeastern halo, with respect to the northwestern halo.

The good statistics of the EPIC~pn data allowed a further subdivision of the halo into smaller regions.
\bold{The spectra of all halo regions are shown in Figs.~\ref{fig:specNWhalo} and \ref{fig:specSEhalo}.
A representative example (region~7) of one of these halo spectra shows prominent oxygen lines at 0.57~keV (\ion{O}{VII}) and 0.65~keV (\ion{O}{VIII}) and also an iron line at $\sim$0.8~keV (\ion{Fe}{XVII}).}

The halo is not uniform in its spectral properties on smaller scales.
The northwestern halo is softer in the east than in the west, while the southeastern halo is softer further away from the disc (see HR1 maps in  Fig.~\ref{fig:HR_C}).
Additionally, \bold{\colblue{there is a marginal indication of a hardening of the emission along the direction of the northwestern outflow (regions 3 and 6, significance: }1.0--1.6$\,\sigma$ in HR1 and 0.40--0.80$\,\sigma$ in HR2).}

The total intrinsic luminosity for the diffuse emission, corrected for the area of the removed point\bold{\colblue{-like}} sources, in the northwestern halo is 8.4\ergs{38} (0.2--1.5~keV), compared to 2.3\ergs{38} in the southeastern halo.
To calculate electron densities, we assumed a volume for the emitting region.
We modeled the northwestern halo with a cylinder with a radius of 4.3~kpc and a height of 4.2~kpc, plus a cylindrical segment with a height of 4.2~kpc, a radius of 4.3~kpc and a width in the southeast-northwest direction of 3.0~kpc (to model region~1).
This gives a volume of 298~kpc$^3$ or $8.7\times 10^{66}$~cm$^3$.
For the southeastern halo we assumed a cylinder with a radius of 3.5~kpc and a height of 2.0~kpc, plus a cylindrical segment with a height of 2.0~kpc, a radius of 3.5~kpc and a width in the southeast-northwest direction of 3.0~kpc (region~25), resulting in a volume of 113~kpc$^3$ or $3.3\times 10^{66}$~cm$^3$.
To calculate densities and the total mass in the emission regions, we corrected the volumes for the cut-out point\bold{\colblue{-like}} sources.
Using the emission measure of the fit (cf.\ the documentation of the apec model in XSPEC), the resulting densities are $3.2\,\eta^{-0.5}\times10^{-3}$~cm$^{-3}$ and $4.7\,\eta^{-0.5}\times10^{-3}$~cm$^{-3}$ for the northwestern and southeastern halo, respectively.
$\eta$ is the volume filling factor ($\eta\le1$).
With solar abundances from \cite{WA2000}, this implies total masses of $3.3\,\eta^{-0.5}\times 10^7$~M$_{\sun}$ and $1.8\,\eta^{-0.5}\times 10^7$~M$_{\sun}$ for the northwestern and southeastern halo, respectively.

\onecolumn
\begin{small}
\begin{landscape}
\begin{longtable}{llcccccccccc}
\caption{\bold{Spectral fits to the disk regions.}}
\label{tab:DiskFits} \\

\hline \hline
Region/model                                & $\chi^2_{\nu}\,(\chi^2/\nu)$ & \bold{\colblue{BIC}}                       & \nh                       & kT                              & norm                              & \nh                               & kT                                        & norm                           & $\Gamma$                         & norm                 \\
                                            &                              &                           & ($10^{22}$ cm$^{-2}$)     & (keV)                           &($10^{-5}$)                        & ($10^{22}$ cm$^{-2}$)             & (kev)                                     & ($10^{-5}$)                    &                                  & ($10^{-5}$)          \\
\hline
\endfirsthead

\multicolumn{12}{c}{{\tablename} \thetable{} -- Continued} \\
\hline \hline
Region/model                                & $\chi^2_{\nu}\,(\chi^2/\nu)$ & \bold{\colblue{BIC}}                       & \nh                       & kT                              & norm                              & \nh                               & kT                                        & norm                           & $\Gamma$                         & norm                 \\
                                            &                              &                           & ($10^{22}$ cm$^{-2}$)     & (keV)                           &($10^{-5}$)                        & ($10^{22}$ cm$^{-2}$)             & (kev)                                     & ($10^{-5}$)                    &                                  & ($10^{-5}$)          \\
\endhead

\hline
\multicolumn{12}{l}{{Continued on next page\ldots}}\\
\endfoot

\hline
\multicolumn{12}{l}{$^1$n.c.: not constrained.}\\
\multicolumn{12}{l}{
\begin{minipage}[t]{23.5cm}
\footnotesize NOTE: All errors are 90\% confidence for a number of interesting parameters equal to the number of free parameters in the model (2, 4, 6, 7, 8, or 9 free parameters depending on the model).
Thin thermal plasma models are apec, absorption models are tbabs.
All models have a fixed foreground absorption of $1.3 \times 10^{20}$~\cm-2. 
The preferred model is printed in bold font.
\end{minipage}
}
\endlastfoot

Region 9     \comment{C10}\\*
$\quad$ po                                  & 2.8 (143.4/52)               & 60.7                      & $\dots$                   & $\dots$                         & $\dots$                           & $\dots$                           & $\dots$                                   & $\dots$                        & $1.7_{-0.1}^{+0.1}$              & $1.8_{-0.2}^{+0.2}$  \\*
$\quad$ apec                                & 4.5 (232.9/52)               & 86.9                      & $\dots$                   & $0.37_{-0.03}^{+0.03}$          & $1.1_{-0.2}^{+0.2}\times 10^5$    & $\dots$                           & $\dots$                                   & $\dots$                        & $\dots$                          & $\dots$         \\*
$\quad$ apec+po                             & 1.2 (60.3/50)                & 21.9                      & $\dots$                   & $0.36_{-0.06}^{+0.09}$          & $1.8_{-0.7}^{+0.7}$               & $\dots$                           & $\dots$                                   & $\dots$                        & $1.3_{-0.3}^{+0.3}$              & $1.1_{-0.3}^{+0.3}$  \\*
$\quad$ apec+apec                           & 1.2 (58.9/51)                & 19.8                      & $\dots$                   & $0.37_{-0.07}^{+0.14}$          & $1.6_{-0.6}^{+0.5}$               & $\dots$                           & $7.74_{-5.27}^{+\mbox{n.c.}^1}$           & $4.9_{-1.2}^{+1.4}$            & $\dots$                          & $\dots$    \\*
$\quad$ apec+apec+po                        & 0.9 (44.3/48)                & 13.2                      & $\dots$                   & $0.24_{-0.09}^{+0.09}$          & $1.2_{-0.8}^{+1.2}$               & $\dots$                           & $0.72_{-0.21}^{+0.28}$                    & $1.1_{-0.9}^{+0.6}$            & $0.96_{-0.34}^{+0.44}$           & $0.81_{-0.49}^{+0.43}$  \\
\hline
\end{longtable}
\end{landscape}
\end{small}
\twocolumn

\onecolumn
\begin{small}
\begin{longtable}{llcccccccc}
\caption{\bold{Spectral fits to the northwestern halo regions.}}
\label{tab:HaloFits}  \\

\hline  \hline 
Region/model                       &  $\chi^2_{\nu}\,(\chi^2/\nu)$ & \bold{\colblue{BIC}}                      &  kT 		     &  norm   	  	&  kT			   &  norm		      &  $\Gamma$		 &  norm				  \\
                                   &                               &                          &  (keV)		     &($10^{-5}$)			&  (kev)		   &  ($10^{-5}$)	      & 			 &  ($10^{-5}$) 		   \\
\hline
\endfirsthead

\multicolumn{10}{c}{{\tablename}  \thetable{}  --  Continued}  \\
\hline  \hline 
Region/model                       &  $\chi^2_{\nu}\,(\chi^2/\nu)$ & \bold{\colblue{BIC}}                      &  kT 		     &  norm				&  kT			   &  norm		      &  $\Gamma$		 &  norm				  \\
                                   &                               &                          &  (keV)		     &($10^{-5}$)			&  (kev)		   &  ($10^{-5}$)	      & 			 &  ($10^{-5}$) 		   \\
\endhead

\hline
\multicolumn{10}{l}{{Continued on next page\ldots}}\\
\endfoot

\hline
\multicolumn{10}{l}{$^1$n.c.: not constrained.}\\
\multicolumn{10}{l}{
\begin{minipage}[t]{17cm}
\footnotesize NOTE: All errors are 90\% confidence for a number of interesting parameters equal to the number of free parameters in the model (2, 4, or 6 free parameters depending on the model).
Thin thermal plasma models are apec, absorption models are tbabs.
All models have a fixed foreground absorption of $1.3 \times 10^{20}$~\cm-2. 
The preferred model is printed in bold font.
\end{minipage}
}
\endlastfoot

\multicolumn{2}{l}{Northwestern halo}\\
\hline
Region  1\\*
$\quad$  po                        &  1.1 (25.9/23)                & 7.3  	                  & $\dots$		  & $\dots$			    & $\dots$		      & $\dots$ 		& $2.0_{-0.3}^{+0.3}$	  & $6.1_{-1.3}^{+1.4}$  \\*
$\quad$  apec                      &  3.9 (88.6/23)                & 38.1 	                  & $0.25_{-0.02}^{+0.02}$  & $1.1_{-0.2}^{+0.3}\times 10^5$  & $\dots$		      & $\dots$ 		& $\dots$		  & $\dots$ \\*
$\quad$  apec+po                   &  1.1 (22.1/21)                & 9.8     		              & $0.19_{-0.19}^{+0.21}$  & $1.3_{-1.3}^{+2.9}$ 	    & $\dots$		      & $\dots$ 		& $1.9_{-0.6}^{+0.6}$	  & $5.6_{-2.0}^{+1.8}$  \\*
$\quad$  \textbf{apec+apec}        &  1.0 (21.1/21)                & 8.6     		              & $0.10_{-0.02}^{+0.02}$  & $19_{-8.3}^{+8.5}$  	    & $0.35_{-0.05}^{+0.06}$  & $5.6_{-1.5}^{+1.6}$	& $\dots$		  & $\dots$		    \\*
$\quad$  apec+apec+po              &  0.9 (16.3/19)                & 8.6     		              & $0.09_{-0.07}^{+0.02}$  & $20_{-20}^{+21}$		    & $0.29_{-0.29}^{+0.20}$  & $2.6_{-2.6}^{+3.1}$	& $0.55_{-\mbox{n.c.}^1}^{+1.75}$ & $5.3_{-3.6}^{+4.0}$ \\*
\hline
Region  2\\
\hline
\end{longtable}
\end{small}
\twocolumn

\begin{table}
\begin{minipage}[t]{88mm}
\caption{\bold{Flux and luminosity values in the extraction regions.}}
\label{tab:fluxlum}
\centering
\renewcommand{\footnoterule}{}  % to avoid a line before footnotes
\begin{tabular}{lcc}
\hline
\hline
Region                        & $F_X$  \comment{L=F*7.966E50 (2.58Mpc}             & $L_X$               \\
                              & (erg s$^{-1}$ cm$^{-2}$)                           & (erg s$^{-1}$)      \\
\hline
\multicolumn{2}{l}{Northwestern halo}\\
\hline
1                             & $1.7 _{-0.7}^{+0.5}\times 10^{-13}$ (0.2--1.0~keV) & $1.4\times 10^{38}$ \\
2                             & $8.4 _{-3.7}^{+2.7}\times 10^{-14}$ (0.2--1.0~keV) & $6.7\times 10^{37}$ \\
3                             & $1.1 _{-0.4}^{+0.3}\times 10^{-13}$ (0.2--1.0~keV) & $8.8\times 10^{37}$ \\
4                             & $8.7 _{-3.3}^{+2.3}\times 10^{-14}$ (0.2--1.0~keV) & $6.9\times 10^{37}$ \\
5                             & $1.0 _{-0.4}^{+0.3}\times 10^{-13}$ (0.2--1.0~keV) & $8.0\times 10^{37}$ \\
\hline
\end{tabular}
\end{minipage}
\end{table}

%%%%%%%%%%%%%%%%%%%%%%%%%%%%%%%%%%%%%%%%%%%%%%%%%%%%%%%%%%%%%%%%%%%%%%%%%%%%%%%%%%%%%%%%%%%%%%%%%%%%%%%%%%%%%%%%%%%%%%
%%%%%%%%%%%%%%%%%%%%%%%%%%%%%%%%%%%%%%%%%%%%%%%%%%%%%%%%%%%%%%%%%%%%%%%%%%%%%%%%%%%%%%%%%%%%%%%%%%%%%%%%%%%%%%%%%%%%%%

\section{Discussion}

\subsection{The extent of the diffuse emission of \n253}
Extended emission from the soft northwestern halo was first reported from \ein\ observations \citep[][]{F1988}.
Later, observations with \ros\ also discovered the southeastern halo in X-rays \citep[\eg][]{PV2000}.
The \ros\ images in the soft band trace the emission in the outer halo to projected distances of up to 9~kpc, both in the northwest and the southeast direction.
With \xmm, the emission is detected out to 9.0~kpc to the northwest and 6.3~kpc to the southeast. 
This difference in the southeastern halo can be explained by the high \ros\ sensitivity extending down to 0.1~keV.
The useful \xmm\ EPIC~pn range is limited to 0.2~keV.
This makes a big difference as there are many strong lines from \ion{O}{IV}, \ion{Ne}{VIII}, \ion{Mg}{IX}, \ion{Mg}{X}, \ion{Si}{IX}, and \ion{Si}{X} in the energy band between 0.1 and 0.2~keV. 
For a thermal plasma at a temperature of $\sim$0.1~keV, these lines are even stronger than the \ion{O}{VII} and \ion{O}{VIII} lines, and about 60\% of the total flux in the energy band from 0.1 to 2.0~keV originates from lines below 0.2~keV.
The southeastern halo shows softer emission than the northwestern halo, therefore, the effect is strongest in the southeastern halo.

Also in the disc the extent of the emission is different.
The \ros\ images trace the soft emission $\sim$6.8~kpc towards the northeast and $\sim$5.3~kpc towards the southwest.
With \xmm, the disc emission has an extent of $\sim$7.2~kpc and $\sim$6.3~kpc to the northeast and southwest, respectively. 
The disc spectra are harder than the halo spectra, and therefore the higher \xmm\ sensitivity at energies $>$0.4~keV comes into play.

\subsection{Is the diffuse emission in the disc really from hot interstellar gas?}

\begin{figure*}
\centering
\fig{\includegraphics[width=\textwidth]{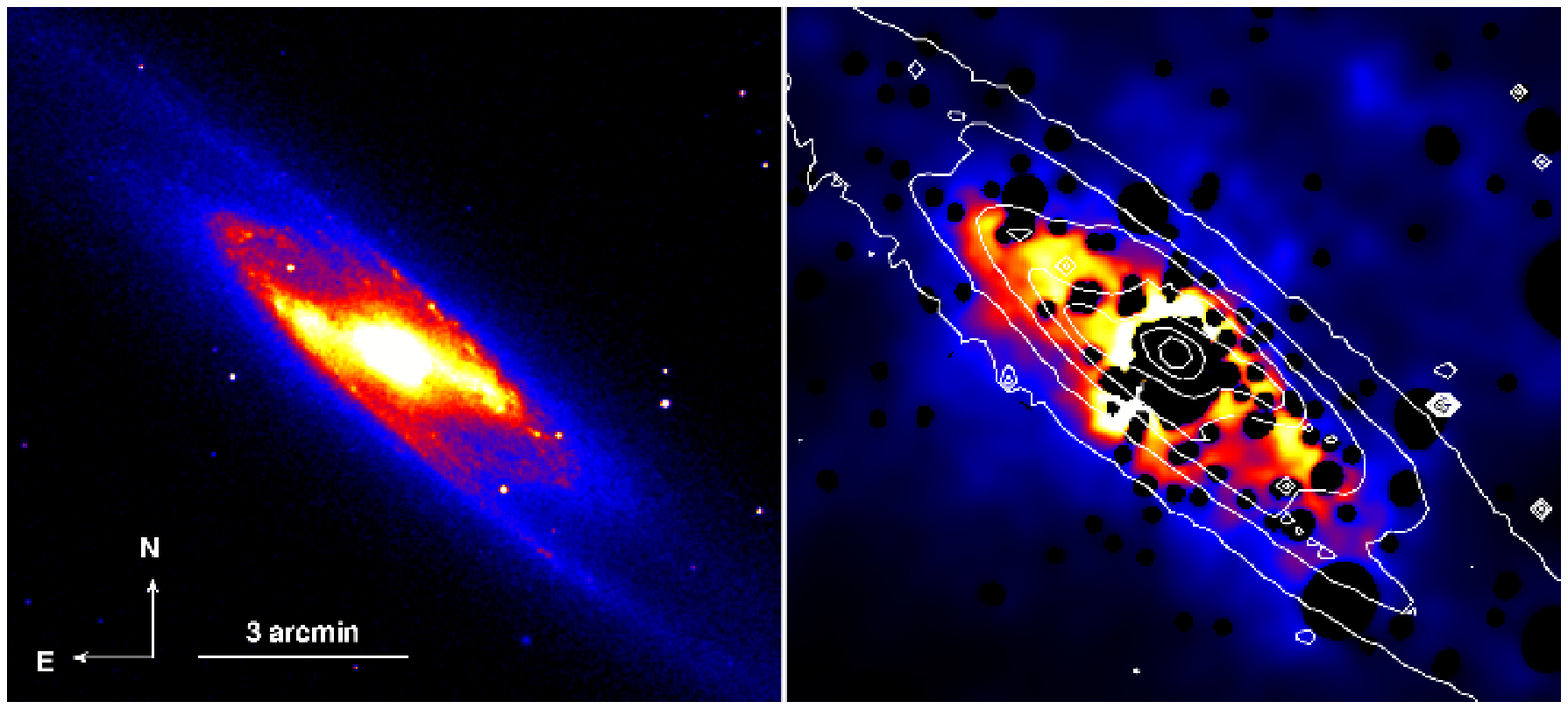}}
\caption{Comparison between the K-band \citep[left, 2MASS,][]{JC2003} and X-ray (right) morphology. The white contours overplotted on the X-ray image represent the K-band brightness levels. Both images are on a linear colour scale, and on the same spatial scale as indicated in the K-band image.}
\label{fig:XrayVsKband}
\end{figure*}

\bold{In this section we investigate the possibility that the observed diffuse emission originates in fact not from a truly diffuse hot interstellar gas component, but from a large population of weak stellar-type X-ray sources that shows the same characteristics.}

\bold{This effect was first discovered in the Milky Way's ridge X-ray emission \citep[\eg][]{RS2006}, where evidence was found that the bulk of the Galactic ridge X-ray emission is composed of weak X-ray sources, mostly cataclysmic variables and coronally active binary systems, with a luminosity of most of these sources of less than \oergs{31}.
Also in other galaxies, \cite{RC2007} found that the apparently diffuse emission is consistent with the emission from an old stellar population like in the Milky Way.}

Following the method by \cite{RC2007}, we used K-band observations, to infer the emissivity of the diffuse X-ray component per unit stellar mass.
We derived the near-infrared luminosity and stellar mass of \n253, using the total K-band magnitude of 3.772 \citep{JC2003}, the distance modulus of 27.06, corrected for interstellar extinction of 0.007 \citep{SF1998}, and the colour-dependent K-band mass-to-light ratio from \cite{BJ2001}, $\log(M_*/L_{\rm{K}})=-0.692+0.652\times(B-V)$, with $(B-V)=-0.16$ \citep{CG2003}.
This yielded a total K-band luminosity $L_{\rm{K}}=1.6\times10^{11}$~$L$\subsun\ and a total stellar mass $M_*=2.6\times10^{10}$~$M$\subsun.
With a X-ray luminosity of \n253\ of 2.0\ergs{39} (0.5--10~keV), the emissivity of the diffuse X-ray component per unit stellar mass then resulted in 

\begin{equation*}
\frac{L_{0.5-10\,\,\mathrm{keV}}}{M_*}=7.6^{+0.9}_{-0.3} \,(\pm 2.3)\times10^{28}\,\,\mathrm{erg}\,\, \mathrm{s}^{-1}\,\, M^{-1}_{\odot}.
\end{equation*}

The errors are statistical errors on the measured X-ray flux.
Additionally, we assumed an uncertainty of $\sim$30\% (given in parentheses), which might be associated with the $L_{\mathrm{K}}$ to $M_*$ conversion \citep{BM2003}.
The emissivity of \n253 should only be considered a lower limit.
We cut out a quite large region in the centre of \n253 and corrected for this by filling the hole with the average flux of the disc.
Therefore, the obtained X-ray luminosity as well as the emissivity are probably \bold{underestimated}.

From the luminosity and other properties of the Galactic ridge X-ray emission \citep[\eg][]{RS2006} and from direct measurements of the luminosity function of sources in the solar neighbourhood \citep{SR2006}, the combined 0.5--10~keV emissivity of cataclysmic variables and coronally active stars has been estimated as $L_{\mathrm{X}}/M_*\sim1.2\pm0.3\times10^{28}\,\,\mathrm{erg}\,\,\mathrm{s}^{-1}\,\,M^{-1}_{\odot}$.
The value derived for \n253\ is larger than the value for the Milky Way, indicating the presence of a hot gaseous component.

An even stronger argument is the following: if the diffuse X-ray emission is produced by an old stellar population, then their morphologies should be similar.
A comparison of the X-ray emission with the 2MASS K-band image \citep{JC2003} is shown in Fig.~\ref{fig:XrayVsKband}.
We found that the X-ray morphology does not match the K-band morphology, therefore the diffuse emission is indeed not simply due to an old stellar population, but has to have a truly diffuse component.

\subsection{Spectral fits and variations in the halo}
As it was mentioned already in earlier publications, there is an ambiguity in the spectral fits between a pure multi-temperature thermal plasma model and a combination of thermal plasmas plus a power law component \citep[e.g.][]{DP2000,SH2002}.
This ambiguity in the halo emission still exists with the \xmm\ data.
Fits to the halo spectra with a thermal plasma plus a power law model (see Table~\ref{tab:HaloFits}) resulted in similar $\chi^2_{\nu}$, as for a multi-temperature thermal plasma model.
A power law component from point sources could be excluded, since we were careful to remove any point source contribution.
Another source for non-thermal emission could be synchrotron emission from cosmic ray electrons that are advected with the superwind or are accelerated locally in internal wind shocks.
\bold{This is possibly what we observe in \colblue{the spectrum of }region 6 (and spectra containing region 6). \colblue{This region }lies along the northwestern outflow axis.
Here the spectral fit improves significantly by adding a power law component to the two thermal plasmas}.
A comparison of the X-ray emission to the 330~MHz and 1.4~GHz radio emission \citep{CH1992} showed that the radio emission is more extended, and does not show the horn structure that we see in X-rays.
Because of this inequality, we prefer the multi-temperature thermal plasma \bold{over the single temperature model plus power law model for most of} the X-ray halo emission at the moment.

%%
%% DB: here is my contribution to extend the NEI discussion as required by the referee
%%
\bold{The appearance of a ``multi-temperature'' halo is a natural consequence of the plasma driven out of collisional ionisation equilibrium (CIE) towards a non-equilibrium ionisation (NEI) state in a galactic outflow. 
It has to be borne in mind that the processes of ionisation, recombination and adiabatic cooling occur all  on different time scales. 
Whereas the former two are controlled by microphysics the latter is due to the hydrodynamics of the flow. 
As has been described by \eg \cite{BS1994} adiabatic cooling can become the shortest in an accelerating outflow, like the one occurring in a starburst driven wind, leading to delayed 
recombination of the plasma. 
This means essentially the presence of highly ionised species in a plasma flow with low kinetic electron temperature, similar to the ions in the solar wind. 
Releasing the energy stored in these highly ionised species results in diffuse soft X-ray emission and some spectral hardening. 
Observing this with the limited spectral resolution of \chandra\ or \xmm\ CCDs, reveals a multi-temperature halo, although physically the plasma does by no means have ``different temperatures'', but is characterised by a non-equilibrium distribution of ions, which by itself is the result of the thermodynamic path the system has taken and reflects directly the hydrodynamic flow characteristics}.
A currently ongoing analysis with non-equilibrium models \citep[\eg][]{BS1999} might also be able to explain the observations \bold{and is the subject of a forthcoming paper}.

The northwestern halo shows significant hardness variations in HR1, as opposed to the findings by \cite{SH2002}.
We checked if this can be caused by a different energy band selection, but the result is independent whether we use the bands from \cite{SH2002} (0.3--0.6~keV and 0.6--1.0~keV) or our own.
These hardness variations might also be a sign of NEI X-ray emission.

\subsection{Temperatures, abundances and column densities}
\label{sec:abund}
The X-ray emission from \n253 has been observed before with several other X-ray observatories. 
Especially the early observatories did not allow to separate the point sources from the diffuse emission since the point spread function was quite large.
Hence, only a combined fit of the emission from the halo, the disc, and the nuclear region was possible.
Temperatures of multi-temperature models ranged between 0.1 and 0.3~keV for the low, and between 0.6 and 0.7~keV for the high temperature component \citep{DW1998, WH2000, DP2000}.
Reported abundances were mostly highly subsolar and therefore unphysical for a supposedly metal enriched starburst galaxy plasma.

Only the X-ray observatories like \xmm, \chandra, and to some degree \ros\ allow us to separate the halo from the disc emission and to remove contribution by point sources via a spatial selection.
From \ros\ data, \cite{PV2000} inferred a foreground absorbed two-temperature thermal model with temperatures of 0.13 and 0.62~keV for the northwestern halo emission.
No highly subsolar abundances were required.
The disc emission could be explained by a 0.7~keV thermal plasma and an additional thermal plasma (kT=0.2~keV) in front of the disc coming from a coronal component.

The first \xmm\ results by \cite{PR2001} required a two-component model for the disc emission with temperatures of 0.13 and 0.4~keV plus residual harder emission, possibly from unresolved point sources.
For the nuclear region three temperatures were needed (0.6, 0.9, and 6~keV).
Both models used solar abundances.
No analysis of the halo emission was presented in their paper.

The best spatial resolution is provided by the \chandra\ observatory.
Results on the diffuse disc and halo emission were first published by \cite{SH2002}.
For the halo emission they needed a multi-temperature model (apec) with at least two temperatures of 0.24 and 0.71~keV (the latter with quite large errors) and with a foreground absorption of $5.3\times 10^{20}$~\cm-2.
A power law ($\Gamma=3.3$) plus a thermal plasma (kT=0.24~keV) gave a similarly good fit.
A combination of other thermal models (mekal) or non-equilibrium models (vnei) did not result in better fits.
The diffuse emission from the disc was fitted with the same models, however, the temperatures were lower than in the halo, with 0.17 and 0.56~keV, respectively.
The foreground absorption yields $4.7\times 10^{20}$~\cm-2.
In all cases unphysically sub-solar abundances had to be assumed.

The temperature values for the halo emission, as found by our analysis, are lower than the ones from previous observations.
Our soft component is about 0.10~keV, which is still compatible with the \ros\ results.
However, the hard component is only $\sim$0.31~keV for the northwestern halo and $\sim$0.29~keV for the southeastern halo.
A higher temperature was not necessary in any of our fits.
A possible explanation could be the way the spectra were background subtracted.
We used a sophisticated method (see App.~\ref{app:bkgspec}) that uses the local background at the border of the field of view, where no emission from \n253\ is expected, while other authors used \eg blank-sky observations \citep{SH2002}.
Using a background from different times and different fields on the sky can lead to systematic effects in the background substraction.
A background region with a higher contribution of the local bubble could, for example, lead to an over-correction, especially at very soft energies ($<$0.5~keV).

In the disc we found temperatures between 0.1 and 0.3~keV and between 0.3 and \bold{0.9}~keV, for the soft and the hard component, respectively.
This is consistent with earlier results.

We also tried to constrain the metal abundances in our fits.
However the errors on the obtained values are so large, that we are not able to give well constrained abundances (northwestern halo: $Z=0.3^{+4.7}_{-0.2}$~$Z$\subsun, southeastern halo: $Z=0.4^{+4.6}_{-0.3}$~$Z$\subsun, disc: $Z=1.0^{+0.9}_{-0.7}$~$Z$\subsun).
Since we do not expect highly subsolar abundances in an environment which is enriched with metals from the starburst via the superwind and galactic fountains, we fixed the abundances in our analysis to solar.
This is entirely consistent with the above values.
A reason for the low abundances, found with different instruments, could be that due to a lower spatial resolution and or/and sensitivity more point sources contribute to the final spectrum, increasing the continuum flux.
The ratio of line emission to continuum flux is therefore decreased, which mimics the spectral shape of a plasma with low metal abundances.
A similar effect can be achieved when a NEI spectrum is fitted with CIE models.
Also, a too simplistic model could be the reason, combining regions with different temperatures.

In the disc, \bold{\colblue{two}} of the regions required an extra absorption component in the spectral model.
The additional column densities range between \bold{0.76\hcm{21} and 5.8\hcm{21}}.
Direct radio measurements of the \ion{H}{I} column density showed lower values than we derived from the X-ray data.
An interpolation of the \ion{H}{I} maps by \cite{PC1991} and \cite{KW1995} resulted in \bold{($\sim$2.4 and 3.4)\comment{, and 4.8}\hcm{21} for regions 9 and 16,\comment{ and 17} respectively}.
However, the \ion{H}{I} value for the region including the nucleus of \n253 (region 16) is affected by \ion{H}{I} absorption, so the resulting column density can only be considered as a lower limit.
Additional absorption is expected from molecular hydrogen.
\cite{MH1995} derived the H$_2$ column density in the direction of the nucleus of \n253 to 3.7\hcm{23}. 
Taking this value as an upper limit for the column density in the disc regions around the nucleus, the column densities derived from X-ray spectra are within the limits from radio observations.

\subsection{X-ray versus UV morphology}

\begin{figure}
\centering
\fig{\includegraphics[width=88mm]{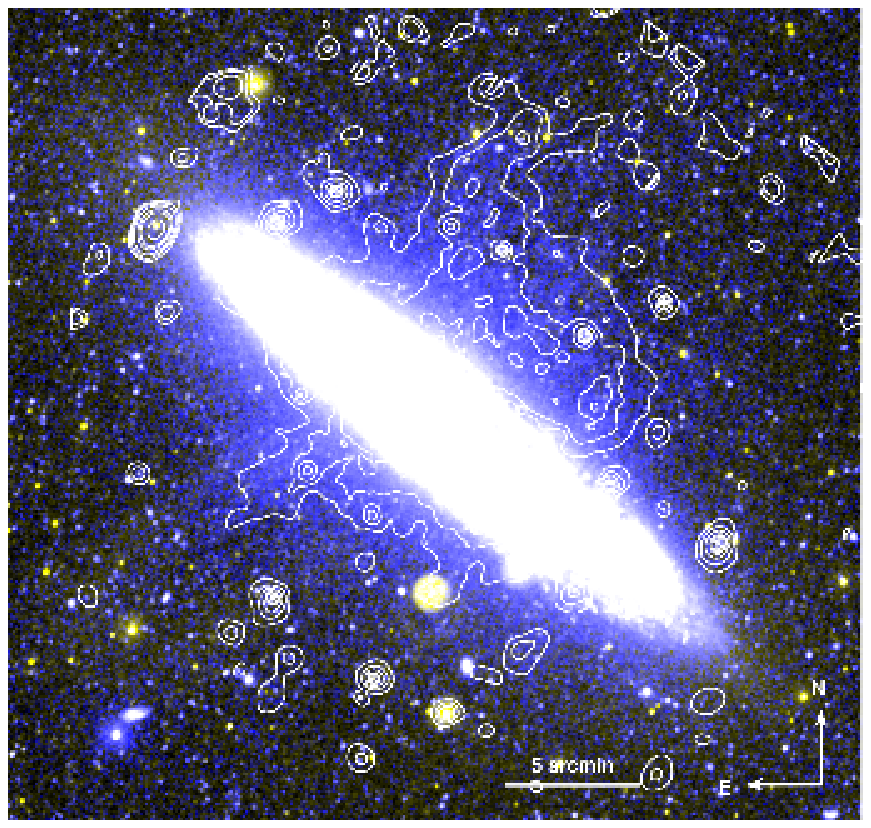}}
\caption{Two-colour UV image of \n253 with NUV (175--280~nm) in yellow and FUV (135--175~nm) in blue. The intensity was rescaled to emphasise the faint, diffuse emission. Overplotted are the 0.2--0.5~keV X-ray contours.}
\label{fig:UV}
\end{figure}

Fig.~\ref{fig:UV} shows the X-ray contours from the energy band 0.2--0.5~keV overplotted on a two-colour UV image, taken with the GALEX observatory (Galaxy Evolution Explorer, a UV space telescope) on 2003 October 13.
For the northwestern halo there is quite a good agreement between the FUV and X-ray emission regions.
The FUV emission traces the western horn to a distance of $\sim$7.5~kpc above the disc, as well as the broad base emission in soft X-rays quite well.
In the southeastern halo, again, the UV and the soft X-ray emission show the western horn structure, where the UV horn extends to about 7~kpc away from the disc.
However, the FUV horn is slightly offset by $\sim$700~pc to the northeast with respect to the X-ray horn.
Images obtained with the Optical Monitor onboard \xmm\ are not sensitive enough to show the extraplanar UV emission.

\cite{HH2005} proposed the following model for the origin of the UV emission:
Since the UV luminosities are too high to to be produced by continuum and line emission from photoionised or shock-heated gas, the UV emission could be explained by dust in the outflow that scatters the stellar continuum from the starburst into our line of sight.
They also found that the UV halo emission, as seen with GALEX, correlates with the H$\alpha$ emission, which could originate from gas that is photoionised by UV photons from the starburst.
The UV and H$\alpha$ emission would originate in the same cold regions in the halo.

How does the warm gas that is responsible for the \ha\ and UV emission get out into the halo?
There are two possibilities: either it has already been there from the beginning in the form of a cold and maybe clumpy halo component, or it was transported by the superwind and galactic fountains from the disc out into the halo.

There are models, where it is possible to drag up clouds of cold gas into the halo \citep[\eg model 3 of][]{SH2002}.
In a sheet surrounding these clouds, X-ray emission could be produced by shocks or in conductive or turbulent mixing interfaces on the cloud surface.
This model would also account for the non-uniformity of the X-ray emission as seen in the \xmm\ images (Fig.~\ref{fig:PN_images} and \ref{fig:PN_RGB_image}).
However, the model cannot explain the displacement of the UV emission in the southeastern halo, since in the model the clouds are located within or at the inner border of the superwind.

Could the dust even survive this transport from the disc into the halo embedded in a hot plasma environment?
\cite{DS1979} give the sputtering time for a spherical dust grain of radius $a$ embedded in a plasma of hydrogen with temperatures between \oexpo{6} and \oexpo{9}~K and the density $n_{\mathrm{H}}$ as
\begin{equation}
t_{\mathrm{sput}}\sim 10^6 \left(\frac{a}{\mu\mathrm{m}}\right)\left(\frac{n_{\mathrm{H}}}{\mathrm{cm}^{-3}}\right)^{-1}\,\mathrm{yr}.
\end{equation}
For $n_{\mathrm{H}}$ between $2.5\times10^{-2}$~cm$^{-3}$  in the outflow close to the centre \citep{BP2007} and $3.2\times10^{-3}$~cm$^{-3}$ out in the northwestern halo, and a grain size of $a$=0.1~$\mu$m, $t_{\mathrm{sput}}$ varies between 4.0 and 31~Myr.
So to reach a height above the disc of 7.5~kpc in less than 31~Myr, an average velocity of at least 240~km/s is required.
This is well compatible with measurements of outflow velocities in different wavelengths, that range from 260~km/s \citep[Na D absorption,][]{HL2000} to about 400--600~km/s \citep[\ha, \ion{N}{II}, \ion{S}{II}, and \ion{O}{II} emission,][]{U1978,DB1970}.
Therefore it is quite possible that the dust survives the transport from the disc out into the halo.

Another model to explain the UV and X-ray morphology \citep[\eg model 5 of][]{SH2002} requires a thick disc component, through which the superwind emerges into the halo.
On the contact surface between the hot superwind fluid and the cold thick disc material we get a heated layer through shocks and turbulent mixing where the X-rays are produced, surrounded on the outside by a colder layer where the UV emission originates.
The thick disc component was originally created by lifting material up from the disc through the star formation activity \citep[simulations by][]{RB1995}.
This model would easily explain the UV displacement from the X-rays, however we would only get a hollow cone with X-ray emission.
The latter is not what we see in the \xmm\ observations.
Though a mix of both models would be able to explain the observed morphology.

\bold{For completeness, the three other models that \cite{SH2002} present are:}

\bold{(i) Shocked clouds in the halo (model 2).
In this model preexisting neutral clouds are overrun by the superwind, which can explain the often asymmetrical patchiness as a result of the spatial distribution of the clouds.
However, no \ion{H}{I} structures outside the disc have been found by \cite{PC1991} and \cite{KW1995}.}

\bold{(ii) Hot swept-up shell of halo gas (model 4).
In this model a superbubble bounded by a shock expands into the low-density halo medium.
The X-ray radiation originates from the shocked ambient gas following the morphology of the surface of the superbubble.
Possible UV/\ha\ emission could originate from dense cool disc gas entrained in the wind.
Especially this last aspect is not supported by the observations.}

\bold{(iii) Cooling radiation from the wind (model 1).
As presented by \cite{SH2002} this model cannot explain the UV or \ha\ emission, and also does not explain the horn shaped morphology of both the X-ray and UV emission.}

The magnetohydrodynamics ISM simulations of \cite{AB2005} also shows a clumpy halo structure, characterised by turbulent mixing layers, which could explain the UV and X-ray filamentary structure. 
In some regions, the magnetic field forms loops surrounded by shells which may exhibit enhanced UV emission.

%%%%%%%%%%%%%%%%%%%%%%%%%%%%%%%%%%%%%%%%%%%%%%%%%%%%%%%%%%%%%%%%%%%%%%%%%%%%%%%%%%%%%%%%%%%%%%%%%%%%%%%%%%%%%%%%%%%%%%
%%%%%%%%%%%%%%%%%%%%%%%%%%%%%%%%%%%%%%%%%%%%%%%%%%%%%%%%%%%%%%%%%%%%%%%%%%%%%%%%%%%%%%%%%%%%%%%%%%%%%%%%%%%%%%%%%%%%%%

\section{Summary}
We have characterised the diffuse emission in \n253.
The disc extends 13.6~kpc along the major axis and shows emission between 0.2 and 10~keV.
The spectrum could be modelled with two thermal plasmas ($T_{\mathrm{cold}}=0.1-0.3$~keV and $T_{\mathrm{hot}}=0.3-0.9$~keV) with solar abundances plus a power law component and galactic foreground absorption.
The power law component may indicate an unresolved contribution from X-ray binaries in the disc.
The total luminosity of the diffuse emission in the disc is 2.4\ergs{39} (0.2-10.0~keV).
We found clear evidence for hot plasma in the disc.
The diffuse emission does not originate completely from an old stellar population.

The halo resembles a horn structure which reaches out to a projected height of $\sim$9~kpc perpendicular to the disc.
The halo emission on smaller scales seems not to be uniformly distributed, but shows a filamentary structure.
The southeastern halo is softer than the northwestern halo.
To model the spectra in the halo we needed two thermal plasmas ($T_{\mathrm{cold}}\sim0.1$~keV and $T_{\mathrm{hot}}\sim0.3$~keV) with solar abundances plus galactic foreground absorption, \bold{which is an indication for the non-equilibrium ionisation state of the halo plasma}.
The total luminosity of the diffuse emission is 8.4\ergs{38} and 2.3\ergs{38} (0.2-1.5~keV) in the northwestern and southeastern halo, respectively.
Densities computed to $3.2\,\eta^{-0.5}\times10^{-3}$~cm$^{-3}$ and $4.7\,\eta^{-0.5}\times10^{-3}$~cm$^{-3}$, with the volume filling factor $\eta$.
With solar abundances this implies total masses of $3.3\,\eta^{-0.5}\times 10^7$~M$_{\sun}$ and $1.8\,\eta^{-0.5}\times 10^7$~M$_{\sun}$ for the northwestern and southeastern halo, respectively.

A comparison between X-ray and UV emission showed that both originate from the same regions.
The UV emission is more extended in the southeastern halo, where it seems to form a shell around the X-ray emission.
%%%%%%%%%%%%%%%%%%%%%%%%%%%%%%%%%%%%%%%%%%%%%%%%%%%%%%%%%%%%%%%%%%%%%%%%%%%%%%%%%%%%%%%%%%%%%%%%%%%%%%%%%%%%%%%%%%%%%%
%%%%%%%%%%%%%%%%%%%%%%%%%%%%%%%%%%%%%%%%%%%%%%%%%%%%%%%%%%%%%%%%%%%%%%%%%%%%%%%%%%%%%%%%%%%%%%%%%%%%%%%%%%%%%%%%%%%%%%

\begin{acknowledgements}
\bold{\colblue{We thank the referee for providing constructive comments and help in improving the contents of this paper}.
We thank G. Szokoly for providing us with the image of \n253 from the Wide Field Imager, which is based on observations made with ESO Telescopes at the La Silla and Paranal Observatory.
We also thank Stefano Andreon for introducing us to the Bayesian Information Criterion}.
The \xmm\ project is supported by the Bundesministerium f\"ur Wirtschaft und Technologie/Deutsches Zen\-trum f\"ur Luft- und Raumfahrt (BMWI/DLR, FKZ 50 OX 0001), and the Max-Planck Society.
This research has made use of the NASA/IPAC Extragalactic Database (NED) which is operated by the Jet Propulsion Laboratory, California Institute of Technology, under contract with the National Aeronautics and Space Administration.
This research has made use of the SIMBAD database, operated at CDS, Strasbourg, France.
The GALEX data presented in this paper were obtained from the Multimission Archive at the Space Telescope Science Institute (MAST). 
STScI is operated by the Association of Universities for Research in Astronomy, Inc., under NASA contract NAS5-26555. 
Support for MAST for non-HST data is provided by the NASA Office of Space Science via grant NAG5-7584 and by other grants and contracts.
MB acknowledges support from the BMWI/DLR, FKZ 50 OR 0405.
\end{acknowledgements}

\bibliographystyle{aa}
\bibliography{papers}

\begin{thebibliography}{57}
\expandafter\ifx\csname natexlab\endcsname\relax\def\natexlab#1{#1}\fi

\bibitem[{{Bauer} {et~al.}(2007){Bauer}, {Pietsch}, {Trinchieri},
  {Breitschwerdt}, {Ehle}, \& {Read}}]{BP2007}
{Bauer}, M., {Pietsch}, W., {Trinchieri}, G., {et~al.} 2007, \aap, 467, 979

\bibitem[{{Bell} \& {de Jong}(2001)}]{BJ2001}
{Bell}, E.~F. \& {de Jong}, R.~S. 2001, \apj, 550, 212

\bibitem[{{Bell} {et~al.}(2003){Bell}, {McIntosh}, {Katz}, \&
  {Weinberg}}]{BM2003}
{Bell}, E.~F., {McIntosh}, D.~H., {Katz}, N., \& {Weinberg}, M.~D. 2003, \apjs,
  149, 289

\bibitem[{{Breitschwerdt} \& {Schmutzler}(1994)}]{BS1994}
{Breitschwerdt}, D. \& {Schmutzler}, T. 1994, \nat, 371, 774

\bibitem[{{Breitschwerdt} \& {Schmutzler}(1999)}]{BS1999}
{Breitschwerdt}, D. \& {Schmutzler}, T. 1999, \aap, 347, 650

\bibitem[{{Cappi} {et~al.}(1999){Cappi}, {Persic}, {Bassani}, {Franceschini},
  {Hunt}, {Molendi}, {Palazzi}, {Palumbo}, {Rephaeli}, \& {Salucci}}]{CP1999}
{Cappi}, M., {Persic}, M., {Bassani}, L., {et~al.} 1999, \aap, 350, 777

\bibitem[{{Carilli} {et~al.}(1992){Carilli}, {Holdaway}, {Ho}, \& {de
  Pree}}]{CH1992}
{Carilli}, C.~L., {Holdaway}, M.~A., {Ho}, P.~T.~P., \& {de Pree}, C.~G. 1992,
  \apjl, 399, L59

\bibitem[{{Comer{\'o}n} {et~al.}(2003){Comer{\'o}n}, {G{\'o}mez}, \&
  {Torra}}]{CG2003}
{Comer{\'o}n}, F., {G{\'o}mez}, A.~E., \& {Torra}, J. 2003, \aap, 400, 137

\bibitem[{{Dahlem} {et~al.}(2000){Dahlem}, {Parmar}, {Oosterbroek}, {Orr},
  {Weaver}, \& {Heckman}}]{DP2000}
{Dahlem}, M., {Parmar}, A., {Oosterbroek}, T., {et~al.} 2000, \apj, 538, 555

\bibitem[{{Dahlem} {et~al.}(1998){Dahlem}, {Weaver}, \& {Heckman}}]{DW1998}
{Dahlem}, M., {Weaver}, K.~A., \& {Heckman}, T.~M. 1998, \apjs, 118, 401

\bibitem[{{de Avillez} \& {Breitschwerdt}(2005)}]{AB2005}
{de Avillez}, M.~A. \& {Breitschwerdt}, D. 2005, \aap, 436, 585

\bibitem[{{Demoulin} \& {Burbidge}(1970)}]{DB1970}
{Demoulin}, M.~H. \& {Burbidge}, E.~M. 1970, \apj, 159, 799

\bibitem[{{den Herder} {et~al.}(2001){den Herder}, {Brinkman}, {Kahn},
  {Branduardi-Raymont}, {Thomsen}, {Aarts}, {Audard}, {Bixler}, {den Boggende},
  {Cottam}, {Decker}, {Dubbeldam}, {Erd}, {Goulooze}, {G{\"u}del}, {Guttridge},
  {Hailey}, {Janabi}, {Kaastra}, {de Korte}, {van Leeuwen}, {Mauche},
  {McCalden}, {Mewe}, {Naber}, {Paerels}, {Peterson}, {Rasmussen}, {Rees},
  {Sakelliou}, {Sako}, {Spodek}, {Stern}, {Tamura}, {Tandy}, {de Vries},
  {Welch}, \& {Zehnder}}]{HB2001}
{den Herder}, J.~W., {Brinkman}, A.~C., {Kahn}, S.~M., {et~al.} 2001, \aap,
  365, L7

\bibitem[{{Dickey} \& {Lockman}(1990)}]{DL1990}
{Dickey}, J.~M. \& {Lockman}, F.~J. 1990, \araa, 28, 215

\bibitem[{{Draine} \& {Salpeter}(1979)}]{DS1979}
{Draine}, B.~T. \& {Salpeter}, E.~E. 1979, \apj, 231, 77

\bibitem[{{Fabbiano}(1988)}]{F1988}
{Fabbiano}, G. 1988, \apj, 330, 672

\bibitem[{{Fabbiano} \& {Trin\-chieri}(1984)}]{FT1984}
{Fabbiano}, G. \& {Trin\-chieri}, G. 1984, \apj, 286, 491

\bibitem[{{Freeman} {et~al.}(2002){Freeman}, {Kashyap}, {Rosner}, \&
  {Lamb}}]{FK2002}
{Freeman}, P.~E., {Kashyap}, V., {Rosner}, R., \& {Lamb}, D.~Q. 2002, \apjs,
  138, 185

\bibitem[{{Freyberg} {et~al.}(2004){Freyberg}, {Briel}, {Dennerl}, {Haberl},
  {Hartner}, {Pfeffermann}, {Kendziorra}, {Kirsch}, \& {Lumb}}]{FB2004}
{Freyberg}, M.~J., {Briel}, U.~G., {Dennerl}, K., {et~al.} 2004, in X-Ray and
  Gamma-Ray Instrumentation for Astronomy XIII. Edited by Flanagan, Kathryn A.;
  Siegmund,Oswald H. W. Proceedings of the SPIE, Volume 5165., ed. K.~A.
  {Flanagan} \& O.~H.~W. {Siegmund}, 112--122

\bibitem[{{Gehrels}(1986)}]{G1986}
{Gehrels}, N. 1986, \apj, 303, 336

\bibitem[{{Heckman} {et~al.}(2000){Heckman}, {Lehnert}, {Strickland}, \&
  {Armus}}]{HL2000}
{Heckman}, T.~M., {Lehnert}, M.~D., {Strickland}, D.~K., \& {Armus}, L. 2000,
  \apjs, 129, 493

\bibitem[{{Hoopes} {et~al.}(2005){Hoopes}, {Heckman}, {Strickland}, {Seibert},
  {Madore}, {Rich}, {Bianchi}, {Gil de Paz}, {Burgarella}, {Thilker},
  {Friedman}, {Barlow}, {Byun}, {Donas}, {Forster}, {Jelinsky}, {Lee},
  {Malina}, {Martin}, {Milliard}, {Morrissey}, {Neff}, {Schiminovich},
  {Siegmund}, {Small}, {Szalay}, {Welsh}, \& {Wyder}}]{HH2005}
{Hoopes}, C.~G., {Heckman}, T.~M., {Strickland}, D.~K., {et~al.} 2005, \apjl,
  619, L99

\bibitem[{{Jansen} {et~al.}(2001){Jansen}, {Lumb}, {Altieri}, {Clavel}, {Ehle},
  {Erd}, {Gabriel}, {Guainazzi}, {Gondoin}, {Much}, {Munoz}, {Santos},
  {Schartel}, {Texier}, \& {Vacanti}}]{JL2001}
{Jansen}, F., {Lumb}, D., {Altieri}, B., {et~al.} 2001, \aap, 365, L1

\bibitem[{{Jarrett} {et~al.}(2003){Jarrett}, {Chester}, {Cutri}, {Schneider},
  \& {Huchra}}]{JC2003}
{Jarrett}, T.~H., {Chester}, T., {Cutri}, R., {Schneider}, S.~E., \& {Huchra},
  J.~P. 2003, \aj, 125, 525

\bibitem[{{Jeffreys}(1961)}]{J1961}
{Jeffreys}, H. 1961, {Theory of probability, 3rd edn} (Oxford Univ. Press,
  Oxford)

\bibitem[{{Koribalski} {et~al.}(1995){Koribalski}, {Whiteoak}, \&
  {Houghton}}]{KW1995}
{Koribalski}, B., {Whiteoak}, J.~B., \& {Houghton}, S. 1995, Publications of
  the Astronomical Society of Australia, 12, 20

\bibitem[{{Liddle}(2004)}]{L2004}
{Liddle}, A.~R. 2004, \mnras, 351, L49

\bibitem[{{Mason} {et~al.}(2001){Mason}, {Breeveld}, {Much}, {Carter},
  {Cordova}, {Cropper}, {Fordham}, {Huckle}, {Ho}, {Kawakami}, {Kennea},
  {Kennedy}, {Mittaz}, {Pandel}, {Priedhorsky}, {Sasseen}, {Shirey}, {Smith},
  \& {Vreux}}]{MB2001}
{Mason}, K.~O., {Breeveld}, A., {Much}, R., {et~al.} 2001, \aap, 365, L36

\bibitem[{{Mauersberger} {et~al.}(1995){Mauersberger}, {Henkel}, \&
  {Chin}}]{MH1995}
{Mauersberger}, R., {Henkel}, C., \& {Chin}, Y.-N. 1995, \aap, 294, 23

\bibitem[{{Monet} {et~al.}(2003){Monet}, {Levine}, {Canzian}, {Ables}, {Bird},
  {Dahn}, {Guetter}, {Harris}, {Henden}, {Leggett}, {Levison}, {Luginbuhl},
  {Martini}, {Monet}, {Munn}, {Pier}, {Rhodes}, {Riepe}, {Sell}, {Stone},
  {Vrba}, {Walker}, {Westerhout}, {Brucato}, {Reid}, {Schoening}, {Hartley},
  {Read}, \& {Tritton}}]{ML2003}
{Monet}, D.~G., {Levine}, S.~E., {Canzian}, B., {et~al.} 2003, \aj, 125, 984

\bibitem[{{Mukherjee} {et~al.}(1998){Mukherjee}, {Feigelson}, {Jogesh Babu},
  {Murtagh}, {Fraley}, \& {Raftery}}]{MF1998}
{Mukherjee}, S., {Feigelson}, E.~D., {Jogesh Babu}, G., {et~al.} 1998, \apj,
  508, 314

\bibitem[{{Pence}(1980)}]{P1980}
{Pence}, W.~D. 1980, \apj, 239, 54

\bibitem[{{Pietsch}(1992)}]{P1992}
{Pietsch}, W. 1992, in Physics of Nearby Galaxies: Nature or Nurture?, ed.
  T.~X. {Thuan}, C.~{Balkowski}, \& J.~{Tran Thanh van}, 67

\bibitem[{{Pietsch} {et~al.}(2001){Pietsch}, {Roberts}, {Sako}, {Freyberg},
  {Read}, {Borozdin}, {Branduardi-Raymont}, {Cappi}, {Ehle}, {Ferrando},
  {Kahn}, {Ponman}, {Ptak}, {Shirey}, \& {Ward}}]{PR2001}
{Pietsch}, W., {Roberts}, T.~P., {Sako}, M., {et~al.} 2001, \aap, 365, L174

\bibitem[{{Pietsch} {et~al.}(2000){Pietsch}, {Vogler}, {Klein}, \&
  {Zinnecker}}]{PV2000}
{Pietsch}, W., {Vogler}, A., {Klein}, U., \& {Zinnecker}, H. 2000, \aap, 360,
  24

\bibitem[{{Protassov} {et~al.}(2002){Protassov}, {van Dyk}, {Connors},
  {Kashyap}, \& {Siemiginowska}}]{PD2002}
{Protassov}, R., {van Dyk}, D.~A., {Connors}, A., {Kashyap}, V.~L., \&
  {Siemiginowska}, A. 2002, \apj, 571, 545

\bibitem[{{Ptak} {et~al.}(1997){Ptak}, {Serlemitsos}, {Yaqoob}, {Mushotzky}, \&
  {Tsuru}}]{PS1997}
{Ptak}, A., {Serlemitsos}, P., {Yaqoob}, T., {Mushotzky}, R., \& {Tsuru}, T.
  1997, \aj, 113, 1286

\bibitem[{{Puche} {et~al.}(1991){Puche}, {Carignan}, \& {van Gorkom}}]{PC1991}
{Puche}, D., {Carignan}, C., \& {van Gorkom}, J.~H. 1991, \aj, 101, 456

\bibitem[{{Read} {et~al.}(1997){Read}, {Ponman}, \& {Strickland}}]{RP1997}
{Read}, A.~M., {Ponman}, T.~J., \& {Strickland}, D.~K. 1997, \mnras, 286, 626

\bibitem[{{Revnivtsev} {et~al.}(2007){Revnivtsev}, {Churazov}, {Sazonov},
  {Forman}, \& {Jones}}]{RC2007}
{Revnivtsev}, M., {Churazov}, E., {Sazonov}, S., {Forman}, W., \& {Jones}, C.
  2007, \aap, 473, 783

\bibitem[{{Revnivtsev} {et~al.}(2006){Revnivtsev}, {Sazonov}, {Gilfanov},
  {Churazov}, \& {Sunyaev}}]{RS2006}
{Revnivtsev}, M., {Sazonov}, S., {Gilfanov}, M., {Churazov}, E., \& {Sunyaev},
  R. 2006, \aap, 452, 169

\bibitem[{{Rosen} \& {Bregman}(1995)}]{RB1995}
{Rosen}, A. \& {Bregman}, J.~N. 1995, \apj, 440, 634

\bibitem[{{Sazonov} {et~al.}(2006){Sazonov}, {Revnivtsev}, {Gilfanov},
  {Churazov}, \& {Sunyaev}}]{SR2006}
{Sazonov}, S., {Revnivtsev}, M., {Gilfanov}, M., {Churazov}, E., \& {Sunyaev},
  R. 2006, \aap, 450, 117

\bibitem[{{Schlegel} {et~al.}(1998){Schlegel}, {Finkbeiner}, \&
  {Davis}}]{SF1998}
{Schlegel}, D.~J., {Finkbeiner}, D.~P., \& {Davis}, M. 1998, \apj, 500, 525

\bibitem[{{Schwarz}(1978)}]{S1978}
{Schwarz}, U.~J. 1978, \aap, 65, 345

\bibitem[{{Smith} {et~al.}(2001){Smith}, {Brickhouse}, {Liedahl}, \&
  {Raymond}}]{SBL2001}
{Smith}, R.~K., {Brickhouse}, N.~S., {Liedahl}, D.~A., \& {Raymond}, J.~C.
  2001, \apjl, 556, L91

\bibitem[{{Snowden} {et~al.}(2004){Snowden}, {Collier}, \& {Kuntz}}]{SC2004}
{Snowden}, S.~L., {Collier}, M.~R., \& {Kuntz}, K.~D. 2004, \apj, 610, 1182

\bibitem[{{Strickland} {et~al.}(2004{\natexlab{a}}){Strickland}, {Heckman},
  {Colbert}, {Hoopes}, \& {Weaver}}]{SH2004A}
{Strickland}, D.~K., {Heckman}, T.~M., {Colbert}, E.~J.~M., {Hoopes}, C.~G., \&
  {Weaver}, K.~A. 2004{\natexlab{a}}, \apjs, 151, 193

\bibitem[{{Strickland} {et~al.}(2004{\natexlab{b}}){Strickland}, {Heckman},
  {Colbert}, {Hoopes}, \& {Weaver}}]{SH2004B}
{Strickland}, D.~K., {Heckman}, T.~M., {Colbert}, E.~J.~M., {Hoopes}, C.~G., \&
  {Weaver}, K.~A. 2004{\natexlab{b}}, \apj, 606, 829

\bibitem[{{Strickland} {et~al.}(2002){Strickland}, {Heckman}, {Weaver},
  {Hoopes}, \& {Dahlem}}]{SH2002}
{Strickland}, D.~K., {Heckman}, T.~M., {Weaver}, K.~A., {Hoopes}, C.~G., \&
  {Dahlem}, M. 2002, \apj, 568, 689

\bibitem[{{Str{\"u}der} {et~al.}(2001){Str{\"u}der}, {Briel}, {Dennerl},
  {Hartmann}, {Kendziorra}, {Meidinger}, {Pfeffermann}, {Reppin}, {Aschenbach},
  {Bornemann}, {Br{\"a}uninger}, {Burkert}, {Elender}, {Freyberg}, {Haberl},
  {Hartner}, {Heuschmann}, {Hippmann}, {Kastelic}, {Kemmer}, {Kettenring},
  {Kink}, {Krause}, {M{\"u}ller}, {Oppitz}, {Pietsch}, {Popp}, {Predehl},
  {Read}, {Stephan}, {St{\"o}tter}, {Tr{\"u}mper}, {Holl}, {Kemmer}, {Soltau},
  {St{\"o}tter}, {Weber}, {Weichert}, {von Zanthier}, {Carathanassis}, {Lutz},
  {Richter}, {Solc}, {B{\"o}ttcher}, {Kuster}, {Staubert}, {Abbey}, {Holland},
  {Turner}, {Balasini}, {Bignami}, {La Palombara}, {Villa}, {Buttler},
  {Gianini}, {Lain{\'e}}, {Lumb}, \& {Dhez}}]{SBD2001}
{Str{\"u}der}, L., {Briel}, U., {Dennerl}, K., {et~al.} 2001, \aap, 365, L18

\bibitem[{{Turner} {et~al.}(2001){Turner}, {Abbey}, {Arnaud}, {Balasini},
  {Barbera}, {Belsole}, {Bennie}, {Bernard}, {Bignami}, {Boer}, {Briel},
  {Butler}, {Cara}, {Chabaud}, {Cole}, {Collura}, {Conte}, {Cros}, {Denby},
  {Dhez}, {Di Coco}, {Dowson}, {Ferrando}, {Ghizzardi}, {Gianotti}, {Goodall},
  {Gretton}, {Griffiths}, {Hainaut}, {Hochedez}, {Holland}, {Jourdain},
  {Kendziorra}, {Lagostina}, {Laine}, {La Palombara}, {Lortholary}, {Lumb},
  {Marty}, {Molendi}, {Pigot}, {Poindron}, {Pounds}, {Reeves}, {Reppin},
  {Rothenflug}, {Salvetat}, {Sauvageot}, {Schmitt}, {Sembay}, {Short},
  {Spragg}, {Stephen}, {Str{\"u}der}, {Tiengo}, {Trifoglio}, {Tr{\"u}mper},
  {Vercellone}, {Vigroux}, {Villa}, {Ward}, {Whitehead}, \& {Zonca}}]{TA2001}
{Turner}, M.~J.~L., {Abbey}, A., {Arnaud}, M., {et~al.} 2001, \aap, 365, L27

\bibitem[{{Ulrich}(1978)}]{U1978}
{Ulrich}, M.-H. 1978, \apj, 219, 424

\bibitem[{{Vogler} \& {Pietsch}(1999)}]{VP1999}
{Vogler}, A. \& {Pietsch}, W. 1999, \aap, 342, 101

\bibitem[{{Weaver} {et~al.}(2000){Weaver}, {Heckman}, \& {Dahlem}}]{WH2000}
{Weaver}, K.~A., {Heckman}, T.~M., \& {Dahlem}, M. 2000, \apj, 534, 684

\bibitem[{{Weaver} {et~al.}(2002){Weaver}, {Heckman}, {Strickland}, \&
  {Dahlem}}]{WH2002}
{Weaver}, K.~A., {Heckman}, T.~M., {Strickland}, D.~K., \& {Dahlem}, M. 2002,
  \apjl, 576, L19

\bibitem[{{Wilms} {et~al.}(2000){Wilms}, {Allen}, \& {McCray}}]{WA2000}
{Wilms}, J., {Allen}, A., \& {McCray}, R. 2000, \apj, 542, 914

\end{thebibliography}

%%%%%%%%%%%%%%%%%%%%%%%%%%%%%%%%%%%%%%%%%%%%%%%%%%%%%%%%%%%%%%%%%%%%%%%%%%%%%%%%%%%%%%%%%%%%%%%%%%%%%%%%%%%%%%%%%%%%%%
%%%%%%%%%%%%%%%%%%%%%%%%%%%%%%%%%%%%%%%%%%%%%%%%%%%%%%%%%%%%%%%%%%%%%%%%%%%%%%%%%%%%%%%%%%%%%%%%%%%%%%%%%%%%%%%%%%%%%%

\appendix

\section{EPIC~pn images}
\label{app:images}
We developed an algorithm to create vignetting corrected and adaptively smoothed EPIC~pn images.
This procedure is based on the observation itself.
Specifically, closed filter observations were not used in addition.
In the following we will describe the algorithm step-by-step.

The basis for this procedure is a cleaned event file and an out-of-time event file.
This cleaning included screening for high background and also removing bad pixels and bad columns.
If we want to use more than one observation, all steps have to be performed for all observations separately, before the products are merged.

We will here concentrate on the creation of three images, which can be combined to a RGB colour image at the end of the procedure.
In the following all steps are to be done for all three energy bands, unless stated otherwise.

For the desired energy band, we extracted an image from the event file.
This image was then corrected for out-of-time events, via subtraction of an image that was extracted from the out-of-time event file and rescaled with the out-of-time event fraction.
Next, we corrected for the detector background (electronic noise, high energy particles) by subtracting the detector background surface brightness from the image.
This value was determined from the corners of the detector which are outside of the field of view of the telescope.
We here assumed that the detector background is uniform across the whole detector \citep[for energies above $\sim$7.2~keV this is no longer a good approximation, see][]{FB2004}.

We created vignetting corrected exposure maps and masks using the {\tt SAS}-tasks {\tt eexpmap} and {\tt emask}, respectively, which will be used to account for differences in the exposure times in the images, and to mask the images to regions with an acceptable minimum exposure time.
Before we could smooth the image adaptively, we had to create a template with smoothing kernels.
This template guarantees that images in different energy bands are smoothed with the same kernel size.
Therefore we added up the images in the different energy bands.
With the task {\tt asmooth}, we created the template using the merged image, the merged mask, and an exposure map.

Now we have all the necessary products to smooth the single images.
We smoothed the images in the different energy bands with the smoothing template, the vignetting corrected exposure map, and with the corresponding mask.
This step included the vignetting correction via the exposure map, and a masking of the image to a region with an acceptable minimum exposure time.

In a final step we used {\tt ds9} to create the RGB colour image.
The resulting images of \n253 and a combined RGB colour image of the lowest three energy bands is shown in Fig.~\ref{fig:PN_images} and \ref{fig:PN_RGB_image}.

%%%%%%%%%%%%%%%%%%%%%%%%%%%%%%%%%%%%%%%%%%%%%%%%%%%%%%%%%%%%%%%%%%%%%%%%%%%%%%%%%%%%%%%%%%%%%%%%%%%%%%%%%%%%%%%%%%%%%%
\section{Background spectra}
\label{app:bkgspec}

\begin{figure*}[t]
\begin{minipage}{\textwidth}
\begin{equation}
\label{EQ:background}  
  \begin{split}
    B(E)&=\underbrace{S^{\mathrm{OOT}}_{\mathrm{obs}}(E)\times f}_{\mathrm{Out-of-Time\,\,\,events}}+
    \underbrace{S_{\mathrm{det}}(E)\frac{R_{\mathrm{obs}}}{R_{\mathrm{det}}}\frac{t_{\mathrm{obs}}}{t_{\mathrm{det}}}}_{\mathrm{detector\,\,\,background}}-
    \underbrace{S^{\mathrm{OOT}}_{\mathrm{det}}(E)\frac{R_{\mathrm{obs}}}{R_{\mathrm{det}}}\frac{t_{\mathrm{obs}}}{t_{\mathrm{det}}}\times f}_{\mathrm{detector\,\,Out-of-Time\,\,events}}+\\
    &+\underbrace{\frac{V(E,\theta_S)}{V(E,\theta_B)}\frac{A_S}{A_B} \times
    \left(
    B_{\mathrm{obs}}(E)- 
    \underbrace{B^{\mathrm{OOT}}_{\mathrm{obs}}(E)\times f}_{\mathrm{Out-of-Time\,\,\,events}}-
    \underbrace{B_{\mathrm{det}}(E)\frac{R_{\mathrm{obs}}}{R_{\mathrm{det}}}\frac{t_{\mathrm{obs}}}{t_{\mathrm{det}}}}_{\mathrm{detector\,\,\,background}}+
    \underbrace{B^{\mathrm{OOT}}_{\mathrm{det}}(E)\frac{R_{\mathrm{obs}}}{R_{\mathrm{det}}}\frac{t_{\mathrm{obs}}}{t_{\mathrm{det}}}\times f}_{\mathrm{detector\,\,Out-of-time\,\,events}} \right)}_{\mathrm{sky\,\,background}}\\
  \end{split}
\end{equation}
\end{minipage}
\end{figure*}

The conventional way to create a background spectrum is to select a region from the same observation where there is no emission from the source.
Additionally, the region should be close to the source.
This way, the spectral background should have the same characteristics as the background at the source region.
In \n253, a region which suffices the first criterion can be found at the border of the field of view in the southwestern part of the detector.
The second criterion, however, is not satisfied.
The background region may show a different detector background, and additionally the vignetting is different.

\begin{table}[b]
\centering
\caption{Rejected CCD rows due to MIPs per time unit in the used observations.}
\begin{tabular}{ccc}
\hline\noalign{\smallskip}
\hline\noalign{\smallskip}
Obs ID & Filter & rejected line counter value\\
\hline
0122320707 & Closed & 181.7\\
0125960101 & Medium & 190.2\\
\hline
0152020101 & Thin & 141.4\\
0160362801 & Closed & 120.4\\
\hline
\end{tabular}
\label{tab:RejLineCount}
\end{table}

Since we were interested in determining the characteristics of emission with low surface brightness, that extends over a large region, where the background is (probably) the dominant component, we needed a very accurate estimate of the background.
Given the very soft nature of the emission, we cannot use blank sky observations that were taken in regions of the sky where the foreground \nh\ is different (not to mention other uncertainties due to different detector settings, particle radiation levels, etc.).
Here we describe a method to use a local estimate of the sky background that takes properly into account vignetting and detector background issues.

To remove the detector background, we used archival observations which were taken in the same mode as the \n253\ observations, but where the filter wheel was closed.
To avoid effects due to changes in the detector settings, or changes of the detector performance due to other reasons, we chose the closed observations to be as close as possible in time to the \n253\ observations.
The closed observations we used for observation 1 and 4 are: revolution 59, obs.\ id.\ 0122320701, exposure S003 (50.5~ks) and revolution 732, obs.\ id.\ 0160362801, exposure S005 (38.6~ks), respectively.
To ensure, that there are as little as possible differences between the source observation and the closed observation, we removed bad columns and bad pixels both in the \n253\ and closed observation.
Additionally, the closed observations may have been taken when the spacecraft was exposed to a different particle radiation level than the one present during observations 1 or 4.
The \xmm\ house keeping file contains information on how many CCD rows per time unit were rejected due to a possible minimum ionising particle (MIP) event, which is a direct estimator of the average radiation level.
We used these values (see Table~\ref{tab:RejLineCount}), to rescale the count rate of the closed observations.

We used Out-of-Time spectra from the source and background region to correct for contribution from Out-of-Time events.
When one subtracts a closed observation spectrum from a Out-of-Time corrected spectrum, one actually removes the Out-of-Time spectrum of the detector background twice.
This is corrected in our method by adding again the Out-of-Time spectra of the detector background.

We corrected the background region spectrum for Out-of-Time events and the detector background and applied the vignetting correction in each energy bin as a function of off-axis angle of the source and background spectrum.
This gave us the sky background spectrum.

\begin{figure*}
  \centering
  \fig{\includegraphics[width=8.9cm]{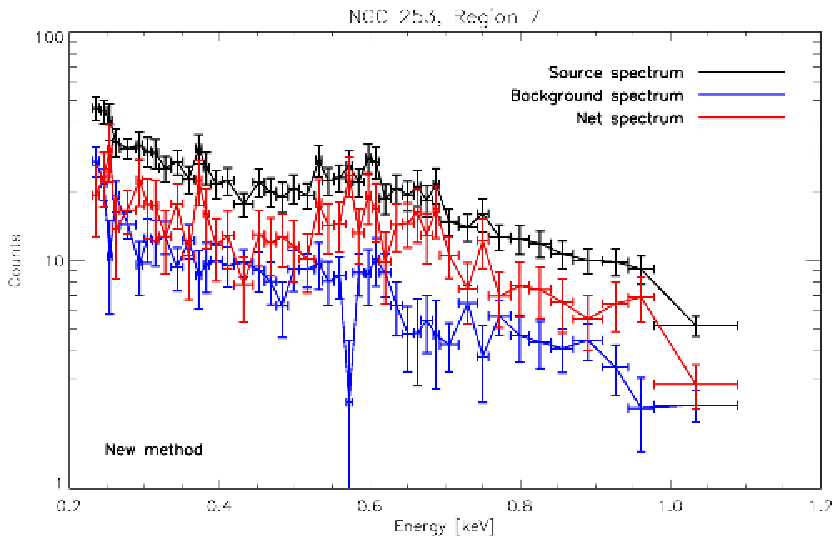}
  \includegraphics[width=8.9cm]{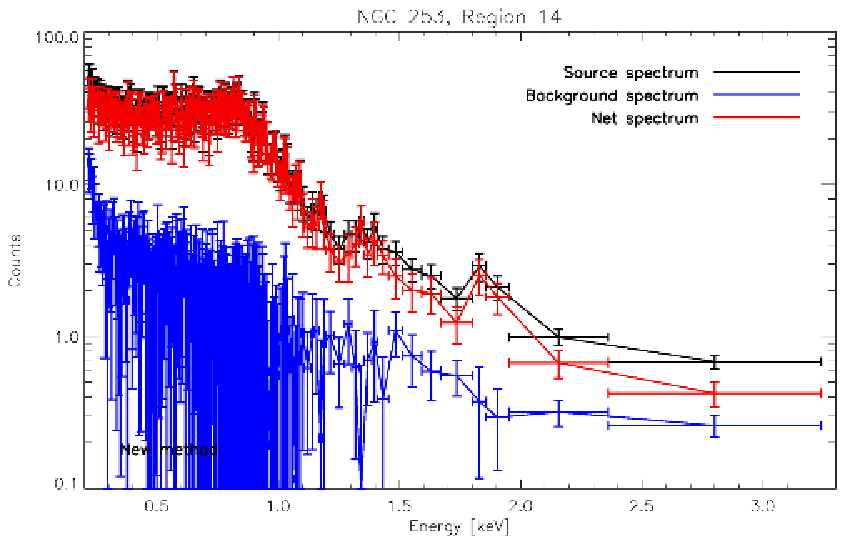}
  \includegraphics[width=8.9cm]{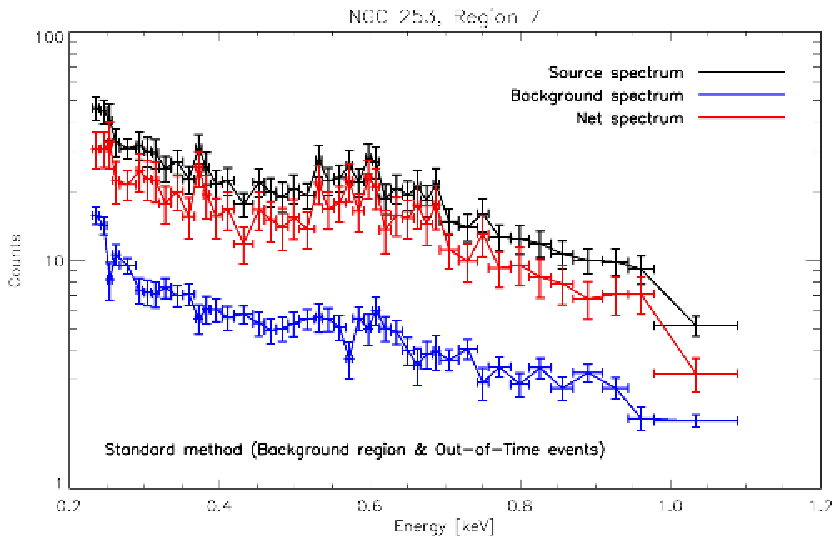}
  \includegraphics[width=8.9cm]{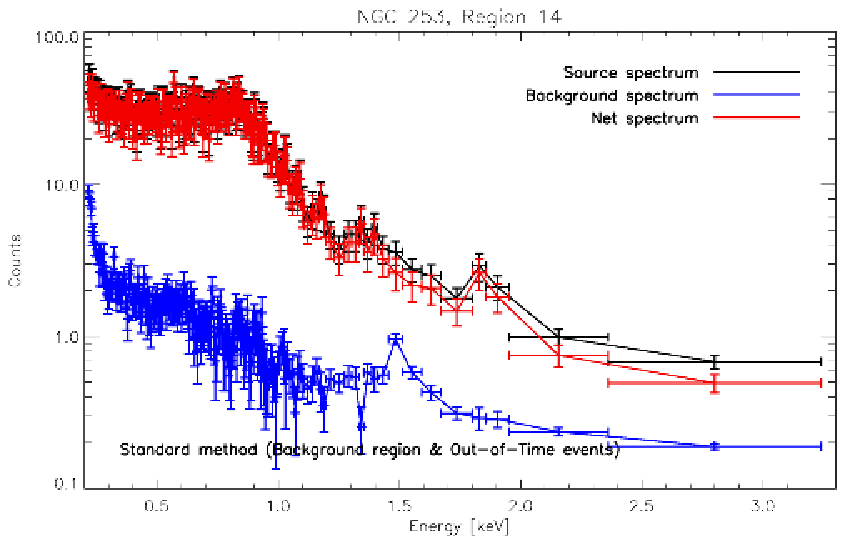}}
  \caption{Comparison between the background substraction on two examples\comment{ (same source regions as in Fig~\ref{fig:spectra})}. {\it (left):} the spectrum of region~7, {\it (right):} the spectrum of region~14. The top panel shows the new method, as described in this paper, the bottom panel shows the conventional method, where the raw background spectrum is used, and a correction for Out-of-Time events has been applied. We only show the spectra of observation 4 here, since these have the better statistics.}
  \label{fig:bkgspec}
\end{figure*}

In all of the above steps, different exposure times and areas in the extraction regions have been accounted for.
Since some of the components in the final background spectrum do have low number statistics, we used the conservative approximation to Poissonian errors $\sigma_N\approx 1+\sqrt{0.75+N}$ \citep{G1986}.
To avoid unjustified large errors, we roughly binned the spectrum before calculating errors.
\bold{This binning to a significance of $>3\,\sigma$ in each bin is performed on the conventionally background subtracted source spectrum (no vignetting correction and no closed observations) and guarantees that the final} background subtracted spectrum then has a significance in each bin of a least 3~$\sigma$.
The errors were propagated properly and were included in the file with the final background spectrum.
This spectrum can be used with XSPEC as a background spectrum.

The whole method can be summarised by Eq.~\ref{EQ:background} with the following symbols:
\begin{itemize}
\item[--] $B(E)$ is the counts at energy $E$ in the background spectrum
\item[--] $B_{\mathrm{obs}}(E)$ is the counts in the \n253\ observation
\item[--] $S_{\mathrm{det}}(E)$ is the counts from the detector background spectrum in the source region
\item[--] $B_{\mathrm{obs}}(E)$ is the counts in the detector background spectrum in the background region
\item[--] $S^{\mathrm{OOT}}_{\mathrm{obs}}(E)$ are the counts in the Out-of-Time spectra in the source region
\item[--] $B^{\mathrm{OOT}}_{\mathrm{obs}}(E)$ are the counts in the Out-of-Time spectra in the background region
\item[--] $t_{\mathrm{obs}}$ is the exposure time in the \n253\ observation
\item[--] $t_{\mathrm{det}}$ is the exposure time in the closed observation
\item[--] $R_{\mathrm{obs}}$ is the rejected line counter values (see Table~\ref{tab:RejLineCount}) in the \n253\ observation
\item[--] $R_{\mathrm{det}}$ is the rejected line counter values (see Table~\ref{tab:RejLineCount}) in the closed observation
\item[--] $A_S$ is the area in the source region
\item[--] $A_B$ is the area in the background region
\item[--] $V(E,\theta_S)$ is the vignetting value in the source region, depending on the offset angle $\theta$ and the energy $E$
\item[--] $V(E,\theta_B)$ is the vignetting value in the background region, depending on the offset angle $\theta$ and the energy $E$
\item[--] $f$ is the fraction of Out-of-Time events in the corresponding mode of the observation
\end{itemize}

\begin{figure*}
  \fig{\includegraphics[width=8.9cm]{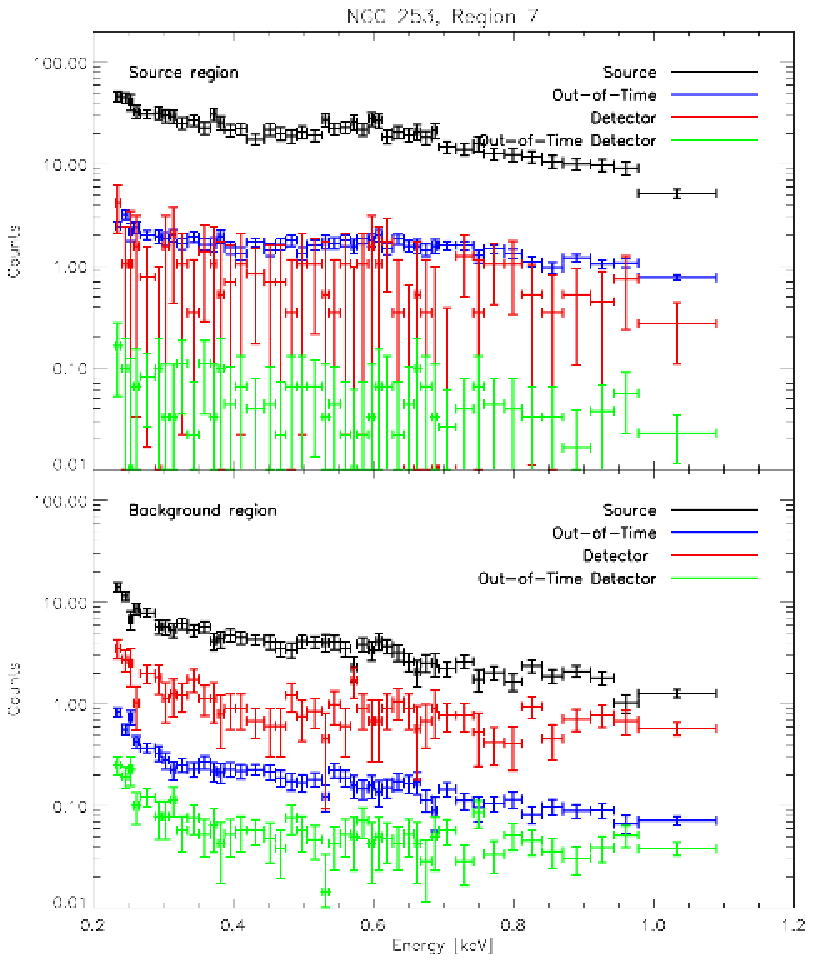}
  \includegraphics[width=8.9cm]{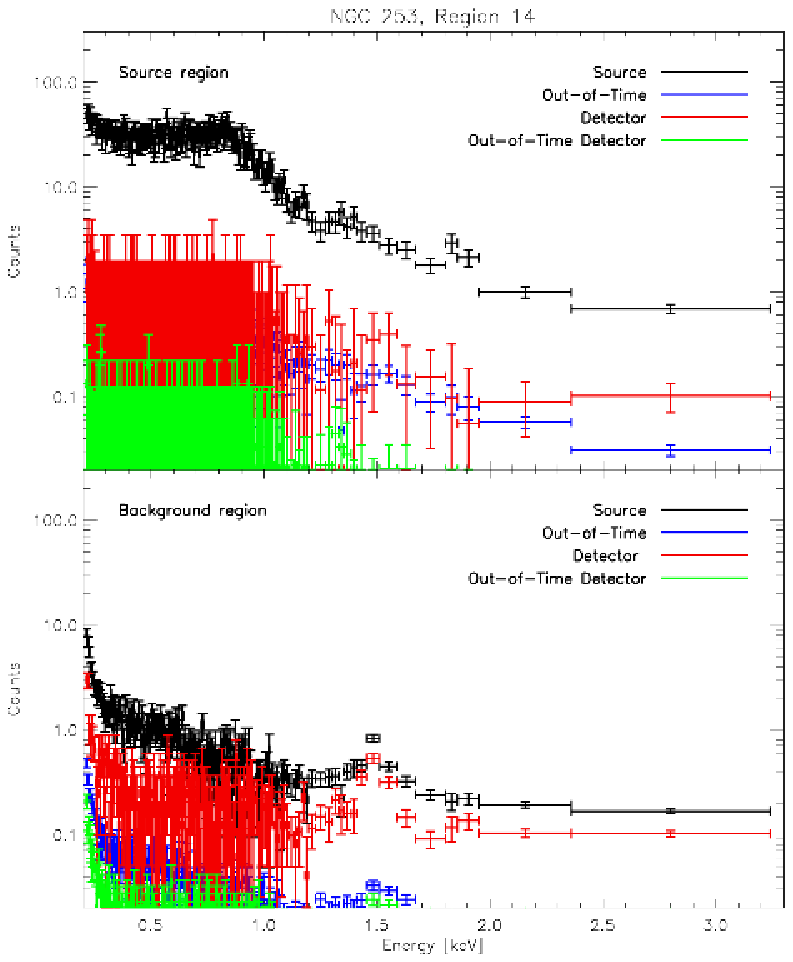}}
  \caption{The single components that are part of the total background spectrum compared to the source spectrum. {\it (left):} Region~7, {\it (right):} Region~14, {\it (top):} components from the source region, {\it (bottom):} components from the background region. The single components were corrected for areas, exposure time, and radiation level, with respect to the source spectrum in the source region, but no vignetting correction was applied yet.}
  \label{fig:bkgspec_comp}
\end{figure*}

A comparison between this new method and the conventional method, that does not use the vignetting correction nor the closed observations, is shown in Fig.~\ref{fig:bkgspec} for two example spectra, both in observations 1 and 4.
The single background components in the source and background region in observation 4 are shown in Fig.~\ref{fig:bkgspec_comp}.
All figures show counts integrated over the extraction region.
The counts in the background region were rescaled to the source region area to be able to compare them to the source spectrum. 
Also, the counts in the closed observation were rescaled to the exposure time and radiation level in the source observation.

The differences between the new and the conventional method in terms of the resulting best fits are the following:
In the majority of the tested cases, an additional power law component with $\Gamma\sim0$ is required for the fit in the spectrum, obtained with the conventional method.
The temperatures are consistent between both methods, but the resulting flux levels in the conventional method are higher.
Differences in total flux values range between 2\% and 22\%.
The effect between the two methods is highest in regions with low surface brightness.
Here the background dominates and a correct treatment is crucial.
As an example, the difference in flux in region~7 (low surface brightness) is 15\% and 22\%, for observations 1 and 4, respectively.
Whereas in region~14 (high surface brightness), the differences are 2\% and 3\%.

\end{document}